\documentclass[aps, prl, superscriptaddress, reprint, twocolumn]{revtex4-2}
\usepackage[T1]{fontenc}
\usepackage[utf8]{inputenc}
\usepackage{graphicx}

\usepackage{amsmath, amssymb, amstext, amsfonts, amsthm, braket}
\usepackage{mathtools}
\usepackage{dsfont}
\usepackage{graphicx}
\usepackage[dvipsnames]{xcolor}
\usepackage{float}
\usepackage{verbatim}
\usepackage{ragged2e}
\usepackage[caption=false, justification=justified]{subfig}
\DeclareCaptionJustification{justified}{\justifying}
\usepackage{placeins}
\usepackage{standalone}
\usepackage{tikz}
\usepackage{enumitem}
\usepackage{microtype}

\usepackage{subdepth}   
\usepackage[colorlinks=true, linkcolor = BlueViolet, citecolor = BlueViolet, urlcolor = BlueViolet]{hyperref}

\DeclareMathOperator{\Tr}{Tr}
\DeclareMathOperator{\swap}{SWAP}
\DeclareMathOperator{\vecspan}{span}

\usepackage{stix2} 

\newcommand{\edwin}[1]{\textcolor{purple}{edwin: #1}}

\newcommand{\edit}[1]{{#1}}

\begin{document}
\title{Bidirectional teleportation using scrambling dynamics: a practical protocol}

\author{Amit Vikram}
\email[Corresponding author:]{AmitVikram.Anand@colorado.edu}
\affiliation{JILA, University of Colorado and National Institute of Standards and Technology, and Department of Physics, University of Colorado, Boulder, Colorado 80309, USA}
\affiliation{Center for Theory of Quantum Matter and Department of Physics, University of Colorado, Boulder, Colorado 80309, USA}

\author{Edwin Chaparro}
\affiliation{JILA, University of Colorado and National Institute of Standards and Technology, and Department of Physics, University of Colorado, Boulder, Colorado 80309, USA}
\affiliation{Center for Theory of Quantum Matter and Department of Physics, University of Colorado, Boulder, Colorado 80309, USA}

\author{Muhammad Miskeen Khan}
\affiliation{JILA, University of Colorado and National Institute of Standards and Technology, and Department of Physics, University of Colorado, Boulder, Colorado 80309, USA}
\affiliation{Center for Theory of Quantum Matter and Department of Physics, University of Colorado, Boulder, Colorado 80309, USA}
\affiliation{Department of Electrical and Computer Engineering,
Saint Louis University, St. Louis, Missouri, 63103, USA}

\author{Andrew Lucas}
\affiliation{Center for Theory of Quantum Matter and Department of Physics, University of Colorado, Boulder, Colorado 80309, USA}

\author{Chris Akers}
\affiliation{Center for Theory of Quantum Matter and Department of Physics, University of Colorado, Boulder, Colorado 80309, USA}

\author{Ana Maria Rey}
\affiliation{JILA, University of Colorado and National Institute of Standards and Technology, and Department of Physics, University of Colorado, Boulder, Colorado 80309, USA}
\affiliation{Center for Theory of Quantum Matter and Department of Physics, University of Colorado, Boulder, Colorado 80309, USA}

\begin{abstract}

We show that quantum information scrambling can enable a generic SWAP gate between collective degrees of freedom in systems without universal local control. Our protocol combines \edit{a} Hayden–Preskill \edit{type} recovery scheme,  associated with the black hole information paradox, with quantum teleportation and runs them in parallel and in opposite directions, enabling bidirectional exchange of quantum states through global interactions alone. This approach cleanly distinguishes the roles of information spreading, entanglement, and chaos for  enabling both coherent state transfer and recovery. We propose an experimental realization using the Dicke model, which can be realized in optical cavity and trapped-ion platforms, highlighting the  utility of holography in designing practical quantum gates.
\end{abstract}

\maketitle

\textit{Introduction---} 
The holographic principle, most concretely realized in the AdS/CFT correspondence~\cite{Maldacena1997}, posits a surprising equivalence between a higher-dimensional gravitational system and a lower-dimensional quantum many-body system living on its boundary. This duality has reshaped our understanding of quantum gravity and revealed deep structural connections between gravitational dynamics and quantum information processing~\cite{Preskill2000,Harlow2016, TakayanagiEssay}. One of the most influential examples is the Hayden--Preskill thought experiment~\cite{HaydenPreskill}, which models how information thrown into an ``old'' black hole, well past the midpoint of its evaporation time~\cite{PageSubsystem, PageTime}, can be rapidly scrambled among its microscopic degrees of freedom, yet still be efficiently recoverable from a small amount of outgoing Hawking radiation. This phenomenon highlights two remarkable lessons: complex quantum dynamics can efficiently redistribute
quantum information, and scrambled quantum information can remain operationally useful when accompanied by an appropriate decoding protocol.


These insights have motivated a broader program utilizing scrambling ~\cite{HaydenPreskill, SekinoSusskind, LashkariFastScrambling, BentsenGuLucasScrambling, dynamicalqspeedlimit, dynamicalqfastscrambling}  for more general applications of fundamental interest including  benchmarking, demonstrations of quantum computational advantage ~\cite{GoogleScrambling,Zhu2022,BoixoSupremacy, Aharonov2023},   metrology~\cite{ButterflyMetrology, Zhang2025,Li2023} and  teleportation protocols to differentiate unitary scrambling from decoherence~\cite{YoshidaKitaev, YoshidaYao, Verifiedscrambling, ScramblingQutrits}
or to diagnose whether a system can admit a holographic description~\cite{GaoJafferisWall, Brown2023}. Yet, despite intense study, the  immediate technological potential of such holography-inspired protocols remains largely unexplored. 

Here, we demonstrate that scrambling itself can mediate a bidirectional quantum information transfer~\cite{bidirectionalteleportation1, bidirectionalteleportation2} --- effectively implementing a SWAP gate~\cite{BQT_SWAP1, BQT_SWAP2}, which coherently exchanges quantum states between two subsystems --- in generic settings where no explicit SWAP operations or universal local control exist.
This is of significant practical relevance, as up to date, no practical method is known for implementing SWAP-like gates using collective dynamics alone, i.e. without local control of qubits, except for highly polarized initial states whose interactions are effectively bilinear~\cite{Khan2025, WamplerCooperPolarizedSWAP}.

We realize this SWAP operation by \edit{introducing a black hole-inspired information recovery protocol closely related to the} Hayden--Preskill protocol~\cite{HaydenPreskill, YoshidaKitaev}, \edit{and combining it with an adaptation of the conventional decoder~\cite{YoshidaKitaev} for the Hayden-Preskill protocol to} the standard quantum teleportation scheme~\cite{BennettTeleportation}, operating them in parallel and  within a single circuit. This  generates  a collective-spin SWAP based solely on natural scrambling dynamics where one can  isolate the distinct roles of information spreading and entanglement generation, and also distinguish them from chaos (sensitivity to parameters) through an operational protocol. 

Finally, we show a concrete numerical realization based on the Dicke model~\cite{HeppLieb1973,Dicke1954,Ritsch2013}, a paradigmatic platform for collective quantum dynamics exhibiting rapid entanglement growth, chaotic behavior, and strong sensitivity of collective observables, as characterized for example by out-of-time-order correlators~\cite{LewisSwan2019,ChvezCarlos2019,Buijsman2017,Kirkova2022}. Dicke-like physics has been realized across a broad range of experimental platforms, including optical cavity systems, superconducting circuits, ultracold atoms, trapped ions, and cavity–tweezer architectures~\cite{Black2003,Maschler2005,Domokos2002,Zhang2018,Fink2009,Baumann2010,Nagy2010,Brennecke2013,Klinder2015,Leonard2017,Kollar2017,Landini2018,Morales2019,SafaviNaini2018,Aedo2018,Sutherland2019,Bohnet2016,LewisSwan2021,Yan2023}, and also proposed as a possible Hayden-Preskill decoder~\cite{Cheng2020PRResearch}. The finding  that the SWAP protocol is implementable under experimentally realistic conditions illustrates that holographic thinking is not merely
metaphorical, it can directly inform a viable  route for approximate, hardware-native quantum gates that are otherwise prohibitively difficult to engineer. 

\begin{figure}[!t]
    \centering
    \includegraphics[width=\linewidth]{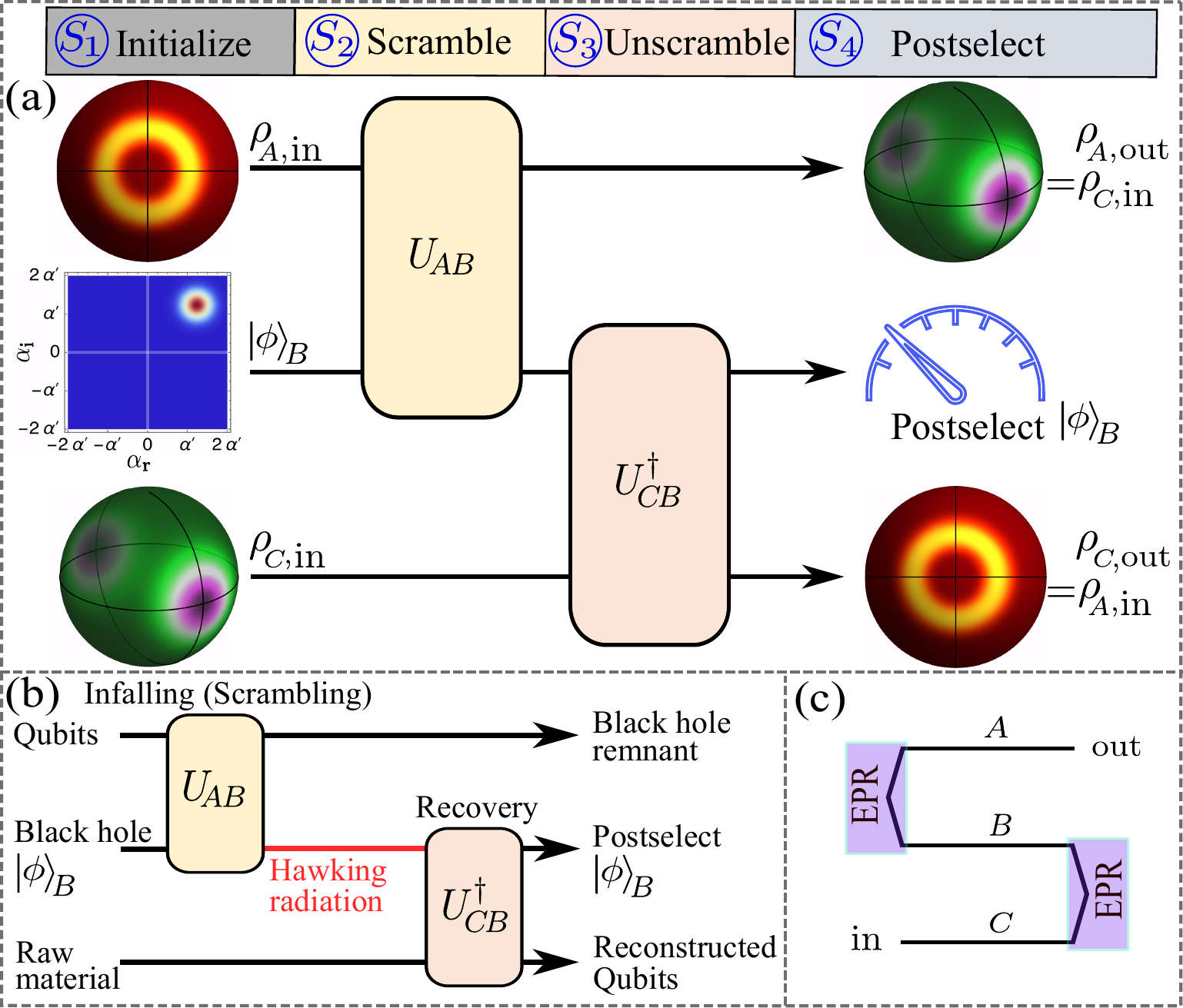}
    \caption{Bidirectional teleportation 
    enabled by scrambling.
(a) The full protocol with 3 systems $A,B,C$, with $d_A=d_C$. Systems $A$ and $C$ start in $\rho_{A,\mathrm{in}}$ and $\rho_{C,\mathrm{in}}$, while $B$ is prepared in $|\phi\rangle_B$. A scrambling unitary $U_{AB}$ spreads information in $A$ into $AB$ entangling them. Applying $U^\dagger_{CB}$ and postselecting $B$ onto $|\phi\rangle_B$ heralds success, yielding $\rho_{A,\mathrm{out}}=\rho_{C,\mathrm{in}}$ and $\rho_{C,\mathrm{out}}=\rho_{A,\mathrm{in}}$ with probability $p\sim 1/d_A^2$. (b) Black hole analogue ($A \to C$): recovery from Hawking radiation via decoding, implemented here by reverse evolution and postselection on $B$. (c) Standard quantum teleportation ($C \to A$): $U_{AB}$ supplies an entangled resource and the measurement on $B$ acts as an effective Bell projection given the $A \to C$ mechanism succeeds, teleporting information $C$ to $A$.}
    \label{fig:SWAP_protocol}
\end{figure}

\textit{Overview of the protocol---} 
Our protocol is depicted in Fig.~\ref{fig:SWAP_protocol}. Our goal is to SWAP arbitrary states in $A$ and $C$, of equal dimension $d_A=d_C$, using a larger system $B$ of dimension $d_B \gg d_A^2$ as an auxiliary. The first step is to initialize $A$ and $C$ in arbitrary states, and the auxiliary system $B$ in a specific state $\lvert \phi\rangle_B$.
In the second step, we have a ``scrambling'' unitary $U_{AB}$ that has a dual role: (1) It spreads the initial information in $A$ into some nonlocal degrees of freedom across $AB$, and (2) It generates a high degree of entanglement between $A$ and $B$, which can be used as a resource state for eventual quantum teleportation. These two ``scrambling mechanisms''
enable the two directions of information transfer in our protocol. If $B$ is sufficiently large, then $B$ contains ``most'' of the initial information in $A$ at this stage,
allowing a high-fidelity recovery of $A$'s original state from $B$ without access to $A$. In the third step, $U_{CB}^\dagger$ (the inverse of $U_{AB}$ in the first step) attempts to ``unscramble'' the information in $B$ into $C$. Without accessing the information in $A$, it produces a complex ensemble of states in $CB$ (of ``size'' determined by $d_A$), including the initial state in $AB$ together with a generically random combination of other states. The final step, projection onto ${_B}\langle \phi\rvert$ in the $B$ system, again has a dual role: (1) it removes the generic states in the ensemble in $CB$, restoring the original state in $A$ to $C$, and (2) it performs the equivalent of an Einstein-Podolsky-Rosen (EPR) measurement in the conventional teleportation protocol~\cite{BennettTeleportation, NielsenChuang} between $B$ and $C$, thus teleporting the information in $C$ to $A$.
Our SWAP protocol may therefore be summarized as:
\begin{equation}
    \left({{_B}\langle}\phi\rvert U_{CB}^{\dagger} U_{AB}\lvert \phi\rangle_B\right) \lvert \psi\rangle_A\lvert \chi\rangle_C \simeq \sqrt{p} \lvert \chi\rangle_A \lvert \psi\rangle_C,
\end{equation}
where $U$ is \textit{any} unitary that exhibits the relevant scrambling mechanisms (information spreading and entanglement).
We  note that this gate is a \textit{probabilistic} gate, with $p$ denoting the probability of the outcome $\lvert \phi\rangle_B$ in the final projective measurement. Typically, we will see that $p \simeq 1/d_A^2$, and therefore $d_B$ can be made quite large to improve the fidelity of our operation without sacrificing probability. This is the same postselection probability that one would obtain in standard teleportation protocols without sufficient control to perform error correction on the output qubit, as in many complex experimental platforms~\cite{BennettTeleportation, NielsenChuang, TeleportationReview1, TeleportationReview2, HighDimEntanglementReview, HighDimEntanglement1, HighDimEntanglement2}.

We  note some important connections to previous work~\cite{BennettTeleportation, HaydenPreskill, YoshidaKitaev, RenyiEntropyMeasurementTwoGates}(see Appendix~\ref{app:previouswork} for more details). The main simplification achieved by our protocol is to perform an effective reduction of Hayden-Preskill (HP) \edit{type} information transfer protocols (conventionally implemented with 5 subsystems, \edit{as in the Yoshida-Kitaev realization}~\cite{YoshidaKitaev, Verifiedscrambling}), associated with black hole physics, to the \textit{same} $3$-system circuit as the conventional quantum teleportation protocol. In addition, instead of requiring precisely-engineered EPR pairs as in these  protocols, which are prohibitively difficult to prepare~\cite{HighDimEntanglement1, HighDimEntanglement2, HighDimEntanglementReview} e.g. for collective degrees of freedom~\cite{ Hammerer2010, Khan2025}, we use the same scrambling unitary that transfers information in the HP\edit{-like} protocol to generate entanglement for teleportation. This allows us to implement both of the otherwise  unidirectional information transfer protocols \textit{in parallel} on the same hardware, while  achieving as a whole a  bidirectional SWAP gate that is more generic and requires less fine-tuned engineering of states and interactions
than either unidirectional protocol. 
As shown in Fig.~\ref{fig:SWAP_protocol}, the $A\to C$ direction can be pictured in terms of a model of information recovery from Hawking radiation in black hole physics [panel (b)] \edit{that differs from Hayden-Preskill in requiring no pre-emitted Hawking radiation}, while the $C \to A$ direction \edit{adapts the Yoshida-Kitaev decoder from the conventional Hayden-Preskill scheme to} resemble the standard quantum teleportation scheme [panel (c)]. Separately, Ref.~\cite{RenyiEntropyMeasurementTwoGates} proposes that a similar circuit to Fig.~\ref{fig:SWAP_protocol}, averaged over initial states with additional projections, may be used to \textit{measure} the generation of R\'{e}nyi entropy by an arbitrary unitary $U_{AB}$, without identifying or requiring information transfer.



\textit{Scrambling mechanisms---} Our protocol works via two different ``scrambling'' mechanisms:
(1) $A\to C$ : Orthogonal subspace encoding (formalizing information spreading), and
(2) $C \to A$ : Entanglement-based quantum teleportation.
These mechanisms are most easily illustrated for the SWAP of pure states when $A$ and $C$ are single-qubit systems
with computational basis states $\lvert \uparrow\rangle$, $\lvert \downarrow\rangle$, and $B$ has a large Hilbert space.  The more general case with mixed initial states and other system sizes
is discussed in the supplement~\cite{SupplementalMaterial}, where these mechanisms are related to operator entanglement in time~\cite{HosurQiRobertsYoshida} and space, and shown to determine recovery for any given initial state without state averages as required in previous analyses~\cite{YoshidaKitaev, YoshidaYao}. After the scrambling interaction $U$, the output state in the $AB$ system can always be written in the form:
\begin{equation}
    U_{AB} \lvert \psi\rangle_A\lvert\phi\rangle_B\lvert \chi\rangle_C \!=\! \frac{1}{\sqrt{2}}\!\left(\lvert \uparrow\rangle_A\lvert \phi^{\uparrow}(\psi)\rangle_B \!+\! \lvert \downarrow\rangle_A\lvert \phi^{\downarrow}(\psi)\rangle_B\right)\! \lvert \chi\rangle_C,
    \label{eq:scrambledstate_qubit}
\end{equation}
where $\lvert \phi^{\uparrow}(\psi)\rangle_B, \lvert \phi^{\downarrow}(\psi)\rangle_B$ are not \textit{necessarily} orthonormal states (only achieving orthonormality for maximal entanglement). Generically, they span a specific $2$-dimensional subspace $\mathcal{H}_{B}(\psi) \equiv \vecspan\lbrace \lvert \phi^{\uparrow}(\psi)\rangle_B, \lvert \phi^{\downarrow}(\psi)\rangle_B\rbrace \subseteq \mathcal{H}_B$ of $B$; once we trace out $A$, the role of the scrambler is effectively to map the initial spin state $\lvert \psi\rangle_A$ to this subspace.

Now, our goal is to restore $\lvert \psi\rangle$ in $C$ from $\mathcal{H}_{B}(\psi)$, when $C$ is initialized to an arbitrary initial state. Assuming that the $\lvert \phi^{\uparrow,\downarrow}(\psi)\rangle$ are distinct states, an arbitrary pure state in $\mathcal{H}_C \otimes \mathcal{H}_{B}(\psi)$ can be expanded in a (not necessarily orthonormal) basis of ``quasi-Bell'' states of the form (with $s=0,1$):
\begin{align}
    \lvert \Phi^{0s}(\psi)\rangle_{CB} &= \frac{1}{\sqrt{2}}\lvert \uparrow\rangle_C\lvert \phi^{\uparrow}(\psi)\rangle_B + \frac{(-1)^s}{\sqrt{2}}\lvert \downarrow\rangle_C\lvert \phi^{\downarrow}(\psi)\rangle_B, \nonumber \\
    \lvert \Phi^{1s}(\psi)\rangle_{CB} &= \frac{1}{\sqrt{2}}\lvert \uparrow\rangle_C\lvert \phi^{\downarrow}(\psi)\rangle_B + \frac{(-1)^s}{\sqrt{2}}\lvert \downarrow\rangle_C\lvert \phi^{\uparrow}(\psi)\rangle_B.
    \label{eq:quasiBellstates_qubits}
\end{align}
This reduces to the standard maximally entangled basis of Bell states when the $\lvert \phi^{\uparrow,\downarrow}(\psi)\rangle_B$ are perfectly orthonormal. We note that $\lvert \Phi^{00}\rangle_{CB}$ is precisely the state in Eq.~\eqref{eq:scrambledstate_qubit} with $A$ replaced by $C$; acting with the reversed scrambler on this state successfully transfers $A$ to $C$, $U^{\dagger}_{CB} \lvert \Phi^{00}\rangle_{CB} = \lvert \psi\rangle_C\lvert\phi\rangle_B$. Up to this point, we do not require any ``scrambling'' properties of $U$. 
However, 
when reduced to $CB$, Eq.~\eqref{eq:scrambledstate_qubit} generally gives a \textit{mixture} of the quasi-Bell states in Eq.~\eqref{eq:quasiBellstates_qubits}. When acting with $U^\dagger_{CB}$ on these states (other than $\lvert \Phi^{00}\rangle_{CB}$), we will typically obtain a complicated and intractable output state.

This can be resolved through our first scrambling mechanism: \textit{orthogonal subspace encoding}. This is the assumption that a state $\lvert \psi^{\perp}\rangle_A$ that is orthogonal to our initial state $\lvert \psi\rangle_A$ is mapped by the scrambler to a subspace $\mathcal{H}_B(\psi^\perp)$ that is essentially orthogonal to $\mathcal{H}_B(\psi)$, when $B$ is initialized in the same state $\lvert \phi\rangle_B$.
In other words, given the expression for the scrambled state in Eq.~\eqref{eq:scrambledstate_qubit}, we want that the states  $\lvert \phi^{\uparrow,\downarrow}(\psi)\rangle_B$ are essentially orthogonal to $\lvert \phi^{\uparrow, \downarrow}(\psi^\perp)\rangle_B$
(see supplement~\cite{SupplementalMaterial} for details). This means that given a product basis for $\mathcal{H}_A \otimes \mathcal{H}_B$,
$
    \lbrace \lvert \psi\rangle, \lvert \psi^{\perp}\rangle\rbrace_A \otimes \lbrace \lvert \phi\rangle, \ldots \lvert \phi_\ell^{\perp}\rangle,\ldots\rbrace_B,
$
where the $\lvert \phi_\ell^{\perp}\rangle_B$ are orthogonal to $\lvert \phi\rangle_B$, the only combinations involving $\lvert \psi^{\perp}\rangle$ that can generate the quasi-Bell states in Eq.~\eqref{eq:quasiBellstates_qubits} (which are based entirely on the subspace $\mathcal{H}_{B}(\psi)$ instead of $\mathcal{H}_B(\psi^\perp)$) are the ones that do not involve $\lvert \phi\rangle_B$. Consequently, when acting with $U^{\dagger}$ on the quasi-Bell states, we should get (for some arbitrary states $\lvert \ell\rangle_C$),
\begin{equation}
    U^{\dagger}_{CB}\lvert \Phi^{rs}(\psi)\rangle_{CB} = c_{rs} \lvert \psi\rangle_C\lvert\phi\rangle_B + \sum_{\ell} c^{\ell}_{rs} \lvert \ell\rangle_C\lvert \phi^{\perp}_\ell\rangle_B.
\end{equation}
Projecting this state onto $\lvert \phi\rangle_B$ recovers $\lvert \psi\rangle_C$ 
with a postselection probability that is determined by the coefficients $\lvert c_{rs}\rvert^2$, successfully transferring information from $A$ to $C$.

For information transfer from $C$ to $A$, we require that the two states $\lvert \phi^{\uparrow,\downarrow}(\psi)\rangle_B$ within the subspace $\mathcal{H}_{B}(\psi)$ themselves form a nearly orthonormal basis within this subspace. This corresponds to maximal entanglement generation between $A$ and $B$, in which case the quasi-Bell states in Eq.~\eqref{eq:quasiBellstates_qubits} approach a genuine orthonormal set of Bell states as in the standard quantum teleportation protocol. We can then think of the scrambling step as preparing the Bell state $\lvert \Phi^{00}(\psi)\rangle_{AB}$, and a projection onto the same Bell state in $CB$ would successfully teleport information from $C$ to $A$; this also determines the postselection probability to be $p=1/4$ (the inverse of the number of Bell states). More generally, when $A$ has more than two levels, we would have $p \simeq 1/d_A^2$.  However, the ${_{CB}\langle}\Phi^{00}(\psi)\rvert$ projection would na\"{i}vely correspond to evolution by $U^{\dagger}$ and a projection onto the full initial state of $AB$ in $CB$:
\begin{equation}
    {_{CB}\langle}\Phi^{00}(\psi)\rvert = {_C\langle}\psi\rvert {_B\langle}\phi\rvert U^{\dagger},
    \label{eq:Bell00_projection}
\end{equation}
which would in general destroy the information in the output branch of $C$. We can avoid this provided that the $A\to C$ mechanism succeeds. If the projection onto $\lvert \phi\rangle_B$ already guarantees that the output state in $C$ is essentially the input state in $A$, it automatically implements the Bell state projection in Eq.~\eqref{eq:Bell00_projection} without actually acting on $C$.
In this way, the initial state in $C$ is transferred to $A$ without disturbing the final state in $C$.

To evaluate the performance of our protocol in a general setting, we quantify how well the final state on $AC$ reproduces the ideal outcome of a SWAP operation. Specifically, we use a fidelity measure that compares the output of the protocol with the state that would result from a perfect SWAP. For an input density operator $\rho_{AC}$, we define
\begin{equation}
    \mathcal{F}[\rho_{AC}] =
    \frac{\Tr_{AC}\!\left[\rho^{\text{out}}_{AC}\,\swap(\rho_{AC})\right]}
    {\sqrt{\Tr_{AC}\!\left[(\rho^{\text{out}}_{AC})^2\right]\;
    \Tr_{AC}\!\left[\rho_{AC}^2\right]}} ,
    \label{eq:fidelities}
\end{equation}
where $\rho_{AC}^{\text{out}}$ denotes the state produced by the protocol, and $\swap(\rho_{AC})$ denotes the effect of a perfect SWAP gate on $\rho_{AC}$; this measure lies between $0$ (for orthogonal states) and $1$ (for identical states).

 To demonstrate  ``universality'' of the protocol,  we first study its performance for the case  when $U_{AB}$ is a random unitary matrix, which represents the broadest universality class of dynamics associated with ``quantum chaos''~\cite{Mehta, Haake, ChaosComplexityRMT}.  Such Haar-random unitaries  
provide a useful theoretical proxy for \textit{generic, strongly scrambling dynamics}, capturing the typical statistical behavior expected of complex many-body evolutions~\cite{Mehta,Haake,ChaosComplexityRMT}. Using standard typicality arguments~\cite{SupplementalMaterial}, we estimate
\begin{equation}
    \mathcal{F} \simeq \frac{1}{1 + d_A^2/d_B}, \qquad
    p \simeq \frac{1}{d_A^2}.
    \label{eq:HaarFidelity}
\end{equation}
When the Hilbert space dimension of subsystem $B$ is large compared to that of $A$, the fidelity approaches unity while $p$ remains finite, $p \simeq 1/d_A^2$. Eq.~\eqref{eq:HaarFidelity} therefore establishes that maximally scrambling dynamics are sufficient to implement a high-fidelity SWAP between collective degrees of freedom.

Crucially, our protocol does not rely on the ability to implement such Haar-random dynamics. Rather, any sufficiently complex unitary with high enough information spreading and entanglement generation capabilities can  realize the desired SWAP functionality. This could include unitary  evolution for a time $t$, generated by a
simpler scrambling Hamiltonian $H$
native to an experimental platform, $U_{AB}(t) = e^{-iH_{AB} t}$. Below we provide  an example of such a capability by considering the  experimentally  accessible Dicke  model.  

\begin{figure}[!t]
\centering
\includegraphics[width=1.01\linewidth]{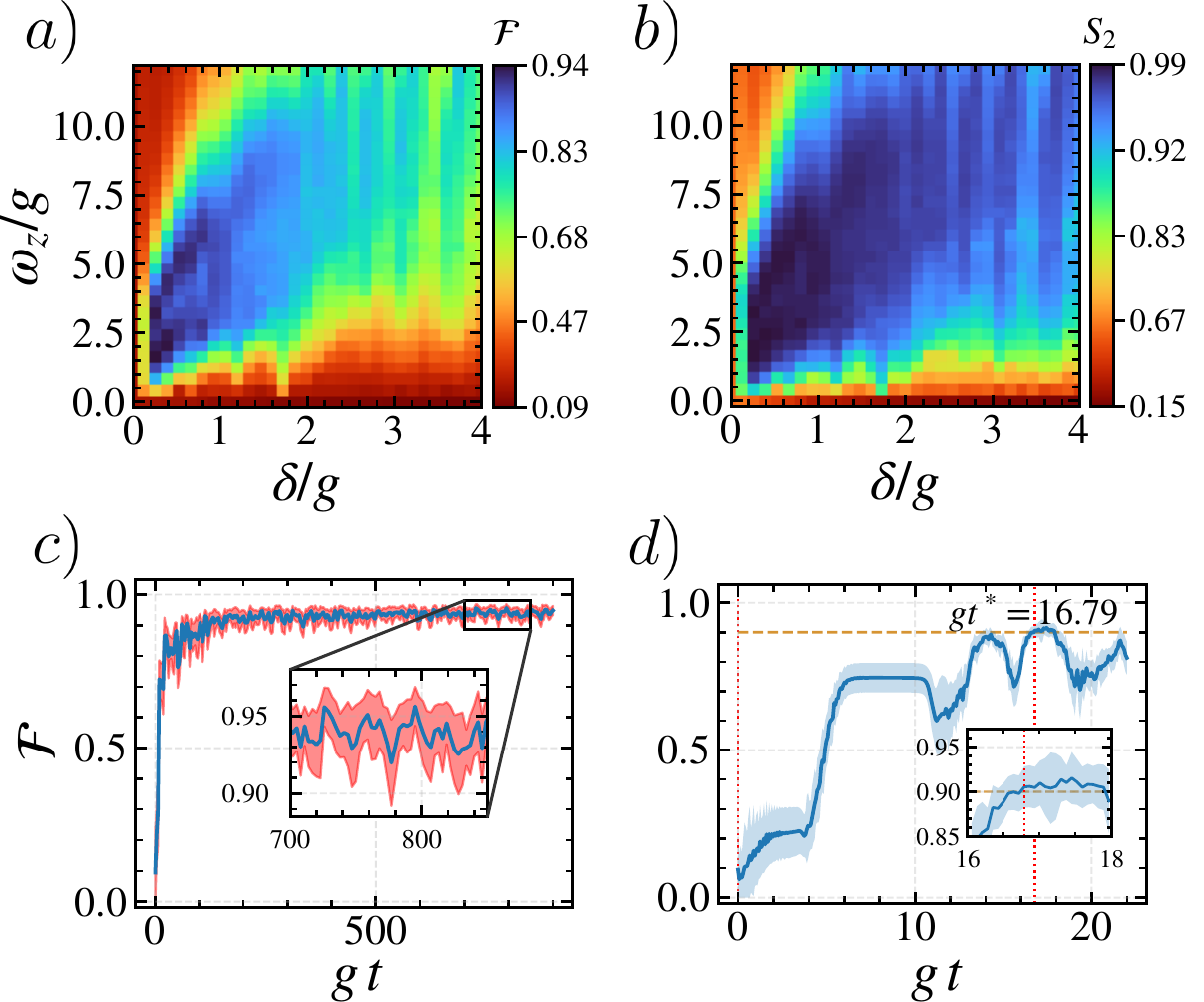}
\caption{Protocol performance in the Dicke model with $N = 4$ and $\phi = 30$, for $30$ pure initial states in $A$ that are uniformly (Haar) sampled from the Hilbert space \cite{SupplementalMaterial}.
(a) Dependence of fidelity $\mathcal{F}$ on the parameters $\delta/g$ and $\omega_z/g$, averaged over initial states and over times $700\leq gt \leq 850$ (chosen within a regime in which $\mathcal{F}$ remains close to a steady value).
(b) Entanglement between $A$ and $B$ quantified by the normalized second R\'enyi entropy~\cite{HorodeckiEntanglementReview} $S_2(A)\equiv -\log_{d_A}\overline{\Tr_A(\rho_A^2)}$ with $d_A=N+1$, where $\overline{\Tr_A(\rho_A^2)}$ is the purity in $A$ averaged over initial states over the same time interval; regions of large $\mathcal{F}$ in (a) are seen to correlate with large $S_2$. 
(c) $\mathcal{F}$ as a function of $gt$ in $A$ at $\delta/g=0.267$ and $\omega_z/g=3.20$, corresponding to maximal $\mathcal{F}$ in (a), showing the attainment of a steady regime  of large $\mathcal{F} \sim 0.94$ (inset: the time interval used for averaging in (a)); fluctuations in $\mathcal{F}$ between initial states of up to one standard deviation and non-negative values are shaded. (d) $\mathcal{F}$ vs. $gt$ prior to the steady state regime, showing transient regions of acceptable fidelity $\mathcal{F} \geq 0.9$, for the parameters $\delta/g = 0.40$ and $\omega_z/g = 3.78$ corresponding to the earliest such time $gt^\ast = 16.79$ (depicted in Fig.~\ref{fig:transients} by a red star in Appendix B).
}
\label{fig:fourpanel}
\end{figure}


\textit{Experimental proposal}---
Our protocol can be implemented in state-of-the-art trapped ion platforms with  three  \textit{collective}  sub-ensembles  of spin-$1/2$  ions,  which we label $l=A, C,M$  each independently addressable by a drive with Rabi frequency $\Omega_l$ and detuning $\Delta_l$. All the ions are  coupled via standard spin-boson interactions generated by a set of Raman beams  to a shared center of mass bosonic mode \cite{Monroe2021,Leibfried2003,Khan2025}, $B$, described by the creation and annihilation operators $b_B^\dagger$, $b_B$. We take $A$ and $C$ to have the same number of ions $N_A=N_C= N$, and work in their fully collective manifold with dimension $d_A =d_C= N+1$. We will additionally need an auxiliary spin subsystem $M$, with $N_M$ ions to perform the final projective measurement in $B$. The corresponding  spin--boson Hamiltonian can be written as 
\begin{equation}
 H
=\delta b_B^\dagger b_B
+\sum_{l=A,C,M} \Big(\Omega_l S_l^x+\Delta_l S_l^z\
+\frac{2 \widetilde{g}_l}{\sqrt{N_T}}\bigl[ b_B+ b^\dagger_B\bigr] S_l^z\Big),
\label{eq:H_total_lab}
\end{equation}
where $N_T=2N+N_M$ is the total  number of ions, $S_l^{j}$ are the collective spin components in $l$ for $j = x,y,z$. Here  $\frac{2\widetilde{g}_l}{\sqrt{N_T}}$ is the spin-boson coupling strength for each of the subensembles $l$;  we define $g_l = \widetilde{g}_l\sqrt{N_l/N_T}$ and take $g_A=g_C\equiv g$.
We initialize $M$ in the state $\ket{\downarrow}^{\otimes N_M}$ and set $\Omega_M=0$, so that it remains effectively frozen with $\langle S_M^z\rangle =-N_M/2$, and decoupled from the $ABC$ dynamics until the final measurement. The  displacement of the mode $b_B\to b_B - \frac{g_M \sqrt{N_M}}{\delta}$,  can  be compensated by adjusting $\Delta_{A},\Delta_{C}$ appropriately and displacing the initial state in B \cite{SupplementalMaterial}.
By separately controlling $\Omega_{A,C} $ and $\Delta_{A,C}$, 
this  Hamiltonian can be tuned  to realize the Dicke model between $A$ and $B$, 
\begin{equation}
    \label{eq:Dicke_model_hamiltonian}
    H_{AB}^{\rm Dicke}= \omega_z S_A^z + \delta b^\dagger_B b_B - \frac{2g_A}{\sqrt{N}} S_A^x (b_B + b_B^\dagger),
\end{equation} 
with $\Omega_A=\omega_z$, while keeping ensemble $C$ highly decoupled thus realizing $U_{AB}(t)=e^{-it H_{AB}^{\rm Dicke}}$. 
Likewise,   we can also  implement $- H_{CB}^{\rm Dicke}$ and thus
$U_{CB}^\dagger (t)=e^{it H_{CB}^{\rm Dicke}}$
while keeping ensemble $A$ effectively undisturbed (see supplement \cite{SupplementalMaterial}).

To realize our protocol, we need to initialize $A$  and $C$ in  input states $\rho_A$, and $\rho_C$, and  the bosonic mode in a known reference state $\ket{\phi}_B$. 
For definiteness, we take $\lvert \phi\rangle_B= e^{ (\phi b_B^\dagger - \phi^* b_B)}\ket{0}_B$ to be a coherent state~\cite{ShankarQM, Agarwal2012} of the boson mode  displaced by  $\phi$, which implicitly determines an effective $d_B$ as the dimension of the Hilbert space effectively accessible from this state. We then evolve for a duration $t$ to implement $U_{AB}(t)$.
The ensemble  $C$ is effectively idle during this step. 
For the reversal step, we can implement an approximate backward evolution on the $C$--$B$ subsystem for the same duration $t$, realizing $U_{CB}^\dagger(t)$ as explained above.

Implementing a direct bosonic projective measurement is difficult in many platforms. In our case, projection onto $\lvert \phi\rangle_B$ can be effected by transferring the information about occupancy (after a displacement by $-\phi$) to the measurement spins $M$ (with $N_M$ large compared to the range of occupancy of the boson mode), and then directly implementing a native projection on $M$~\cite{SupplementalMaterial} (see Ref.~\cite{Chen2021} for other strategies).

 Figure \ref{fig:fourpanel}(a,c) illustrate the optimal achievable fidelity of the SWAP protocol in the Dicke model, obtained from numerical simulations over a broad range of parameters and averaged over a set of random pure initial states. We consider a system with 
$N=N_A=N_C=4$ spins in ensembles $A$ and 
$C$ respectively, and an initial bosonic coherent state with 
$\phi=30 $. To elucidate the role of correlations, we also plot in  Figure \ref{fig:fourpanel}(b) an ``averaged'' R\'{e}nyi entropy~\cite{HorodeckiEntanglementReview} of the reduced spin state after tracing out the bosonic mode, computed after the application of the $U_{AB}$ unitary, 
$S_2(A)\equiv -\log_{d_A}\overline{\Tr_A(\rho_A^2)}$, where $\overline{\Tr_A(\rho_A^2)}$ is the initial-state-averaged purity in $A$ with $\rho_A=\Tr_B(\rho_{AB})$. $S_2(A)$ quantifies the degree of entanglement generated between the spins and the boson during the protocol. We observe SWAP fidelities as high as  $\mathcal{F} \sim 0.94$ in the parameter regime 
$\omega_z\sim \delta\sim g$ where spin and bosonic parameter  scales are energetically comparable and the dynamics generates strong correlated spin–boson excitations. In contrast, in regimes where 
$|\delta | \gg |\omega_z|>g$ or $|  \omega_z| \gg |\delta|>g$
 the Dicke model becomes effectively integrable due to the suppression of either bosonic or spin excitations; correspondingly, we find significantly reduced SWAP fidelities, accompanied by low spin–boson entanglement.

Although precise simulations with larger spin numbers become numerically intractable, we expect similar performance to persist provided that the relevant scrambling mechanisms remain active and the effective bosonic Hilbert space explored during the dynamics satisfies 
$d_B\gg d_A^2$, assuming that experimental coherence times and control over time reversal are sufficient to reach the optimal operating conditions. This requirement highlights a key practical consideration: the minimal time needed to achieve a target fidelity. While optimal fidelities are reached at long evolution times where spin–boson correlations are maximal, realistic experimental implementations are limited by decoherence and other imperfections. This motivates an analysis of the shortest timescale required to reach e.g. $\mathcal{F}>0.9$.
From numerical simulations with 
similar ion numbers, Figure \ref{fig:fourpanel}(d), we find that fidelities above this threshold can be achieved on timescales of order 
$gt\sim 16.8$ which are compatible with current experimental capabilities (see Appendix \ref{app:experimentalerrors}).

\edit{While the above analysis is for time-independent Hamiltonian dynamics, which may generically be easier in collective spin platforms in the near term, we point out that periodically alternating between two non-commuting Hamiltonians with similar structure to Eq.~\eqref{eq:Dicke_model_hamiltonian} converts the Hamiltonian scrambling problem considered here to a Floquet scrambling problem. Generic complex Floquet dynamics is seen to thermalize to infinite temperature including for thermodynamically large systems~\cite{Regnault2016,PhysRevX.4.041048,lazarides2014floqueteth,ponte2015floqueteth,kim2014floqueteth, GarrattChalker}, corresponding to maximal scrambling via entanglement generation. Indeed, we show numerically in Sec. II of the supplement~\cite{SupplementalMaterial} that introducing phases in the boson term in Eq.~\eqref{eq:Dicke_model_hamiltonian} and periodically alternating between these phases sharply increases the fidelity. Further, though lacking in transients, this Floquet switching easily surpasses the fidelity of early-time transients in the Hamiltonian case in a \textit{sustained} manner, in a comparable but slightly longer time (by a small factor of $\sim 3$). This suggests that implementing Floquet dynamics may be a viable route to scale our protocol to larger systems with high fidelity, in platforms that can effectively mitigate the noise that may be introduced by such alternating.}

A final and crucial consideration is the sensitivity of the protocol to imperfections in the time-reversal operation, i.e., implementing $U^\dagger$ as the precise inverse of $U$. If high SWAP performance relied on strongly chaotic dynamics, even small time-reversal errors could lead to catastrophic degradation. However, numerical simulations incorporating experimentally relevant time-reversal imperfections (see Appendix~\ref{app:experimentalerrors}) show that operating the protocol at the shortest time sufficient to reach high fidelity substantially reduces this sensitivity. Specifically, while long-time operation requires control of the time-reversal imperfection at the level of 
$10^{-5}$, operating at the minimal optimal time relaxes this requirement by approximately two orders of magnitude. This suggests the existence of a favorable experimental window in which the protocol can be practically implemented even in current, non–error-corrected platforms.

\textit{Conclusion---} 
Apart from providing a generic implementation of an essential quantum gate, our protocol constitutes a significant step towards a careful experimental study of scrambling that can separate different mechanisms: each direction of the SWAP is directly sensitive to a specific scrambling mechanism (information spreading for $A \to C$ and entanglement generation for $C \to A$), whereas chaos in the precise sense of \textit{sensitivity to parameters} only plays a role in the sensitivity of the protocol to experimental errors rather than its ideal functionality. Overall, our results introduce new opportunities for practical applications such as bidirectional quantum teleportation of highly entangled states in a many-body setting and new routes toward robust precision metrology that do not require delicate or system-specific reconstruction protocols. 

\begin{acknowledgments}
    \textit{Acknowledgments---} {We thank Diego Fallas Padilla   and James Thompson   for feedback on the manuscript. This work was supported by the Heising-Simons Foundation under Grant 2024-4848,  by the U.S. Department of Energy,
Office of Science, National Quantum Information Science Research Centers, Quantum Systems Accelerator,
by the National Science Foundation under Grant Number PHY-2317149 (Physics Frontier Center), by ARO W911NF24-1-0128, and by NIST.}

\textit{Data availability statement---} The data that support the findings of this study were generated by numerical simulations. The source code and parameters used to generate the simulations are publicly available \cite{ChaparroZenodo}.
\end{acknowledgments}

\begin{appendix}
\setcounter{secnumdepth}{4}

\setcounter{section}{0}




\section{Connections to Hayden-Preskill recovery, black holes, and quantum teleportation}
\label{app:previouswork}

The mechanism of information transfer from $C$ to $A$ is closely related (though not mathematically equivalent) to the probabilistic Yoshida-Kitaev decoder~\cite{YoshidaKitaev} circuit for the Hayden Preskill protocol~\cite{HaydenPreskill}, as discussed in the supplement~\cite{SupplementalMaterial}, whose success probability is also $p \simeq 1/d_A^2$. This protocol was designed as a quantum information thought experiment to model the recovery of information from evaporating black holes, given certain assumptions about quantum gravity~\cite{HaydenPreskill}. A key difference with our proposal is that the original protocol requires at least five systems to capture all aspects of the black hole problem ($3$ resembling $A$ and $2$ resembling $B$) as well as a precise preparation of EPR states, in which form it has been experimentally realized in \cite{Verifiedscrambling, ScramblingQutrits} for small systems (totalling $7$ qubits~\cite{Verifiedscrambling}, with $d_A = 2$ and $d_B = 4$, or $5$ qutrits~\cite{ScramblingQutrits} with $d_A = d_B = 3$). However, both of these aspects severely complicate the implementation of this protocol in spin-boson systems (as is our target) and other experimental platforms, though there have been some theoretical proposals~\cite{Cheng2020PRResearch}. 
Our SWAP protocol involves only $3$ systems ($2$ of type $A$ and $1$ of $B$) and no precisely constructed EPR states, addressing these challenges in the implementation of this protocol.

At an intuitive level, the Yoshida-Kitaev teleportation mechanism~\cite{YoshidaKitaev} essentially amounts to quantum teleportation~\cite{BennettTeleportation, NielsenChuang}, if the scrambling unitary $U$ were replaced by one that generates Bell states~\cite{SupplementalMaterial, dynamicalqentanglement}. In sufficiently simple systems, the outcome of the final Bell state measurement (our steps $3$ and $4$) can be used to modify the initial state. In the language of error correction~\cite{NielsenChuang, KnillLaflamme}, the Bell state measurement on $BC$ measures a syndrome that can subsequently be used to perform ``error correction'' on $A$ to successfully teleport the initial state in $C$ for all outcomes. Once again, in sufficiently complex systems (both in the conventional teleportation protocol and in our case), it is often not clear how to implement such a syndrome-based correction~\cite{HighDimEntanglement1, HighDimEntanglement2, TeleportationReview1}, and one is generically left with a probabilistic protocol that succeeds at the rate $p \simeq 1/d_A^2$. For a usual (e.g. CSS) error-correcting code~\cite{NielsenChuang}, this procedure is analogous to only keeping the outcome if no error is detected, otherwise throwing it away. \edit{However, in smaller system sizes, this probability can be boosted to near $1$ (together with larger fidelity) by applying amplitude amplification (Grover search) to our SWAP protocol, similar to the deterministic Yoshida-Kitaev decoder; this is demonstrated in Sec. IV B of the supplement~\cite{SupplementalMaterial}.}

While our $A$ to $C$ mechanism is formally distinct from quantum teleportation, it can be understood as a variant of the Yoshida-Kitaev teleportation protocol that substitutes entanglement in space (between two subsystems as in regular teleportation~\cite{BennettTeleportation}) with entanglement in time~\cite{HosurQiRobertsYoshida} between an input (past) subsystem and an output (future) subsystem of a scrambler. \edit{As discussed in the supplement~\cite{SupplementalMaterial}, while this direction is closely related to the Hayden-Preskill recovery scheme, it also differs in some notable ways. In particular, where Hayden-Preskill recovery significantly leverages Hawking radiation emitted \textit{before} infalling information (realized on the quantum side by initially prepared EPR pairs) and scrambling via entanglement in space, our protocol corresponds to recovery from a black hole without previously emitted Hawking radiation (as in Fig.~\ref{fig:SWAP_protocol}b) due to scrambling via entanglement in time, more closely realizing the Susskind-Thorlacius \textit{gedankenexperiment}~\cite{SusskindThorlaciusGedanken} that inspired~\cite{SekinoSusskind} the later Hayden-Preskill scheme.}

A key feature of our circuit is that both mechanisms of entanglement in space and time are combined in the same minimal set of subsystems to efficiently implement two-way information transfer in generic platforms. In particular, for models with real Hamiltonians such as the Dicke model, we show in the supplement~\cite{SupplementalMaterial} that both mechanisms must co-exist and imply each other. Therefore, the two-way teleportation protocol can essentially be obtained ``for free'' from one-way teleportation (using our circuit) in a large class of such Hamiltonians. We also show that an even further simplification of this mechanism to two systems $A$ and $B$ is possible in the supplement~\cite{SupplementalMaterial}, though in that case limited to unidirectional teleportation (the recovery of the initial state in $A$ from $B$, after erasing the information in $A$ post scrambling).

\section{Handling experimental errors}
\label{app:experimentalerrors}



Now we will briefly discuss how to choose the operating parameters such as $(g,\delta,\omega_z)$ in Eq.~\eqref{eq:H_total_lab} to minimize experimental errors in our protocol, noting that this choice may in general be different from one that gives the maximum fidelity. There are two key experimental challenges that must be addressed: (1) decoherence over long times, and (2) errors in implementing $U^\dagger$ as the inverse of $U$. Both of these challenges can be broadly addressed by minimizing the time required to apply each step of our protocol.

\begin{figure}[!ht]
\centering
\includegraphics[width=0.8\linewidth]{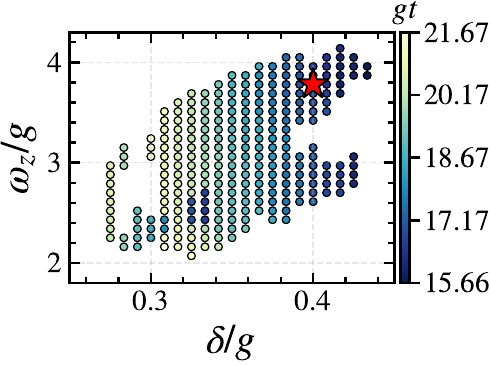}
\caption{Early-time operating point distribution with state averaged fidelity above the target value of $0.9$, taken over 30 uniformly chosen (Haar random) initial states in $A$. The figure depicts the threshold time $gt^\ast$ (color-coded) versus $\delta/g$ and $\omega_z/g$, keeping only parameter regions at which $\mathcal{F}(gt^*)\geq 0.9$. The red star marks the target operating point of Fig.~\ref{fig:fourpanel}d, chosen so that above-threshold fidelity is sustained for three consecutive data points in time ($gt^\ast$ and two successive points each separated by $g\cdot \delta t = 0.15$), embedded in a broad robust region of above-threshold fidelity.}
\label{fig:transients}
\end{figure}

\textit{Decoherence---} For decoherence, this is straightforward: we need to ensure that the overall time required for our protocol is smaller than the decoherence time scale of the relevant experimental platform. It may often turn out to be the case that the time required for \textit{sustained} scrambling~\cite{dynamicalqspeedlimit, dynamicalqfastscrambling} (reaching a scrambled steady state) as illustrated in Fig.~\ref{fig:fourpanel} is much longer than the decoherence time scale in present-day platforms. For example, in trapped-ion platforms, the main source of decoherence is local spin dephasing~\cite{Uys2010}; if $g\simeq2\pi\times4~{\rm kHz}$, $\Gamma_{\rm Rl}\simeq250~{\rm s}^{-1}$ then $g\tau_{\rm dec}\sim g/\Gamma_{\rm Rl}\approx10^2$~\cite{SafaviNaini2018}, which is similar to the timescale required to attain a steady value of the fidelity in Fig.~\ref{fig:fourpanel}.

\begin{figure}[!ht]
  \centering
  \includegraphics[width=\linewidth]{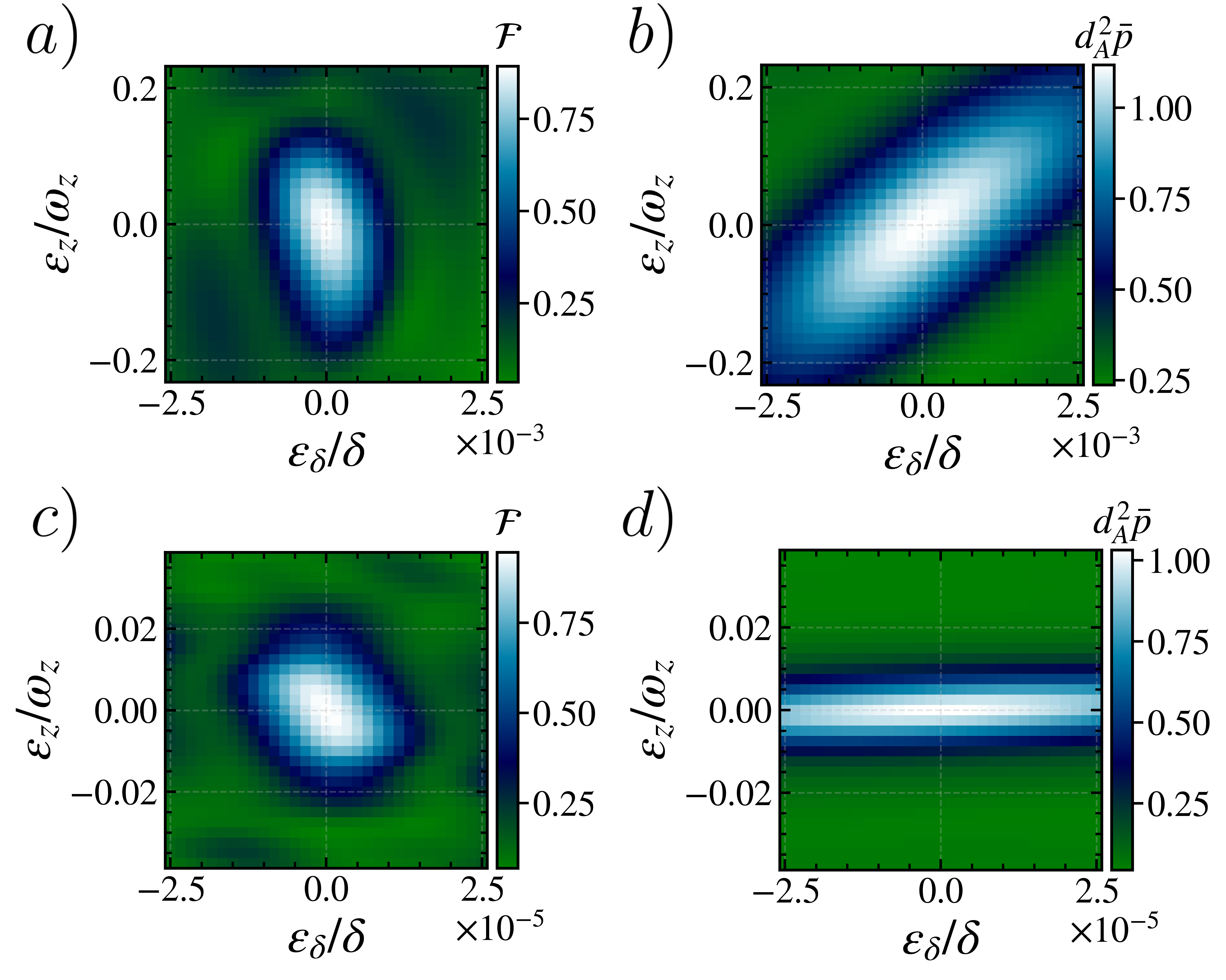}
  \caption{Reversal imperfections parameterized by fractional mismatches $\epsilon_\delta/\delta$ and $\epsilon_z/\omega_z$.
Top row: early-time operating point with parameters as in Fig.~\ref{fig:fourpanel}d.
Bottom row: late-time operating point that maximizes the long-time fidelity with the same parameters as Fig.~\ref{fig:fourpanel}c.
While the early-time regime retains a broad high-fidelity region, the late-time dynamics is markedly more sensitive, with a much narrower tolerance to reversal errors by a relative factor of about $10^{-2}$ compared to the early-time operating point.}
  \label{fig:imperfections}
\end{figure}
We can mitigate this issue by leveraging early-time \textit{transients}, where the fidelity briefly reaches an acceptable level, and by selecting operating parameters that minimize the earliest such time. This is illustrated in Fig.~\ref{fig:fourpanel}d, where we plot $\bar{\mathcal{F}}(t)$ for the red-star parameter set in Fig.~\ref{fig:transients} and define $t^\ast$ as the earliest time at which $\bar{\mathcal{F}}(t)\ge 0.9$ is sustained over a finite time window. Fig.~\ref{fig:transients} maps $gt^\ast$ over $(\delta/g,\omega_z/g)$, restricting to points with $gt^\ast\le 22$ to emphasize early-time dynamics; the star lies within a contiguous region of comparably small $gt^\ast$, indicating robustness to modest parameter variations.

\textit{Reversal imperfections---} While implementing time reversed unitaries $U^\dagger$ has been a considerable challenge in scrambling protocols~\cite{ButterflyMetrology, ReversalButterflyMetrology}, the same strategy of minimizing the time $t$ helps mitigate these errors. For example, if $U^\dagger \to e^{i H_{\text{rev}} t}$, where $H_{\text{rev}}$ differs from $H$ in Eq.~\eqref{eq:Dicke_model_hamiltonian} via some small errors in parameters $\delta \to \delta + \epsilon_\delta$ and $\omega_z \to \omega_z + \epsilon_z$, then we see that each error occurs in $U^\dagger$ via the combinations $\epsilon_\delta t$ and $\epsilon_{z} t$, therefore being magnified by the time $t$. Thus, minimizing the time $t$ is a good strategy for minimizing the effect of these errors according to the energy-time uncertainty principle~\cite{MT, ML, AAMT, LevitinToffoli}, by which a perturbation $H \to H + \epsilon V$ with $V$ having an energy spread $\Delta E_V$ has negligible effect for a time $t \lesssim 1/(\epsilon \Delta E_V)$. To illustrate this effect, the tolerable range of errors $(\epsilon_\delta, \epsilon_z)$ for which the fidelity $\mathcal{F}$ remains large is contrasted between the optimal early-time transient and a long-time sustained scrambling regime in Fig.~\ref{fig:imperfections}, with the former having a much higher tolerance for errors. We also see that $\mathcal{F}$ is more sensitive to $\epsilon_{\delta}$ than $\epsilon_z$, as $\Delta E_{S_z} \sim N_A$ while $\Delta E_{b^\dagger b} \sim \phi$, and we need large $\phi$ (corresponding to a large effective $d_B$) for successful information spreading via scrambling. Alternatively, instead of mitigating reversal errors, one can increase the time $t$ to \textit{improve} sensitivity to these errors, if one is interested in  \textit{metrology applications}  e.g. sensitivity to small differences in parameters between the otherwise largely identical systems $A$ and $C$.  In this way, chaos in the sense of sensitivity to parameters becomes relevant to our protocol.

As an illustration, Fig.~\ref{fig:imperfections} contrasts the sensitivity to reversal errors at an early operating time, using the red-star parameters from Fig.~\ref{fig:transients} ($gt\simeq16.8$, top row), versus deep in the long-time dynamics, using the parameter set from Fig.~\ref{fig:fourpanel}c ($gt\simeq750$, bottom row). The early-time contours remain broad in $(\epsilon_\delta/\delta,\epsilon_z/\omega_z)$, while at long times they collapse and narrow dramatically (note the axis rescaling), revealing orders-of-magnitude tighter tolerances due to the accumulation of phase errors.

\end{appendix}

\bibliography{Scrambling_and_recovery}

@article{bidirectionalteleportation1,
  title={Quantum remote control: teleportation of unitary operations},
  author={Huelga, Susana F and Vaccaro, Joan A and Chefles, Anthony and Plenio, Martin B},
  journal={Phys. Rev. A},
  volume={63},
  number={4},
  pages={042303},
  year={2001},
  publisher={APS},
  url={https://doi.org/10.1103/PhysRevA.63.042303}
}

@article{bidirectionalteleportation2,
  title={Remote control of restricted sets of operations: teleportation of angles},
  author={Huelga, Susana F and Plenio, Martin B and Vaccaro, Joan A},
  journal={Phys. Rev. A},
  volume={65},
  number={4},
  pages={042316},
  year={2002},
  publisher={APS},
  url={https://doi.org/10.1103/PhysRevA.65.042316}
}

@article{BQT_SWAP1,
  title={Quantifying the performance of bidirectional quantum teleportation},
  author={Siddiqui, Aliza U and Wilde, Mark M},
  journal={Canadian Journal of Physics},
  volume={103},
  number={2},
  pages={174--214},
  year={2024},
  publisher={Canadian Science Publishing 123 Slater Street, Suite 610, Ottawa, Ontario~…},
  url={https://doi.org/10.1139/cjp-2023-0233}
}

@article{BQT_SWAP2,
  title={{The SWAP imposter: Bidirectional quantum teleportation and its performance}},
  author={Siddiqui, Aliza U and Wilde, Mark M},
  journal={AVS Quantum Sci.},
  volume={5},
  number={1},
  year={2023},
  publisher={AIP Publishing},
  url={https://doi.org/10.1116/5.0135467}
}

@article{Cheng2020PRResearch,
  title   = {Realizing the {Hayden-Preskill} protocol with coupled {Dicke} models},
  author  = {Cheng, Yanting and Liu, Chang and Guo, Jinkang and Chen, Yu and Zhang, Pengfei and Zhai, Hui},
  journal = {Phys. Rev. Res.},
  volume  = {2},
  pages   = {043024},
  year    = {2020},
  doi     = {10.1103/PhysRevResearch.2.043024}
}

@article{YoshidaKitaev,
  title={Efficient decoding for the {Hayden-Preskill} protocol},
  author={Yoshida, Beni and Kitaev, Alexei},
  journal={arXiv preprint arXiv:1710.03363},
  year={2017},
url={https://doi.org/10.48550/arXiv.1710.03363}
}

@article{YoshidaYao,
  title={Disentangling scrambling and decoherence via quantum teleportation},
  author={Yoshida, Beni and Yao, Norman Y},
  journal={Phys. Rev. X},
  volume={9},
  number={1},
  pages={011006},
  year={2019},
  publisher={APS},
url={https://doi.org/10.1103/PhysRevX.9.011006}
}

@article{HPPetz_Nakayamaetal,
  title={The {Petz} (lite) recovery map for the scrambling channel},
  author={Nakayama, Yasuaki and Miyata, Akihiro and Ugajin, Tomonori},
  journal={Prog. Theor. Exp. Phys.},
  volume={2023},
  number={12},
  pages={123B04},
  year={2023},
  publisher={Oxford University Press},
url={https://doi.org/10.1093/ptep/ptad147}
}

@book{NielsenChuang,
title={Quantum Computation and Quantum Information},
publisher={Cambridge University Press},
author={Nielsen, Michael A. and Chuang, Isaac L.},
year={2010},
doi={10.1017/CBO9780511976667},
url={https://doi.org/10.1017/CBO9780511976667}
}

@article{dynamicalqentanglement,
  title={Enhanced entanglement from quantum ergodicity},
  author={Vikram, Amit},
  journal={arXiv preprint arXiv:2507.08067},
  year={2025},
url={https://doi.org/10.48550/arXiv.2507.08067}
}

@article{BennettTeleportation,
  title={Teleporting an unknown quantum state via dual classical and {Einstein-Podolsky-Rosen} channels},
  author={Bennett, Charles H and Brassard, Gilles and Cr{\'e}peau, Claude and Jozsa, Richard and Peres, Asher and Wootters, William K},
  journal={Phys. Rev. Lett.},
  volume={70},
  number={13},
  pages={1895},
  year={1993},
  publisher={APS},
url={https://doi.org/10.1103/PhysRevLett.70.1895}
}

@article{HosurQiRobertsYoshida,
  title={Chaos in quantum channels},
  author={Hosur, Pavan and Qi, Xiao-Liang and Roberts, Daniel A and Yoshida, Beni},
  journal={J. High Energy Phys.},
  volume={2016},
  number={2},
  pages={1--49},
  year={2016},
  publisher={Springer},
url={https://doi.org/10.1007/JHEP02(2016)004}
}

@article{li2025optimality,
  title={Optimality Condition for the {Petz} Map},
  author={Li, Bikun and Wang, Zhaoyou and Zheng, Guo and Wong, Yat and Jiang, Liang},
  journal={Phys. Rev. Lett.},
  volume={134},
  number={20},
  pages={200602},
  year={2025},
  publisher={APS},
url={https://doi.org/10.1103/PhysRevLett.134.200602}
}

@article{KnillLaflamme,
  title={Theory of quantum error-correcting codes},
  author={Knill, Emanuel and Laflamme, Raymond},
  journal={Phys. Rev. A},
  volume={55},
  number={2},
  pages={900},
  year={1997},
  publisher={APS},
url={https://doi.org/10.1103/PhysRevA.55.900}
}

@article{BarnumKnill,
  title={Reversing quantum dynamics with near-optimal quantum and classical fidelity},
  author={Barnum, Howard and Knill, Emanuel},
  journal={J. Math. Phys.},
  volume={43},
  number={5},
  pages={2097--2106},
  year={2002},
  publisher={American Institute of Physics},
url={https://doi.org/10.1063/1.1459754}
}

@article{Petz,
  title={Sufficiency of channels over von Neumann algebras},
  author={Petz, D{\'e}nes},
  journal={The Quarterly Journal of Mathematics},
  volume={39},
  number={1},
  pages={97--108},
  year={1988},
  publisher={Oxford University Press},
url={https://doi.org/10.1093/qmath/39.1.97}
}

@article{YKrecoveryClifford,
  title={Experimental Decoding Scrambled Quantum Information from the Future},
  author={Huang, Yi-Te and Huang, Siang-Wei and Lin, Jhen-Dong and Miranowicz, Adam and Lambert, Neill and Chen, Guang-Yin and Nori, Franco and Chen, Yueh-Nan},
  journal={arXiv preprint arXiv:2501.16335},
  year={2025}
}

@article{HaydenPreskill,
  title={Black holes as mirrors: quantum information in random subsystems},
  author={Hayden, Patrick and Preskill, John},
  journal={J. High Energy Phys.},
  volume={2007},
  number={09},
  pages={120},
  year={2007},
 publisher={IOP Publishing},
  url = {https://doi.org/10.1088/1126-6708/2007/09/120}
}

@article{HorodeckiEntanglementReview,
  title={Quantum entanglement},
  author={Horodecki, Ryszard and Horodecki, Pawe{\l} and Horodecki, Micha{\l} and Horodecki, Karol},
  journal={Rev. Mod. Phys.},
  volume={81},
  number={2},
  pages={865},
  year={2009},
  publisher={APS},
url={https://doi.org/10.1103/RevModPhys.81.865}
}

@article{SekinoSusskind,
  title={Fast scramblers},
  author={Sekino, Yasuhiro and Susskind, Leonard},
  journal={J. High Energy Phys.},
  volume={2008},
  number={10},
  pages={065},
  year={2008},
  publisher={IOP Publishing},
  url={https://doi.org/10.1088/1126-6708/2008/10/065}
}

@article{SusskindThorlaciusGedanken,
  title={Gedanken experiments involving black holes},
  author={Susskind, Leonard and Thorlacius, Larus},
  journal={Phys. Rev. D},
  volume={49},
  number={2},
  pages={966},
  year={1994},
  publisher={APS},
url={https://doi.org/10.1103/PhysRevD.49.966}
}

@article{dynamicalqspeedlimit,
  title={Exact universal bounds on quantum dynamics and fast scrambling},
  author={Vikram, Amit and Galitski, Victor},
  journal={Phys. Rev. Lett.},
  volume={132},
  number={4},
  pages={040402},
  year={2024},
  publisher={APS},
  url={https://doi.org/10.1103/PhysRevLett.132.040402},
archivePrefix = {arXiv},
arxivId = {2212.14021}
}

@article{dynamicalqfastscrambling,
  title={Proof of a Universal Speed Limit on Fast Scrambling in Quantum Systems},
  author={Vikram, Amit and Shou, Laura and Galitski, Victor},
  journal={Phys. Rev. Lett.},
  volume={136},
  number={15},
  pages={150401},
  year={2026},
  publisher={APS},
  url={https://doi.org/10.1103/y9z4-v641}
}

@article{LashkariFastScrambling,
  title={Towards the fast scrambling conjecture},
  author={Lashkari, Nima and Stanford, Douglas and Hastings, Matthew and Osborne, Tobias and Hayden, Patrick},
  journal={J. High Energy Phys.},
  volume={2013},
  number={4},
  pages={1--33},
  year={2013},
  publisher={Springer},
  url={https://doi.org/10.1007/JHEP04(2013)022}
}

@article{BentsenGuLucasScrambling,
  title={Fast scrambling on sparse graphs},
  author={Bentsen, Gregory and Gu, Yingfei and Lucas, Andrew},
  journal={Proc. Natl. Acad. Sci. U.S.A.},
  volume={116},
  number={14},
  pages={6689--6694},
  year={2019},
  publisher={National Acad Sciences},
  url={https://doi.org/10.1073/pnas.1811033116}
}

@article{DecouplingOneShot,
  title={One-shot decoupling},
  author={Dupuis, Fr{\'e}d{\'e}ric and Berta, Mario and Wullschleger, J{\"u}rg and Renner, Renato},
  journal={Commun. Math. Phys.},
  volume={328},
  number={1},
  pages={251--284},
  year={2014},
  publisher={Springer},
url={https://doi.org/10.1007/s00220-014-1990-4}
}

@article{LewisSwan2021,
  title = {Characterizing the dynamical phase diagram of the Dicke model via classical and quantum probes},
  volume = {3},
  ISSN = {2643-1564},
  url = {http://dx.doi.org/10.1103/PhysRevResearch.3.L022020},
  DOI = {10.1103/physrevresearch.3.l022020},
  number = {2},
  journal = {Physical Review Research},
  publisher = {American Physical Society (APS)},
  author = {Lewis-Swan,  R. J. and Muleady,  S. R. and Barberena,  D. and Bollinger,  J. J. and Rey,  A. M.},
  year = {2021},
  month = jun 
}

@article{Dicke1954,
  title = {Coherence in Spontaneous Radiation Processes},
  volume = {93},
  ISSN = {0031-899X},
  url = {http://dx.doi.org/10.1103/PhysRev.93.99},
  DOI = {10.1103/physrev.93.99},
  number = {1},
  journal = {Physical Review},
  publisher = {American Physical Society (APS)},
  author = {Dicke,  R. H.},
  year = {1954},
  month = jan,
  pages = {99–110}
}

@article{Baumann2010,
  title = {Dicke quantum phase transition with a superfluid gas in an optical cavity},
  volume = {464},
  ISSN = {1476-4687},
  url = {http://dx.doi.org/10.1038/nature09009},
  DOI = {10.1038/nature09009},
  number = {7293},
  journal = {Nature},
  publisher = {Springer Science and Business Media LLC},
  author = {Baumann,  Kristian and Guerlin,  Christine and Brennecke,  Ferdinand and Esslinger,  Tilman},
  year = {2010},
  month = apr,
  pages = {1301–1306}
}

@article{Zhang2018,
  title = {Dicke-model simulation via cavity-assisted Raman transitions},
  volume = {97},
  ISSN = {2469-9934},
  url = {http://dx.doi.org/10.1103/PhysRevA.97.043858},
  DOI = {10.1103/physreva.97.043858},
  number = {4},
  journal = {Physical Review A},
  publisher = {American Physical Society (APS)},
  author = {Zhang,  Zhiqiang and Lee,  Chern Hui and Kumar,  Ravi and Arnold,  K. J. and Masson,  Stuart J. and Grimsmo,  A. L. and Parkins,  A. S. and Barrett,  M. D.},
  year = {2018},
  month = apr 
}

@article{SafaviNaini2018,
  title = {Verification of a Many-Ion Simulator of the Dicke Model Through Slow Quenches across a Phase Transition},
  volume = {121},
  ISSN = {1079-7114},
  url = {http://dx.doi.org/10.1103/PhysRevLett.121.040503},
  DOI = {10.1103/physrevlett.121.040503},
  number = {4},
  journal = {Physical Review Letters},
  publisher = {American Physical Society (APS)},
  author = {Safavi-Naini,  A. and Lewis-Swan,  R. J. and Bohnet,  J. G. and G\"{a}rttner,  M. and Gilmore,  K. A. and Jordan,  J. E. and Cohn,  J. and Freericks,  J. K. and Rey,  A. M. and Bollinger,  J. J.},
  year = {2018},
  month = jul 
}

@article{Maziero2015,
  title = {Random Sampling of Quantum States: a Survey of Methods: And Some Issues Regarding the Overparametrized Method},
  volume = {45},
  ISSN = {1678-4448},
  url = {http://dx.doi.org/10.1007/s13538-015-0367-2},
  DOI = {10.1007/s13538-015-0367-2},
  number = {6},
  journal = {Brazilian Journal of Physics},
  publisher = {Springer Science and Business Media LLC},
  author = {Maziero,  Jonas},
  year = {2015},
  month = sep,
  pages = {575–583}
}

@article{HighDimEntanglementReview,
  title={Advances in high-dimensional quantum entanglement},
  author={Erhard, Manuel and Krenn, Mario and Zeilinger, Anton},
  journal={Nat. Rev. Phys.},
  volume={2},
  number={7},
  pages={365--381},
  year={2020},
  publisher={Nature Publishing Group UK London},
url={https://doi.org/10.1038/s42254-020-0193-5}
}

@article{HighDimEntanglement1,
  title={Experimental high-dimensional quantum teleportation},
  author={Hu, Xiao-Min and Zhang, Chao and Liu, Bi-Heng and Cai, Yu and Ye, Xiang-Jun and Guo, Yu and Xing, Wen-Bo and Huang, Cen-Xiao and Huang, Yun-Feng and Li, Chuan-Feng and others},
  journal={Phys. Rev. Lett.},
  volume={125},
  number={23},
  pages={230501},
  year={2020},
  publisher={APS},
url={https://doi.org/10.1103/PhysRevLett.125.230501}
}

@article{HighDimEntanglement2,
  title={Quantum teleportation in high dimensions},
  author={Luo, Yi-Han and Zhong, Han-Sen and Erhard, Manuel and Wang, Xi-Lin and Peng, Li-Chao and Krenn, Mario and Jiang, Xiao and Li, Li and Liu, Nai-Le and Lu, Chao-Yang and others},
  journal={Phys. Rev. Lett.},
  volume={123},
  number={7},
  pages={070505},
  year={2019},
  publisher={APS},
url={https://doi.org/10.1103/PhysRevLett.123.070505}
}

@article{TeleportationReview1,
  title={Advances in quantum teleportation},
  author={Pirandola, Stefano and Eisert, Jens and Weedbrook, Christian and Furusawa, Akira and Braunstein, Samuel L},
  journal={Nature Photon.},
  volume={9},
  number={10},
  pages={641--652},
  year={2015},
  publisher={Nature Publishing Group UK London},
url={https://doi.org/10.1038/nphoton.2015.154}
}

@article{TeleportationReview2,
  title={Progress in quantum teleportation},
  author={Hu, Xiao-Min and Guo, Yu and Liu, Bi-Heng and Li, Chuan-Feng and Guo, Guang-Can},
  journal={Nat. Rev. Phys.},
  volume={5},
  number={6},
  pages={339--353},
  year={2023},
  publisher={Nature Publishing Group UK London},
url={https://doi.org/10.1038/s42254-023-00588-x}
}

@article{ChaosComplexityRMT,
  title={Chaos, complexity, and random matrices},
  author={Cotler, Jordan and Hunter-Jones, Nicholas and Liu, Junyu and Yoshida, Beni},
  journal={J. High Energy Phys.},
  volume={2017},
  number={11},
  pages={1--60},
  year={2017},
  publisher={Springer},
  url={https://doi.org/10.1007/JHEP11(2017)048}
}

@article{MSSotocBound,
  title={A bound on chaos},
  author={Maldacena, Juan and Shenker, Stephen H and Stanford, Douglas},
  journal={J. High Energy Phys.},
  volume={2016},
  number={8},
  pages={1--17},
  year={2016},
  publisher={Springer},
  url={https://doi.org/10.1007/JHEP08(2016)106}
}

@book{Mehta,
  title={Random matrices},
  author={Mehta, Madan Lal},
  year={2004},
  publisher={Elsevier},
  isbn={978-0-12-088409-4}
}

@book{Haake,
title={Quantum signatures of chaos},
author={Haake, Fritz},
year={2001},
publisher={Springer, Berlin, Heidelberg},
url={https://doi.org/10.1007/978-3-662-04506-0}}

@incollection{MT,
  title={The uncertainty relation between energy and time in non-relativistic quantum mechanics},
  author={Mandelstam, Leonid and Tamm, IG},
  booktitle={Selected papers},
  pages={115--123},
  year={1991},
  publisher={Springer},
  url={https://doi.org/10.1007/978-3-642-74626-0_8}
}

@article{ML,
  title={The maximum speed of dynamical evolution},
  author={Margolus, Norman and Levitin, Lev B},
  journal={Physica D},
  volume={120},
  number={1-2},
  pages={188--195},
  year={1998},
  publisher={Elsevier},
  url={https://doi.org/10.1016/S0167-2789(98)00054-2}
}

@article{Zhu2022,
   author = {Zhu, Qingling and Cao, Sirui and Chen, Fusheng and Chen, Ming-Cheng and Chen, Xiawei and Chung, Tung-Hsun and Deng, Hui and Du, Yajie and Fan, Daojin and Gong, Ming and Guo, Cheng and Guo, Chu and Guo, Shaojun and Han, Lianchen and Hong, Linyin and Huang, He-Liang and Huo, Yong-Heng and Li, Liping and Li, Na and Li, Shaowei and Li, Yuan and Liang, Futian and Lin, Chun and Lin, Jin and Qian, Haoran and Qiao, Dan and Rong, Hao and Su, Hong and Sun, Lihua and Wang, Liangyuan and Wang, Shiyu and Wu, Dachao and Wu, Yulin and Xu, Yu and Yan, Kai and Yang, Weifeng and Yang, Yang and Ye, Yangsen and Yin, Jianghan and Ying, Chong and Yu, Jiale and Zha, Chen and Zhang, Cha and Zhang, Haibin and Zhang, Kaili and Zhang, Yiming and Zhao, Han and Zhao, Youwei and Zhou, Liang and Lu, Chao-Yang and Peng, Cheng-Zhi and Zhu, Xiaobo and Pan, Jian-Wei},
   title = {Quantum computational advantage via 60-qubit 24-cycle random circuit sampling},
   journal = {Science Bulletin},
   volume = {67},
   number = {3},
   pages = {240-245},
   ISSN = {2095-9273},
   DOI = {10.1016/j.scib.2021.10.017},
   url = {https://www.sciencedirect.com/science/article/pii/S2095927321006733},
   year = {2022},
   type = {Journal Article}
}

@article{AAMT,
  title={Geometry of quantum evolution},
  author={Anandan, J and Aharonov, Yakir},
  journal={Phys. Rev. Lett.},
  volume={65},
  number={14},
  pages={1697},
  year={1990},
  publisher={APS},
  url={https://doi.org/10.1103/PhysRevLett.65.1697}
}

@article{LevitinToffoli,
  title={Fundamental limit on the rate of quantum dynamics: the unified bound is tight},
  author={Levitin, Lev B and Toffoli, Tommaso},
  journal={Phys. Rev. Lett.},
  volume={103},
  number={16},
  pages={160502},
  year={2009},
  publisher={APS},
  url={https://doi.org/10.1103/PhysRevLett.103.160502}
}

@article{ButterflyMetrology,
  title={A Universal Protocol for Quantum-Enhanced Sensing via Information Scrambling},
  author={Kobrin, Bryce and Schuster, Thomas and Block, Maxwell and Wu, Weijie and Mitchell, Bradley and Davis, Emily and Yao, Norman Y},
  journal={arXiv preprint arXiv:2411.12794},
  year={2024},
url={https://doi.org/10.48550/arXiv.2411.12794}
}

@article{ReversalButterflyMetrology,
  title={Scrambling Dynamics with Imperfections in a Solvable Model},
  author={LiTenn, Nadie Yiluo and Zhou, Tianci and Swingle, Brian},
  journal={arXiv preprint arXiv:2505.00070},
  year={2025},
url={https://doi.org/10.48550/arXiv.2505.00070}
}

@article{RenyiEntropyMeasurementTwoGates,
  title={Measuring {R{\'e}nyi} entropy using a projected {Loschmidt} echo},
  author={Zhou, Yi-Neng and L{\"o}wenberg, Robin and Sonner, Julian},
  journal={arXiv preprint arXiv:2504.05237},
  year={2025},
url={https://doi.org/10.48550/arXiv.2504.05237}
}

@article{Verifiedscrambling,
  title={Verified quantum information scrambling},
  author={Landsman, Kevin A and Figgatt, Caroline and Schuster, Thomas and Linke, Norbert M and Yoshida, Beni and Yao, Norman Y and Monroe, Christopher},
  journal={Nature},
  volume={567},
  number={7746},
  pages={61--65},
  year={2019},
  publisher={Nature Publishing Group UK London},
url={https://doi.org/10.1038/s41586-019-0952-6}
}

@article{ScramblingQutrits,
  title={Quantum information scrambling on a superconducting qutrit processor},
  author={Blok, Machiel S and Ramasesh, Vinay V and Schuster, Thomas and O’Brien, Kevin and Kreikebaum, John-Mark and Dahlen, Dar and Morvan, Alexis and Yoshida, Beni and Yao, Norman Y and Siddiqi, Irfan},
  journal={Phys. Rev. X},
  volume={11},
  number={2},
  pages={021010},
  year={2021},
  publisher={APS},
  url={https://doi.org/10.1103/PhysRevX.11.021010}
}

@article{QNDspin,
  title={Quantum measurement of a mesoscopic spin ensemble},
  author={Giedke, Geza and Taylor, Jacob M and D’Alessandro, Domenico and Lukin, Mikhail D and Imamo{\u{g}}lu, A},
  journal={Phys. Rev. A},
  volume={74},
  number={3},
  pages={032316},
  year={2006},
  publisher={APS},
url={https://doi.org/10.1103/PhysRevA.74.032316}
}

@article{QNDspinQDOT,
  title={Controlling a mesoscopic spin environment by quantum bit manipulation},
  author={Taylor, JM and Imamoglu, Atac and Lukin, MD},
  journal={Phys. Rev. Lett.},
  volume={91},
  number={24},
  pages={246802},
  year={2003},
  publisher={APS},
url={https://doi.org/10.1103/PhysRevLett.91.246802}
}

@article{SFFmeas,
  title={Monitoring quantum simulators via quantum nondemolition couplings to atomic clock qubits},
  author={Vasilyev, Denis V and Grankin, Andrey and Baranov, Mikhail A and Sieberer, Lukas M and Zoller, Peter},
  journal={PRX Quantum},
  volume={1},
  number={2},
  pages={020302},
  year={2020},
  publisher={APS},
  url={https://doi.org/10.1103/PRXQuantum.1.020302}
}

@article{TezukaSYKScrambling,
  title={{Hayden-Preskill recovery in Hamiltonian systems}},
  author={Nakata, Yoshifumi and Tezuka, Masaki},
  journal={arXiv preprint arXiv:2303.02010},
  year={2023},
url={https://doi.org/10.48550/arXiv.2303.02010}
}

@article{PageSubsystem,
  title={Average entropy of a subsystem},
  author={Page, Don N},
  journal={Phys. Rev. Lett.},
  volume={71},
  number={9},
  pages={1291},
  year={1993},
  publisher={APS},
url={https://doi.org/10.1103/PhysRevLett.71.1291}
}

@article{PageTime,
  title={Time dependence of {Hawking} radiation entropy},
  author={Page, Don N},
  journal={J. Cosmol. Astropart. Phys.},
  volume={2013},
  number={09},
  pages={028},
  year={2013},
  publisher={IOP Publishing},
url={https://doi.org/10.1088/1475-7516/2013/09/028}
}

@article{TakayanagiEssay,
  title = {Essay: Emergent Holographic Spacetime from Quantum Information},
  author = {Takayanagi, Tadashi},
  journal = {Phys. Rev. Lett.},
  volume = {134},
  issue = {24},
  pages = {240001},
  numpages = {9},
  year = {2025},
  month = {Jun},
  publisher = {American Physical Society},
  doi = {10.1103/pg4r-fy8n},
  url = {https://link.aps.org/doi/10.1103/pg4r-fy8n}
}

@article{Maldacena1997,
  title={The large {N} limit of superconformal field theories and supergravity},
  author={Maldacena, Juan M.},
  journal={Advances in Theoretical and Mathematical Physics},
  volume={2},
  pages={231--252},
  year={1998},
  eprint={hep-th/9711200}
}

@article{Preskill2000,
  title = {Quantum information and physics: Some future directions},
  volume = {47},
  ISSN = {1362-3044},
  url = {http://dx.doi.org/10.1080/09500340008244031},
  DOI = {10.1080/09500340008244031},
  number = {2–3},
  journal = {Journal of Modern Optics},
  publisher = {Informa UK Limited},
  author = {Preskill,  John},
  year = {2000},
  month = feb,
  pages = {127–137}
}

@article{Harlow2016,
  title={Jerusalem lectures on black holes and quantum information},
  author={Harlow, Daniel},
  journal={Reviews of Modern Physics},
  volume={88},
  number={1},
  pages={015002},
  year={2016},
  doi={10.1103/RevModPhys.88.015002}
}

@article{GoogleScrambling,
  title={Information scrambling in quantum circuits},
  author={Mi, Xiao and Roushan, Pedram and Quintana, Chris and Mandr{\`a}, Salvatore and Marshall, Jeffrey and Neill, Charles and Arute, Frank and Arya, Kunal and Atalaya, Juan and Babbush, Ryan and others},
  journal={Science},
  volume={374},
  number={6574},
  pages={1479--1483},
  year={2021},
  publisher={American Association for the Advancement of Science},
  url={https://doi.org/10.1126/science.abg5029}
}

@Book{ShankarQM,
  author    = {Shankar, Ramamurti},
  year      = {2012},
  title     = {Principles of quantum mechanics},
  isbn      = {978-1-4757-0576-8},
  publisher = {Springer Science \& Business Media},
  url       = {https://doi.org/10.1007/978-1-4757-0576-8},
}

@article{Li2023,
   author = {Li, Zeyang and Colombo, Simone and Shu, Chi and Velez, Gustavo and Pilatowsky-Cameo, Saúl and Schmied, Roman and Choi, Soonwon and Lukin, Mikhail and Pedrozo-Peñafiel, Edwin and Vuletić, Vladan},
   title = {Improving metrology with quantum scrambling},
   journal = {Science},
   volume = {380},
   number = {6652},
   pages = {1381-1384},
   DOI = {10.1126/science.adg9500},
   url = {https://doi.org/10.1126/science.adg9500},
   year = {2023},
   type = {Journal Article}
}

@article{Zhang2025,
  title = {Harnessing quantum chaos in spin-boson models for all-purpose quantum-enhanced sensing},
  author = {Zhang, Y. and Zu\~niga Castro, J. C. and Lewis-Swan, R. J.},
  journal = {Phys. Rev. Res.},
  volume = {7},
  issue = {1},
  pages = {013227},
  numpages = {14},
  year = {2025},
  month = {Feb},
  publisher = {American Physical Society},
  doi = {10.1103/PhysRevResearch.7.013227},
  url = {https://link.aps.org/doi/10.1103/PhysRevResearch.7.013227}
}

@inproceedings{Aharonov2023,
author = {Aharonov, Dorit and Gao, Xun and Landau, Zeph and Liu, Yunchao and Vazirani, Umesh},
title = {A Polynomial-Time Classical Algorithm for Noisy Random Circuit Sampling},
year = {2023},
isbn = {9781450399135},
publisher = {Association for Computing Machinery},
address = {New York, NY, USA},
url = {https://doi.org/10.1145/3564246.3585234},
doi = {10.1145/3564246.3585234},
booktitle = {Proceedings of the 55th Annual ACM Symposium on Theory of Computing},
pages = {945–957},
numpages = {13},
location = {Orlando, FL, USA},
series = {STOC 2023}
}

@article{BoixoSupremacy,
  title={Characterizing quantum supremacy in near-term devices},
  author={Boixo, Sergio and others},
  journal={Nature Physics},
  volume={14},
  pages={595--600},
  year={2018},
  doi={10.1038/s41567-018-0124-x}
}

@article{GaoJafferisWall,
  title={Traversable wormholes via a double trace deformation},
  author={Gao, Ping and Jafferis, Daniel L. and Wall, Aron C.},
  journal={Journal of High Energy Physics},
  volume={2017},
  number={12},
  pages={151},
  year={2017},
  doi={10.1007/JHEP12(2017)151}
}

@article{Brown2023,
  title = {Quantum Gravity in the Lab. {I}. {Teleportation} by Size and Traversable Wormholes},
  volume = {4},
  ISSN = {2691-3399},
  url = {http://dx.doi.org/10.1103/PRXQuantum.4.010320},

  number = {1},
  journal = {PRX Quantum},
  publisher = {American Physical Society (APS)},
  author = {Brown,  Adam R. and Gharibyan,  Hrant and Leichenauer,  Stefan and Lin,  Henry W. and Nezami,  Sepehr and Salton,  Grant and Susskind,  Leonard and Swingle,  Brian and Walter,  Michael},
  year = {2023},
  month = feb 
}

@article{Fink2009,
  title={Dressed Collective Qubit States and the Tavis--Cummings Model in Circuit QED},
  author={Fink, J. M. and others},
  journal={Physical Review Letters},
  volume={103},
  pages={083601},
  year={2009},
  doi={10.1103/PhysRevLett.103.083601}
}

@article{Ritsch2013,
  title        = {Cold atoms in cavity-generated dynamical optical potentials},
  author       = {Ritsch, Helmut and Domokos, Peter and Brennecke, Ferdinand and Esslinger, Tilman},
  journal      = {Reviews of Modern Physics},
  volume       = {85},
  number       = {2},
  pages        = {553--601},
  year         = {2013},
  doi          = {10.1103/RevModPhys.85.553}
}

@article{Aedo2018,
  title        = {Analog quantum simulation of generalized Dicke models in trapped ions},
  author       = {Aedo, Ibai and Lamata, Lucas},
  journal      = {Physical Review A},
  volume       = {97},
  number       = {4},
  pages        = {042317},
  year         = {2018},
  doi          = {10.1103/PhysRevA.97.042317}
}

@article{HeppLieb1973,
  title = {Equilibrium Statistical Mechanics of Matter Interacting with the Quantized Radiation Field},
  author = {Hepp, K. and Lieb, E. H.},
  journal = {Physical Review A},
  volume = {8},
  pages = {2517--2525},
  year = {1973},
  doi = {10.1103/PhysRevA.8.2517}
}

@article{Black2003,
  title = {Observation of Collective Friction Forces due to Spatial Self-Organization of Atoms: From Rayleigh to Bragg Scattering},
  author = {Black, A. T. and Chan, H. W. and Vuletić, V.},
  journal = {Physical Review Letters},
  volume = {91},
  pages = {203001},
  year = {2003},
  doi = {10.1103/PhysRevLett.91.203001}
}

@article{Maschler2005,
  title = {Cold Atom Dynamics in a Quantum Optical Lattice Potential},
  author = {Maschler, C. and Ritsch, H.},
  journal = {Physical Review Letters},
  volume = {95},
  pages = {260401},
  year = {2005},
  doi = {10.1103/PhysRevLett.95.260401}
}

@article{Domokos2002,
  title = {Collective Cooling and Self-Organization of Atoms in a Cavity},
  volume = {89},
  ISSN = {1079-7114},
  url = {http://dx.doi.org/10.1103/PhysRevLett.89.253003},
  DOI = {10.1103/physrevlett.89.253003},
  number = {25},
  journal = {Physical Review Letters},
  publisher = {American Physical Society (APS)},
  author = {Domokos,  Peter and Ritsch,  Helmut},
  year = {2002},
  month = dec 
}

@article{Nagy2010,
  title = {Dicke-model phase transition in the quantum motion of a Bose-Einstein condensate in an optical cavity},
  author = {Nagy, D{\'a}vid and K{\'o}nya, G{\'a}bor and Szirmai, Gergely and Domokos, Peter},
  journal = {Physical Review Letters},
  volume = {104},
  pages = {130401},
  year = {2010},
  doi = {10.1103/PhysRevLett.104.130401}
}

@article{Brennecke2013,
  title = {Real-time observation of fluctuations at the driven-dissipative Dicke phase transition},
  author = {Brennecke, F. and Mottl, R. and Baumann, K. and Landig, R. and Donner, T. and Esslinger, T.},
  journal = {PNAS},
  volume = {110},
  pages = {11763--11767},
  year = {2013},
  doi = {10.1073/pnas.1306993110}
}

@article{Klinder2015,
  title = {Dynamical phase transition in the open Dicke model},
  author = {Klinder, J. and Keßler, H. and Wolke, M. and Mathey, L. and Hemmerich, A.},
  journal = {PNAS},
  volume = {112},
  pages = {3290--3295},
  year = {2015},
  doi = {10.1073/pnas.1417132112}
}

@article{Leonard2017,
  title = {Supersolid formation in a quantum gas breaking a continuous translational symmetry},
  author = {Léonard, J. and Morales, A. and Zupancic, P. and Esslinger, T. and Donner, T.},
  journal = {Nature},
  volume = {543},
  pages = {87--90},
  year = {2017},
  doi = {10.1038/nature21067}
}

@article{Kollar2017,
  title = {Supermode-density-wave-polariton condensation with a Bose-Einstein condensate in a multimode cavity},
  author = {Kollár, A. J. and Papageorge, A. T. and Vaidya, V. D. and Guo, Y. and Keeling, J. and Lev, B. L.},
  journal = {Nature Communications},
  volume = {8},
  pages = {14386},
  year = {2017},
  doi = {10.1038/ncomms14386}
}

@article{Landini2018,
  title = {Formation of a Spin Texture in a Quantum Gas Coupled to a Cavity},
  author = {Landini, Manuele and Dogra, Nishant and Kroeger, Katrin and Hruby, Lorenz and Donner, Tobias and Esslinger, Tilman},
  journal = {Physical Review Letters},
  volume = {120},
  pages = {223602},
  year = {2018},
  doi = {10.1103/PhysRevLett.120.223602}
}

@article{Morales2019,
  title = {Two-mode Dicke model from nondegenerate polarization modes},
  author = {Morales, Andrea and Dreon, Davide and Li, Xiangliang and Baumg{\"a}rtner, Alexander and Zupancic, Philip and Donner, Tobias and Esslinger, Tilman},
  journal = {Physical Review A},
  volume = {100},
  pages = {013816},
  year = {2019},
  doi = {10.1103/PhysRevA.100.013816}
}

@article{Sutherland2019,
  title = {Analog quantum simulation of superradiance and subradiance in trapped ions},
  author = {Sutherland, R. T.},
  journal = {Physical Review A},
  volume = {100},
  pages = {061405},
  year = {2019},
  doi = {10.1103/PhysRevA.100.061405}
}

@article{Bohnet2016,
  title = {Quantum spin dynamics and entanglement generation with hundreds of trapped ions},
  author = {Bohnet, J. G. and Sawyer, B. C. and Britton, J. W. and Wall, M. L. and Rey, A. M. and Foss-Feig, M. and Bollinger, J. J.},
  journal = {Science},
  volume = {352},
  pages = {1297--1301},
  year = {2016},
  doi = {10.1126/science.aad9958}
}

@article{Yan2023,
  title = {Superradiant and Subradiant Cavity Scattering by Atom Arrays},
  author = {Yan, Z. and Ho, J. and Lu, Y.-H. and Masson, S. J. and Asenjo-Garcia, A. and Stamper-Kurn, D. M.},
  journal = {Physical Review Letters},
  volume = {131},
  pages = {253603},
  year = {2023},
  doi = {10.1103/PhysRevLett.131.253603}
}

@article{Khan2025,
  title = {Generating {Einstein-Podolsky-Rosen} correlations for teleporting collective spin states in a two-dimensional trapped ion crystal},
  volume = {7},
  ISSN = {2643-1564},
  url = {http://dx.doi.org/10.1103/PhysRevResearch.7.L022019},
  number = {2},
  journal = {Phys. Rev. Res.},
  publisher = {American Physical Society (APS)},
  author = {Khan,  Muhammad Miskeen and Chaparro,  Edwin and Sundar,  Bhuvanesh and Carter,  Allison and Bollinger,  John and Molmer,  Klaus and Rey,  Ana Maria},
  year = {2025},
  month = apr 
}

@article{WamplerCooperPolarizedSWAP,
  title={Perfect State Transfer of Mixed States and Purification in Central Spin Systems},
  author={Wampler, Matthew and Cooper, Nigel R},
  journal={arXiv preprint arXiv:2508.12515},
  year={2025},
url={https://doi.org/10.48550/arXiv.2508.12515}
}

@article{Leibfried2003,
  title = {Quantum dynamics of single trapped ions},
  volume = {75},
  ISSN = {1539-0756},
  url = {http://dx.doi.org/10.1103/RevModPhys.75.281},
  DOI = {10.1103/revmodphys.75.281},
  number = {1},
  journal = {Reviews of Modern Physics},
  publisher = {American Physical Society (APS)},
  author = {Leibfried,  D. and Blatt,  R. and Monroe,  C. and Wineland,  D.},
  year = {2003},
  month = mar,
  pages = {281–324}
}

@article{Monroe2021,
  title = {Programmable quantum simulations of spin systems with trapped ions},
  volume = {93},
  ISSN = {1539-0756},
  url = {http://dx.doi.org/10.1103/RevModPhys.93.025001},
  DOI = {10.1103/revmodphys.93.025001},
  number = {2},
  journal = {Reviews of Modern Physics},
  publisher = {American Physical Society (APS)},
  author = {Monroe,  C. and Campbell,  W. C. and Duan,  L.-M. and Gong,  Z.-X. and Gorshkov,  A. V. and Hess,  P. W. and Islam,  R. and Kim,  K. and Linke,  N. M. and Pagano,  G. and Richerme,  P. and Senko,  C. and Yao,  N. Y.},
  year = {2021},
  month = apr 
}

@misc{SupplementalMaterial,
  howpublished = {See the Supplemental Material for a detailed discussion of scrambling mechanisms, analytical results for random matrices, additional numerical details, technical details of the Dicke model, and the implementation of projections to a boson coherent state.},
}

@article{Uys2010,
  title = {Decoherence due to Elastic Rayleigh Scattering},
  volume = {105},
  ISSN = {1079-7114},
  url = {http://dx.doi.org/10.1103/PhysRevLett.105.200401},
  DOI = {10.1103/physrevlett.105.200401},
  number = {20},
  journal = {Physical Review Letters},
  publisher = {American Physical Society (APS)},
  author = {Uys,  H. and Biercuk,  M. J. and VanDevender,  A. P. and Ospelkaus,  C. and Meiser,  D. and Ozeri,  R. and Bollinger,  J. J.},
  year = {2010},
  month = nov 
}

@book{Agarwal2012,
  title = {Quantum Optics},
  ISBN = {9781139035170},
  url = {http://dx.doi.org/10.1017/CBO9781139035170},
  DOI = {10.1017/cbo9781139035170},
  publisher = {Cambridge University Press},
  author = {Agarwal,  Girish S.},
  year = {2012},
  month = nov 
}

@article{LewisSwan2019,
  title = {Unifying scrambling,  thermalization and entanglement through measurement of fidelity out-of-time-order correlators in the Dicke model},
  volume = {10},
  ISSN = {2041-1723},
  url = {http://dx.doi.org/10.1038/s41467-019-09436-y},
  DOI = {10.1038/s41467-019-09436-y},
  number = {1},
  journal = {Nature Communications},
  publisher = {Springer Science and Business Media LLC},
  author = {Lewis-Swan,  R. J. and Safavi-Naini,  A. and Bollinger,  J. J. and Rey,  A. M.},
  year = {2019},
  month = apr 
}

@article{ChvezCarlos2019,
  title = {Quantum and Classical Lyapunov Exponents in Atom-Field Interaction Systems},
  volume = {122},
  ISSN = {1079-7114},
  url = {http://dx.doi.org/10.1103/PhysRevLett.122.024101},
  DOI = {10.1103/physrevlett.122.024101},
  number = {2},
  journal = {Physical Review Letters},
  publisher = {American Physical Society (APS)},
  author = {Chávez-Carlos,  Jorge and López-del-Carpio,  B. and Bastarrachea-Magnani,  Miguel A. and Stránský,  Pavel and Lerma-Hernández,  Sergio and Santos,  Lea F. and Hirsch,  Jorge G.},
  year = {2019},
  month = jan 
}

@article{Buijsman2017,
  title = {Nonergodicity in the Anisotropic Dicke Model},
  volume = {118},
  ISSN = {1079-7114},
  url = {http://dx.doi.org/10.1103/PhysRevLett.118.080601},
  DOI = {10.1103/physrevlett.118.080601},
  number = {8},
  journal = {Physical Review Letters},
  publisher = {American Physical Society (APS)},
  author = {Buijsman,  Wouter and Gritsev,  Vladimir and Sprik,  Rudolf},
  year = {2017},
  month = feb 
}

@article{Kirkova2022,
  title = {Out-of-time-order correlator in the quantum Rabi model},
  volume = {105},
  ISSN = {2469-9934},
  url = {http://dx.doi.org/10.1103/PhysRevA.105.032444},
  DOI = {10.1103/physreva.105.032444},
  number = {3},
  journal = {Physical Review A},
  publisher = {American Physical Society (APS)},
  author = {Kirkova,  Aleksandrina V. and Porras,  Diego and Ivanov,  Peter A.},
  year = {2022},
  month = mar 
}

@article{Hammerer2010,
  title = {Quantum interface between light and atomic ensembles},
  volume = {82},
  ISSN = {1539-0756},
  url = {http://dx.doi.org/10.1103/RevModPhys.82.1041},
  DOI = {10.1103/revmodphys.82.1041},
  number = {2},
  journal = {Reviews of Modern Physics},
  publisher = {American Physical Society (APS)},
  author = {Hammerer,  Klemens and Sørensen,  Anders S. and Polzik,  Eugene S.},
  year = {2010},
  month = apr,
  pages = {1041–1093}
}

@article{Chen2021,
  title = {Quantum computation and simulation with vibrational modes of trapped ions},
  volume = {30},
  ISSN = {1674-1056},
  url = {http://dx.doi.org/10.1088/1674-1056/ac01e3},
  DOI = {10.1088/1674-1056/ac01e3},
  number = {6},
  journal = {Chinese Physics B},
  publisher = {IOP Publishing},
  author = {Chen,  Wentao and Gan,  Jaren and Zhang,  Jing-Ning and Matuskevich,  Dzmitry and Kim,  Kihwan},
  year = {2021},
  month = jun,
  pages = {060311}
}

@article{Regnault2016,
author = {Regnault, Nicolas and Nandkishore, Rahul},
doi = {10.1103/PhysRevB.93.104203},
issn = {24699969},
journal = {Phys. Rev. B},
number = {10},
pages = {104203},
title = {{Floquet thermalization: Symmetries and random matrix ensembles}},
volume = {93},
year = {2016}
}

@article{GarrattChalker,
  title = {Local Pairing of Feynman Histories in Many-Body Floquet Models},
  author = {Garratt, S. J. and Chalker, J. T.},
  journal = {Phys. Rev. X},
  volume = {11},
  issue = {2},
  pages = {021051},
  numpages = {31},
  year = {2021},
  month = {Jun},
  publisher = {American Physical Society},
  doi = {10.1103/PhysRevX.11.021051},
  url = {https://link.aps.org/doi/10.1103/PhysRevX.11.021051}
}

@article{lazarides2014floqueteth,
  title = {Equilibrium states of generic quantum systems subject to periodic driving},
  author = {Lazarides, Achilleas and Das, Arnab and Moessner, Roderich},
  journal = {Phys. Rev. E},
  volume = {90},
  issue = {1},
  pages = {012110},
  numpages = {6},
  year = {2014},
  month = {Jul},
  publisher = {American Physical Society},
  doi = {10.1103/PhysRevE.90.012110},
  url = {https://link.aps.org/doi/10.1103/PhysRevE.90.012110}
}

@article{ponte2015floqueteth,
author = {Ponte, Pedro and Chandran, Anushya and Papi{\'{c}}, Z and Abanin, Dmitry A},
doi = {10.1016/j.aop.2014.11.008},
issn = {0003-4916},
journal = {Ann. Phys.},
pages = {196--204},
title = {{Periodically driven ergodic and many-body localized quantum systems}},
url = {https://www.sciencedirect.com/science/article/pii/S0003491614003212},
volume = {353},
year = {2015}
}

@article{kim2014floqueteth,
  title = {Testing whether all eigenstates obey the eigenstate thermalization hypothesis},
  author = {Kim, Hyungwon and Ikeda, Tatsuhiko N. and Huse, David A.},
  journal = {Phys. Rev. E},
  volume = {90},
  issue = {5},
  pages = {052105},
  numpages = {8},
  year = {2014},
  month = {Nov},
  publisher = {American Physical Society},
  doi = {10.1103/PhysRevE.90.052105},
  url = {https://link.aps.org/doi/10.1103/PhysRevE.90.052105}
}

@article{PhysRevX.4.041048,
  title = {Long-time Behavior of Isolated Periodically Driven Interacting Lattice Systems},
  author = {D'Alessio, Luca and Rigol, Marcos},
  journal = {Phys. Rev. X},
  volume = {4},
  issue = {4},
  pages = {041048},
  numpages = {12},
  year = {2014},
  month = {Dec},
  publisher = {American Physical Society},
  doi = {10.1103/PhysRevX.4.041048},
  url = {https://link.aps.org/doi/10.1103/PhysRevX.4.041048}
}

@misc{ChaparroZenodo,
  author       = {Chaparro, Edwin},
  title        = {{Source code for numerical data in "Bidirectional
                   teleportation using scrambling dynamics: a
                   practical protocol"}
                  },
  month        = sep,
  year         = 2026,
  publisher    = {Zenodo},
  version      = {V.1},
  doi          = {10.5281/zenodo.23050900},
  howpublished = {\url{https://doi.org/10.5281/zenodo.23050900}}
}

\end{document}


\title{Bidirectional teleportation using scrambling dynamics: a practical protocol \\ Supplemental Material}

\author{Amit Vikram}
\affiliation{JILA, University of Colorado and National Institute of Standards and Technology, and Department of Physics, University of Colorado, Boulder, Colorado 80309, USA}
\affiliation{Center for Theory of Quantum Matter and Department of Physics, University of Colorado, Boulder, Colorado 80309, USA}

\author{Edwin Chaparro}
\affiliation{JILA, University of Colorado and National Institute of Standards and Technology, and Department of Physics, University of Colorado, Boulder, Colorado 80309, USA}
\affiliation{Center for Theory of Quantum Matter and Department of Physics, University of Colorado, Boulder, Colorado 80309, USA}

\author{Muhammad Miskeen Khan}
\affiliation{JILA, University of Colorado and National Institute of Standards and Technology, and Department of Physics, University of Colorado, Boulder, Colorado 80309, USA}
\affiliation{Center for Theory of Quantum Matter and Department of Physics, University of Colorado, Boulder, Colorado 80309, USA}
\affiliation{Department of Electrical and Computer Engineering,
Saint Louis University, St. Louis, Missouri, 63103, USA}

\author{Andrew Lucas}
\affiliation{Center for Theory of Quantum Matter and Department of Physics, University of Colorado, Boulder, Colorado 80309, USA}

\author{Chris Akers}
\affiliation{Center for Theory of Quantum Matter and Department of Physics, University of Colorado, Boulder, Colorado 80309, USA}

\author{Ana Maria Rey}
\affiliation{JILA, University of Colorado and National Institute of Standards and Technology, and Department of Physics, University of Colorado, Boulder, Colorado 80309, USA}
\affiliation{Center for Theory of Quantum Matter and Department of Physics, University of Colorado, Boulder, Colorado 80309, USA}

\maketitle

\section{The probabilistic SWAP protocol: analytical details}

To perform a careful analysis of the mechanism behind our SWAP protocol, it is convenient to use the language of maps between operators acting on different Hilbert spaces. A map from operators on $A$ to those on $B$ may be denoted by
\begin{equation}
    \mathcal{M}_{BA}: L(\mathcal{H}_A) \to L(\mathcal{H}_B).
\end{equation}
When transferring information from an input subsystem $A$ to an intermediate subsystem $B$, we will call $\mathcal{M}_{BA}$ an encoding map. If $A$ is an intermediate subsystem, and $B$ an output subsystem, then $\mathcal{M}_{BA}$ will be called a decoding map.

We will look at specific properties that these encoding and decoding maps will have to satisfy to transfer information with high fidelity. Our analysis adds to previous analyses of information scrambling and recovery~\cite{YoshidaKitaev, YoshidaYao} by identifying precise necessary and sufficient conditions for the teleportation of individual states (for which only sufficient conditions have previously been obtained in the form of bounds on fidelity~\cite{YoshidaYao}), as opposed to an effective average over initial states via a coupling to a reference system as considered in previous work. Moreover, from an intuitive standpoint, rather than emphasizing decoupling or mutual information with reference systems as done previously, we emphasize that high fidelity recovery essentially follows from a simple mathematical identity (the definition of conjugates/pullbacks) \textit{in general} (see Sec.~\ref{subsec:pullbacksaredecoders}), and known variants of teleportation/recovery protocols can be understood as special cases of this identity (as discussed in Sec.~\ref{sec:connectionToYK}). We show that a convenient set of efficient decoders in such protocols are essentially the time-reversed versions of encoders, and that previous decoding protocols such as that due to Yoshida and Kitaev~\cite{YoshidaKitaev} contain equivalent variants of such decoders.

More crucially, our analysis allows these decoders to be somewhat more general than time-reversed encoders, in particular allowing the decoder to be configured with a variety of initial states (as opposed to maximally mixed or entangled states in prior protocols). This is necessitated by the bidirectional nature of our protocol. Specifically, when encoding from $A$ to $B$, and decoding from $B$ to $C$, it is necessary in our case to account for \textit{all} possible initial states in $C$ (which we eventually also want to preserve via teleportation to $A$), leading to a variety of decoders --- one for each initial state. In contrast, in the standard setting of recovery in this context, $C$ is initialized to a maximally mixed state, resulting in a unique decoder that precisely corresponds to the time-reversed encoder.

\subsection{The isometric encoding mechanism ($A \to C$)}

\subsubsection{Formal description of the protocol}

\begin{figure}
    \centering
    \includegraphics[width=0.5\linewidth]{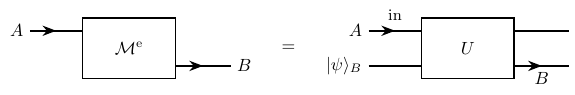}
    \caption{General schematic of the encoding operation, with a specific illustration involving a scrambling unitary. The arrows indicate the \textit{formal} input (inward arrow) and output (outward arrow) subsystems. We will see in later examples that the \textit{formal} input and output subsystems need not necessarily be the same as the practical input (``in'' label) and output (``out'' label) subsystems; the latter more directly label the information to be encoded and the subsystem with the output state. 
    }
    \label{fig:encoder}
\end{figure}

\begin{figure}
    \centering
    \includegraphics[width=0.5\linewidth]{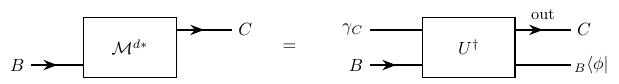}
    \caption{General schematic of the decoding operation, with a specific illustration involving the conjugate of the scrambling unitary. 
    }
    \label{fig:decoder}
\end{figure}

To begin with, let us consider operator transfer from $A$ to $B$, and onward to $C$. We will treat the scrambling interaction as encoding an operator $O_A$ acting on $\mathcal{H}_A$ into an operator acting on $\mathcal{H}_B$, via the (linear, completely positive, trace preserving) \textit{encoding} map $\mathcal{M}^e$ (see Fig.~\ref{fig:encoder}):
\begin{equation}
    \mathcal{M}^{e}_{BA}(O_A) = \Tr_A\left(U \left[ O_A \otimes \lvert\phi\rangle_B\langle\phi\rvert \right]U^\dagger\right).
    \label{eq:encoder}
\end{equation}
Here $U = e^{-i H t}$ is the scrambling unitary and $O_A$ is usually the initial density operator $O_A = \rho_A$ in $A$. While here we have taken this specific example with unitary dynamics for simplicity and direct relevance to our protocol in the main text, our formal analysis also applies when $\mathcal{M}_{BA}$ is the most general completely positive quantum operation~\cite{NielsenChuang} from $A$ to $B$.

We will also introduce a \textit{reverse-decoding} map as an encoding map $\mathcal{M}^d$ from $A$ to $B$ with an additional ``projection'' onto an operator $\gamma_A^\dagger$ in $A$:
\begin{equation}
    \mathcal{M}^{d}_{BA}(O_A) = \Tr_A\left(U \left[O_A \otimes \lvert\phi\rangle_B\langle\phi\rvert \right]U^\dagger \gamma_A^\dagger\right).
    \label{eq:reverse_decoder}
\end{equation}
In our protocol, $\gamma_A$ corresponds to the initial state in $C$ (after mapping $A$ to $C$, which we will do shortly). For the special choice of a maximally mixed state $\gamma_A = \idop_A$ (up to normalization), the reverse-decoder is identical to the encoder, $\mathcal{M}^d = \mathcal{M}^e$. The latter is usually the case for the typical examples of scrambling-recovery protocols in the literature, such as the Yoshida-Kitaev protocol (the connection to which is discussed in more detail in Sec.~\ref{sec:connectionToYK}); this forms a special case of our analysis in that we identify a specific mechanism of recovery without requiring identical encoding and (reverse-)decoding maps.

The reverse-decoding map may be converted into a \textit{decoding} map $\mathcal{M}^{d\ast}$ from $B$ to $C$, which is loosely the ``conjugate'' of $\mathcal{M}^{d}$ (see Fig.~\ref{fig:decoder}):
\begin{equation}
    \mathcal{M}^{d\ast}_{CB} : L(\mathcal{H}_B) \to L(\mathcal{H}_C),
\end{equation}
given in our protocol by the time-reversed version of the reverse-decoder:
\begin{equation}
    \mathcal{M}^{d\ast}_{CB}(O_B) = {{_B}\langle} \phi\rvert U^\dagger ( \gamma_C \otimes O_B) U \lvert \phi\rangle_B,
    \label{eq:decoder}
\end{equation}
justifying the interpretation of $\gamma_C$ as the initial state in $C$. Formally, we define the decoder to be the map that gives the pullback from $B$ to $C$ by the reverse decoder $\mathcal{M}^{d}$, with respect to the trace inner product of operators. This is defined by the requirement that for any operator $O_{1A}$ in $A$ (isomorphic to $O_{1C}$ in $C$) and any operator in $O_{2B}$,
\begin{equation}
    \Tr_B[\mathcal{M}^{d}_{BA}(O_{1A}) O_{2B}^\dagger] = \Tr_C[O_{1C} \mathcal{M}^{d\ast}_{CB}(O_{2B})^\dagger].
    \label{eq:pullback_def}
\end{equation}
It is easy to see that Eq.~\eqref{eq:decoder} satisfies this defining condition given the expression for the reverse decoder in Eq.~\eqref{eq:reverse_decoder}. For simplicity, we will refer to $\mathcal{M}^{d\ast}$ itself as the pullback of $\mathcal{M}^d$ in what follows\footnote{Here, we are relying on the isomorphism between the space of linear functionals on $L(\mathcal{H}_A)$ and the action of trace inner products with a given operator on this space to call this a pullback, i.e., $f_{O_2}(O_1) \equiv \Tr[O_1 O_2^\dagger]$. Strictly, $\mathcal{M}^\ast$ is the formal transpose of $\mathcal{M}$ that gives the pullback $\mathcal{M}^\ast(f_{O_2})$ of $f_{O_2}$ by $\mathcal{M}$, but we want to avoid confusing it with the transpose of operators, $O_A \to O_A^T$ for example; and pullback as a word appears to better capture the intuition behind this ``decoding'' procedure (the decoder pulls back the encoded version from the encoding system to another system isomorphic to the original system).}.

\subsubsection{General principle: Pullbacks are decoders for mutually isometric encoders}
\label{subsec:pullbacksaredecoders}

\begin{figure}
    \centering
    \includegraphics[scale=0.95]{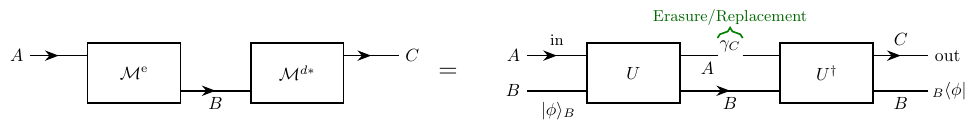}
    \caption{Combining the encoder and decoder for information storage and recovery. While $C$ is formally a different subsystem from $A$, we can also regard it as the same subsystem as $A$ after an erasure operation, or some other (possibly uncontrolled) modification/error on this subsystem. 
    }
    \label{fig:encoderdecoder}
\end{figure}

Having set up our notation, we begin by specializing to one direction of our protocol: information transfer from $A$ to $C$. We can write this as an encoding of $A$ into $B$ followed by a decoding of $B$ into $C$:
\begin{equation}
    \rho^{\text{out}}_{C}(\rho_A) = \mathcal{M}^{d\ast}_{CB}(\mathcal{M}^e_{BA}(\rho_A)).
    \label{eq:outputstate_def}
\end{equation}
Our goal is now to analyze under what circumstances $\rho^{\text{out}}_{C}(\rho_A)$ is sufficiently close to $\rho_A$.

Mathematically, it follows from \eqref{eq:pullback_def} that for some initial state $\rho_A$ and observable $O_A$ in $A$,
\begin{equation}
    \Tr_B[\mathcal{M}^e_{BA}(\rho_A)\mathcal{M}^d_{BA}(O_A)^\dagger] = \Tr_C[\mathcal{M}^{d\ast}_{CB}(\mathcal{M}^e_{BA}(\rho_A)) O_C].
    \label{eq:encoder_decoder_pullback}
\end{equation}
This relation the captures the key principle behind decoding. Loosely, almost as a trivial consequence of Eq.~\eqref{eq:encoder_decoder_pullback}, \textit{if} the encoding of states and reverse-decoding of observables from $A$ to $B$ preserves their expectation values in $B$, which is an \textit{isometry} condition, \textit{then} the encoder-decoder combination transferring states from $A$ to $C$ preserves the expectation values of operators in $C$ in these states, relative to the expectation value of comparable operators in $A$ in the original state.

Formally, given a pair of an encoder $\mathcal{M}^e$ and reverse decoder $\mathcal{M}^d$, the condition for \textit{isometric encoding} is that:
\begin{equation}
    \Tr_B[\mathcal{M}^e_{BA}(O_{1A})\mathcal{M}^d_{BA}(O_{2A})^\dagger] \simeq p_{\lambda} \Tr_A[O_{1A} O_{2A}^\dagger],
    \label{eq:isometric_encoding_pair}
\end{equation}
where $a \simeq b$ is now a stand-in for the exact inequality
\begin{equation}
    \lvert a-b\rvert \leq \epsilon
\end{equation}
for some suitably chosen $\epsilon$. The parameter $p_{\lambda}$ allows for a scale factor that measures the ``loss'' of probability in the encoding-decoding procedure. Without loss of generality, we can define $p_{\lambda}$ by requiring the inequality to be exact for $O_{1A} = O_{2A} = \idop_A$, i.e. using Eq.~\eqref{eq:encoder_decoder_pullback},
\begin{equation}
    p_{\lambda} \equiv \frac{1}{d_A}\Tr_C[\mathcal{M}^{d\ast}_{CB}(\mathcal{M}^e_{BA}(\idop_A))] = \Tr_C[\rho^{\text{out}}_{C}(\rho_A)]_{\rho_A = \idop_A/d_A}.
    \label{eq:plambdadef}
\end{equation}
Despite this loss of probability, the decoder involves a postselection step via a projective measurement onto $\lvert \phi\rangle_B\langle\phi\rvert$: this means that if we keep only those outcomes where the decoder succeeds (the projective measurement gives the outcome $\lvert \phi\rangle_B$), we can restore the probability of recovery to nearly $1$ within this subset, as we will see below. 

For a general initial state, if Eq.~\eqref{eq:isometric_encoding_pair} is satisfied by the encoding and reverse-decoding maps, then it follows from Eq.~\eqref{eq:encoder_decoder_pullback} that 
\begin{equation}
    \Tr_C[\rho^{\text{out}}_{C}(\rho_A) O_C] \simeq p_{\lambda} \Tr_A[\rho_A O_A],
    \label{eq:successful_isometric_teleportation}
\end{equation}
which indicates a successful teleportation from $A$ to $C$ (with the same $\epsilon$ as implied in Eq.~\eqref{eq:isometric_encoding_pair}). 
This transfer of information from $A$ to $C$ is a probabilistic operation, with success probability
\begin{align}
    p \equiv \Tr_C[\rho^{\text{out}}_{C}(\rho_A)] &= \Tr_B[\mathcal{M}^e_{BA}(\rho_A)\mathcal{M}^d_{BA}(\idop_A)^\dagger] \label{eq:postselectionprobability}\\
    \implies p &\simeq p_{\lambda},
    \label{eq:equalpostselection}
\end{align}
where in the last line, we have used the isometric encoding criterion \eqref{eq:isometric_encoding_pair} with $O_C = \idop_C$ and $\Tr_A[\rho_A] = 1$. This shows that our teleportation operation has nearly the same postselection probability $p_{\lambda}$ for essentially all initial states in $A$, provided that isometric encoding Eq.~\eqref{eq:encoder_decoder_pullback} is successful. We will see that typically, for ``maximal'' scramblers, $p_{\lambda} \simeq 1/d_A^2$.
    

We can also define a direct fidelity measure between the input and output states: 
\begin{equation}
    \mathcal{F}_{A \to C}[\rho_A] \equiv \frac{\Tr_C[\rho^{\text{out}}_{C}(\rho_A) \rho_{A\to C}]}{\sqrt{\Tr_C[(\rho^{\text{out}}_{C}(\rho_A))^2]\Tr_A[\rho_A^2]}}.
    \label{eq:AC_fidelity_def}
\end{equation}
Here, $\rho_{A\to C}$ is the input state in $A$ that is formally transferred to $C$ via the implicit isomorphism between $A$ and $C$ (e.g., identifying spin-down states in $A$ with those in $C$), while $\rho^{\text{out}}_{C}(\rho_A)$ is the output state generated by our decoder as in Eq.~\eqref{eq:outputstate_def}. This fidelity has a range of $[0,1]$ as $\rho_A$ and $\rho^{\text{out}}_{C}(\rho_A)$ are positive operators (for the lower limit) and due to the Cauchy-Schwarz inequality (for the upper limit).
According to Eq.~\eqref{eq:successful_isometric_teleportation}, if we set $O_A = \rho_A$, the fidelity is constrained by:
\begin{equation}
    \mathcal{F}_{A \to C}[\rho_A] \geq \sqrt{\frac{\Tr_A[\rho_A^2]}{\Tr_C[(\rho^{\text{out}}_{C}(\rho_A))^2]/p^2}} \left\lvert \frac{p_{\lambda}}{p} - \frac{\epsilon}{p \Tr_A[\rho_A^2]}\right\rvert
    \label{eq:fidelityintermediateconstraint}
\end{equation}
Intuitively, we can keep track of when the fidelity can be close to $1$ through a self-consistency argument: if the output state is the same as the input state up to the postselection probability $p$, we should expect that $\Tr_A[\rho_A^2] \approx \Tr_C[(\rho^{\text{out}}_{C}(\rho_A))^2]/p^2$; moreover, $p_{\lambda}/p \approx 1$ for isometric encoding, which implies that $\mathcal{F}_{A \to C}[\rho_A] \approx 1$ if $\epsilon \approx 0$.

To show this quantitatively, by setting $O_C = \rho_C^{\text{out}}$ in Eq.~\eqref{eq:successful_isometric_teleportation}, we get
\begin{equation}
    \left\lvert\Tr_C[(\rho^{\text{out}}_{C}(\rho_A))^2] - p_{\lambda} \Tr_C[\rho^{\text{out}}_{C}(\rho_A) \rho_{A\to C}]\right\rvert \leq \epsilon.
\end{equation}
Using Eq.~\eqref{eq:AC_fidelity_def} in the second term on the left hand side gives:
\begin{equation}
    \left\lvert\Tr_C[(\rho^{\text{out}}_{C}(\rho_A))^2] - p_{\lambda} \mathcal{F}_{A \to C}[\rho_A] \sqrt{\Tr_C[(\rho^{\text{out}}_{C}(\rho_A))^2]\Tr_A[\rho_A^2]} \right\rvert \leq \epsilon.
\end{equation}
Rearranging factors, we get 
\begin{align}
    \left\lvert\sqrt{\frac{\Tr_A[\rho_A^2]}{\Tr_C[(\rho^{\text{out}}_{C}(\rho_A))^2]/p^2}}-\frac{p}{p_{\lambda} \mathcal{F}_{A \to C}[\rho_A]}\right\rvert &\leq \frac{\epsilon}{p_{\lambda} p \mathcal{F}_{A \to C}[\rho_A] \Tr_A[\rho_A^2]} \frac{\Tr_A[\rho_A^2]}{ \Tr_C[(\rho^{\text{out}}_{C}(\rho_A))^2]/p^2}.
    \label{eq:complicatedpurityconstraint}
\end{align}

Let $\xi = \Tr_A[\rho_A^2] /( \Tr_C[(\rho^{\text{out}}_{C}(\rho_A))^2]/p^2 )$, $\eta = \epsilon/(p^2 \Tr[\rho_A^2])$ and $f = p/(p_{\lambda} F_{A\to C}[\rho_A])$. The essential content of the above inequality is that $\xi \approx f$ when $\eta \approx 0$, which can be expressed more formally through the constraints:
\begin{align}
    \xi &\geq f(1-\eta \xi^2), \nonumber \\
    \xi &\leq f(1+\eta \xi^2).
\end{align}
These constraints are solved by:
\begin{equation}
    \frac{\sqrt{1+4\eta f^2}-1}{2\eta f} \leq \xi \leq \frac{1-\sqrt{1-4\eta f^2}}{2\eta f},
    \label{eq:complicatedpurityconstraint2}
\end{equation}
with the upper bound holding as long as $4 \eta f \leq 1$, or
\begin{equation}
    \epsilon \leq p_{\lambda} p \mathcal{F}_{A \to C}[\rho_A] \Tr_A[\rho_A^2],
\end{equation}
but with no such restrictions for the lower bound. Using this notation, we can also reexpress Eq.~\eqref{eq:fidelityintermediateconstraint} as:
\begin{equation}
    \frac{p^2}{p_{\lambda}^2 f} \geq \xi \left\lvert1 - \frac{p^2\eta}{p_{\lambda}}\right\rvert.
    \label{eq:fidelityintermediateconstraint2}
\end{equation}

Then, inserting the lower bound in Eq.~\eqref{eq:complicatedpurityconstraint2} into Eq.~\eqref{eq:fidelityintermediateconstraint2} gives
\begin{equation}
    \frac{\sqrt{1+4\eta f^2}-1}{2\eta p^2/p_{\lambda}^2} \left\lvert 1 - \frac{p^2\eta}{p_{\lambda}}\right\rvert \leq 1,
    \label{eq:fidelityintermediateconstraint3}
\end{equation}
which, with some rearrangements, becomes
\begin{equation}
    f^2 \leq \frac{\left\lvert 1-\frac{p^2 \eta}{p_{\lambda}}\right\rvert\frac{p^2}{p_\lambda^2} + \eta \frac{p^4}{p_{\lambda}^4}}{\left\lvert 1-\frac{p^2 \eta}{p_{\lambda}}\right\rvert^2}.
\end{equation}
Expressing $f$ in terms of the fidelity, we get
\begin{equation}
    \mathcal{F}_{A \to C}[\rho_A] \geq \frac{\left\lvert 1-\frac{\epsilon}{p_{\lambda} \Tr_A[\rho_A^2]}\right\rvert}{\sqrt{\left\lvert 1-\frac{\epsilon}{p_{\lambda} \Tr_A[\rho_A^2]}\right\rvert + \frac{\epsilon}{p_{\lambda}^2 \Tr_A[\rho_A^2]}}}
\end{equation}
which approaches $1$ for $\epsilon \to 0$.

More quantitatively, we can pick a scale for the error by writing
\begin{equation}
    \epsilon = \widetilde{\varepsilon} p_{\lambda} \Tr_A[\rho_A^2], 
\end{equation}
as by Eq.~\eqref{eq:fidelityintermediateconstraint}, the fidelity is near $1$ when $\widetilde{\varepsilon} \ll 1$ (or more directly, it is $\widetilde{\varepsilon}$ that measures the relative error in the fidelity). In this case, we have
\begin{equation}
    \mathcal{F}_{A \to C}[\rho_A] \geq \frac{\left\lvert 1-\widetilde{\varepsilon}\right\rvert}{\sqrt{\left\lvert 1-\widetilde{\varepsilon}\right\rvert + \frac{\widetilde{\varepsilon}}{p_{\lambda}}}}.
\end{equation}
Here, as (typically) $p_{\lambda} \sim 1/d_A^2$, we need $\widetilde{\varepsilon} \ll 1/d_A^2$ for a successful decoding operation with fidelity $\approx 1$.

\subsubsection{A special case: Orthogonal encoding for pure states}
\label{subsec:orthogonalencoding}

For pure states, a simpler criterion than Eq.~\eqref{eq:isometric_encoding_pair} suffices for successful teleportation from $A$ to $C$. Given any initial state $\Psi_A = \lvert \psi\rangle_A\langle \psi\rvert$ of interest, let $\overline{\Psi}_A = \idop_A - \Psi_A$ be its complementary (orthogonal) subspace in $\mathcal{H}_A$. Then, let us say that $\Psi_A$ is \textit{orthogonally} encoded into $B$, 
\begin{equation}
    \Tr_B[\mathcal{M}^e_{BA}(\Psi_A)\mathcal{M}^d_{BA}(\overline{\Psi}_A)^\dagger] \simeq 0,
    \label{eq:orthogonalencoding2}
\end{equation}
which formalizes the discussion in the main text, and is a special case (necessary but not sufficient condition) of Eq.~\eqref{eq:isometric_encoding_pair}. Then, it follows in analogy with Eq.~\eqref{eq:successful_isometric_teleportation} that, with $\Psi^{\text{out}}_C$ being the output state in $C$,
\begin{equation}
    \Tr_C[\Psi^{\text{out}}_C \overline{\Psi}_C] \simeq 0 \implies \frac{\langle \psi\rvert \Psi^{\text{out}}\lvert \psi\rangle_C}{\Tr_C[\Psi^{\text{out}}_C]} \geq 1-\frac{\epsilon}{\Tr_C[\Psi^{\text{out}}_C]},
    \label{eq:successful_pure_state_teleportation}
\end{equation}
which indicates successful pure state teleportation. This corresponds to Eq.~(4) in the main text, where the undetermined coefficients are now absorbed into $\Tr_C[\Psi^{\text{out}}_C]$.

The success probability here is given by:
\begin{align}
     p = \Tr_C[\Psi^{\text{out}}_C] &= \Tr_B[\mathcal{M}^e_{BA}(\Psi_A)\mathcal{M}^d_{BA}(\idop_A)^\dagger] \nonumber \\
     &= \Tr_B[\mathcal{M}^e_{BA}(\Psi_A)\mathcal{M}^d_{BA}(\Psi_A)^\dagger] + \Tr_B[\mathcal{M}^e_{BA}(\Psi_A)\mathcal{M}^d_{BA}(\overline{\Psi}_A)^\dagger] \nonumber \\
     &\simeq \Tr_B[\mathcal{M}^e_{BA}(\Psi_A)\mathcal{M}^d_{BA}(\Psi_A)^\dagger],
\end{align}
where in the last line we have used Eq.~\eqref{eq:orthogonalencoding2}. To develop some intuition for this expression, let us momentarily consider a special protocol where $\gamma_C = \idop_C/d_C$ is the maximally mixed state, which corresponds to e.g. Yoshida-Kitaev decoding as discussed in Sec.~\ref{sec:connectionToYK}. In this case, we have
\begin{equation}
    \mathcal{M}^d_{BA}(\Psi_A) = \frac{1}{d_A} \mathcal{M}^e_{BA}(\Psi_A),
    \label{eq:AC_specialcase}
\end{equation}
which implies that $p \simeq \mathcal{P}_B/d_A$, where $\mathcal{P}_B$ is the purity in $B$ of the state generated by the scrambling unitary $U$:
\begin{equation}
    \mathcal{P}_B \equiv \Tr_B[\mathcal{M}^e_{BA}(\Psi_A)^2].
    \label{eq:specialcase_purity}
\end{equation}
Therefore, $p \simeq 1/d_A^2$ for maximal entanglement generation between $A$ and $B$ (for which $\mathcal{P}_B = 1/d_A$), which is consistent with other probabilistic teleportation protocols~\cite{BennettTeleportation, NielsenChuang, YoshidaKitaev, TeleportationReview1, TeleportationReview2}.

Correspondingly, it is convenient to discuss a notion of ``cross entanglement'' of the initial state $\Psi_A$ in $B$ between encoding and reverse-decoding dynamics, via the ``cross purity'' defined for a general choice of $\gamma_C$ by:
\begin{equation}
    \widetilde{\mathcal{P}}_B[\gamma_C] \equiv d_A \Tr_B[\mathcal{M}^e_{BA}(\Psi_A)\mathcal{M}^d_{BA}(\Psi_A)^\dagger].
    \label{eq:crosspurity_def}
\end{equation}
In terms of this quantity, we can write the success probability as
\begin{equation}
    p = \frac{1}{d_A} \widetilde{\mathcal{P}}_B[\Psi_A],
    \label{eq:probability_and_cross_entanglement}
\end{equation}
in terms of the cross-entanglement generated for an individual state. If these are different for different pure states, then a mixed state formed as an ensemble of these pure states would acquire different weights after decoding into $C$, determined by their success probability. Moreover, for the telportation to be nontrivially detectable over errors, we also need $p \gg \epsilon$. This can be ensured if $\epsilon \ll 1/d_A^2$ in the event that cross-entanglement reduces to regular entanglement, as $\mathcal{P}_B \geq 1/d_A$.

Orthogonal encoding is therefore not sufficient to teleport mixed states, and we need the full isometric encoding criterion in Eq.~\eqref{eq:isometric_encoding_pair}; in this context, this effectively amounts to the criterion that the post-selection probability is identical for all pure states. In particular, combining Eqs.~\eqref{eq:postselectionprobability} and \eqref{eq:equalpostselection}, we obtain from the isometry condition (after restoring $\epsilon$ and using the triangle inequality):
\begin{equation}
    \left\lvert \Tr_B[\mathcal{M}^e_{BA}(\Psi_A)\mathcal{M}^d_{BA}(\Psi_A)^\dagger] - d_A p_{\lambda}\right\rvert \leq 2\epsilon d_A.
\end{equation}
This shows that a necessary condition for isometric encoding is that all initial states $\Psi_A$ lead to approximately the same amount of ``cross''-entanglement generation between $A$ and $B$ under scrambling dynamics. By Eq.~\eqref{eq:probability_and_cross_entanglement}, this corresponds to $p \approx p_{\lambda}$ for all states. As the purity can be as small as $1/d_A$ (which we generically also expect for the cross purity, given near-maximal entanglement), we require that $\epsilon \ll d_A^{-2}$ for these statements to hold with reasonable accuracy for highly entangling dynamics.

\subsection{Entanglement generation and quantum teleportation ($C \to A$)}
\label{sec:CtoAmechanism}

Now, let us consider the teleportation mechanism for operator transfer from $C$ to $A$. While the general comments on isometric encoding in Sec.~\ref{subsec:pullbacksaredecoders} continue to apply for this direction (with different encoding and decoding operators), our focus here will be to identify the amount of entanglement necessary to successfully transfer operators (see also Sec.~\ref{sec:isometry_to_entanglement} for a more formal connection to the language of isometric encoding). Our analysis will be qualitatively similar to \cite{dynamicalqentanglement}, with quantitative differences due to considering the transfer of density operators here rather than operators acting on pure states.

\subsubsection{Partial reliance on isometric encoding from $A \to C$}

A key assumption for our argument, as outlined in the main text, is that the projection on $\lvert \phi\rangle_B$ at the end of our protocol amounts to an EPR measurement on $BC$. For this to hold, we \textit{require} that the isometric encoding mechanism from $A \to C$ succeeds. To illustrate this with pure initial states in $A$ and $C$, this assumption entails that
\begin{equation}
    {_B}\langle \phi\rvert U_{CB}^\dagger U_{AB} \lvert \psi\rangle_A\lvert \phi\rangle_B \lvert \chi\rangle_C \approx \lvert \psi\rangle{_C}\langle \psi\rvert \left[{_B}\langle \phi\rvert U_{CB}^\dagger U_{AB} \lvert \psi\rangle_A\lvert \phi\rangle_B \lvert \chi\rangle_C\right],
\end{equation}
which automatically holds to $O(\epsilon)$ error if $\rho_C^{\text{out}} \approx \rho_A$ as per the considerations of Sec.~\ref{subsec:pullbacksaredecoders}. If this holds, then defining
\begin{equation}
    \lvert \Phi\rangle_{AB} = U_{AB}\lvert \psi\rangle_A\lvert \phi\rangle_B
\end{equation}
as the entangled state generated by the scrambling unitary $U$ acting on the above product state, we can write the output state of the $C \to A$ part of our protocol as approximately equal to the ideal teleported~\cite{BennettTeleportation, NielsenChuang} state $\rho_{A}^{\text{tel}}(\rho_C)$ using the entangled state $\lvert \Phi\rangle$:
\begin{equation}
    \rho_{A}^{\text{out}}(\rho_C) \approx \rho_{A}^{\text{tel}}(\rho_C) \equiv {_{CB}}\langle \Phi\rvert\ \left(\lvert \Phi\rangle_{AB}\langle \Phi\rvert \otimes \rho_C  \right) \lvert \Phi\rangle_{CB}.
    \label{eq:idealteleportedstate}
\end{equation}
We will therefore work with $\rho_{A}^{\text{tel}}(\rho_C)$ in the remainder of this subsection.

It is also important to consider the case when the initial state  $\rho_A$ in $A$ is mixed (as we want our SWAP protocol to work for mixed states). To reason about this case, it is convenient to purify~\cite{NielsenChuang} this mixed state with the help of an auxiliary subsystem $R$ (at least as large as $A$):
\begin{equation}
    \lvert \Psi\rangle_{RA} = \sum_{k=1}^{d_A} q_k \lvert \psi_k\rangle_R \lvert \psi_k\rangle_A,
\end{equation}
where $\sum_k |q_k|^2 = 1$ and the $\lvert \psi_k\rangle_R$ are orthonormal. The reduced density operator in $A$ is given by $\rho_A = \sum_k |q_k|^2 \lvert \psi_k\rangle_A\langle\psi\rvert$.
Then, for a term with a given $\lvert \psi_k\rangle_R$, the isometric encoding mechanism effectively produces $\lvert \psi_k\rangle_C$ as the output state in $C$.Thus, teleportation proceeds in each term (for the overall pure state in $RABC$) with the entangled state $\lvert \Phi\rangle$ replaced by:
\begin{equation}
    \lvert \Phi_k\rangle_{AB} = U_{AB}\lvert \psi_k\rangle_A\lvert \phi\rangle_B.
\end{equation}
Here, we recall that in practice, we only perform the projection onto $\lvert \phi\rangle_B$ in $B$, while isometric encoding from $A\to C$ automatically implies a projection onto $\lvert \psi_k\rangle_C$ in each term with the corresponding $\lvert \psi_k\rangle_R$. Thus, for each such term, the final projection happens onto the same state ${_{CB}\langle \Phi_k\rvert}$ in $CB$ as the initial state $\lvert \Phi_k\rangle_{AB}$ in $AB$.

In this setting, the total teleported state can be obtained by tracing out the auxiliary subsystem $R$:
\begin{align}
    \rho_{A}^{\text{tel}}(\rho_C) &= \sum_{k,\ell} q_k q_{\ell}^\ast \Tr_R\left[ {_{CB}}\langle \Phi_k\rvert\ \left(\lvert \Phi_k\rangle_{AB}\lvert \psi_k\rangle_R \otimes \rho_C \otimes {_R}\langle \psi_{\ell}\rvert {_{AB}}\langle \Phi_{\ell}\rvert \right) \lvert \Phi_{\ell}\rangle_{CB}\right] \nonumber \\
    &= \sum_k \lvert q_k\rvert^2 {_{CB}}\langle \Phi_k\rvert\ \left(\lvert \Phi_k\rangle_{AB}\langle \Phi_k\rvert \otimes \rho_C  \right) \lvert \Phi_k\rangle_{CB},
\end{align}
which just reduces to a statistical average over the output states obtained for each pure state $\lvert \psi_k\rangle$.
Now it is convenient to apply a formal trick: we relabel $B$ to $B'$ for the factors on the right hand side, i.e. 
\begin{equation}
    \rho_{A}^{\text{tel}}(\rho_C) = \sum_{k} |q_k|^2 {_{CB}}\langle \Phi_k\rvert\ \left(\lvert \Phi_k\rangle_{AB} \otimes \rho_C \otimes {_{AB'}}\langle \Phi_k\rvert \right) \lvert \Phi_k\rangle_{CB'}.
\end{equation}
This is a safe operation as the initial state in $B$ is a pure state $\lvert \phi\rangle_B$, so the ``left'' and ``right'' branches of the above equation do not have any links (operator multiplication) between them that are not already projected onto this pure state. To write this out more explicitly, our trick is to perform the following relabeling for an arbitrary initial state $\rho_{AC}$ in $AC$:
\begin{equation}
    \left[{_B}\langle \phi\rvert U_{CB}^\dagger U_{AB}\lvert \phi\rangle_B\right] \rho_{AC} \left[{_B}\langle \phi\rvert U_{CB}^\dagger U_{AB}\lvert \phi\rangle_B\right] = \left[{_B}\langle \phi\rvert U_{CB}^\dagger U_{AB}\lvert \phi\rangle_B\right] \rho_{AC} \left[{_{B'}}\langle \phi\rvert U_{C{B'}}^\dagger U_{A{B'}}\lvert \phi\rangle_{B'}\right],
\end{equation}
and then treat $B'$ as if it were a completely separate subsystem from $B$. Another trick is to identify $A$ and $C$ through an implicit isomorphism $\idop_{A \leftarrow C}$ (e.g., one that maps spin-up states in $A$ to spin-up states in $C$) so that e.g. $\rho_{C \to A} = \idop_{A \leftarrow C} \rho_C \idop_{C\to A}$ (with $\idop_{C \to A} = \idop_{A \leftarrow C}^\dagger$).
This allows us to trace out $B$ from the factor on the left and $B'$ from the factor on the right, so that we have
\begin{equation}
    \rho_{A}^{\text{tel}}(\rho_C) = \sum_{k} |q_k|^2 \Tr_B\left(\lvert \Phi_k\rangle_{AB} \langle \Phi_k\rvert\right) \rho_{C\to A} \Tr_{B'}\left(\lvert \Phi_{k}\rangle{_{AB'}}\langle \Phi_{k}\rvert \right) .
\end{equation}
Here, it is convenient to identify the reduced density operator of each ``scrambled state'' $\lvert \Phi_k\rangle_{AB}$ in $A$: 
\begin{equation}
    \widetilde{\rho}_{kA} \equiv \Tr_B\left(\lvert \Phi_k\rangle_{AB} \langle \Phi_k\rvert\right),
\end{equation}
so that we can write:
\begin{equation}
    \rho_{A}^{\text{tel}}(\rho_C) = \sum_{k} |q_k|^2 \widetilde{\rho}_{kA} \rho_{C \to A} \widetilde{\rho}_{kA}.
    \label{eq:generalteleportedstate}
\end{equation}
Having set up this teleported state as a simple statistical average over pure initial states in $A$, we will now work with a single pure state in $A$ such as a density operator with $q_k = \delta_{k1}$, and just write $\rho_{C\to A}$ as $\rho_A$ and $\widetilde{\rho}_{1A} = \widetilde{\rho}_{A}$ for simplicity.

\subsubsection{Constraints on entanglement generation}

As before, let us consider an initial state $\rho_C$ in $C$, which is teleported to $\rho_{A}^{\text{tel}}(\rho_C)$ in $A$, and probed using the observable $O_A$. For $\rho_{A}^{\text{tel}}(\rho_C)$ to be comparable to $\rho_C$, we require that
\begin{equation}
    \Tr_A[\rho_{A}^{\text{tel}}(\rho_C) O_A^\dagger] \simeq \mu \Tr_C[\rho_C O_C^\dagger],
\end{equation}
for some constant $\mu$. Substituting Eq.~\eqref{eq:generalteleportedstate} for a single pure state in $A$ i.e. $q_k = \delta_{k1}$ and using the isomorphism between $A$ and $C$, we require that
\begin{equation}
    \Tr_A[\widetilde{\rho}_{A} \rho_A \widetilde{\rho}_{A}  O_A^\dagger] \simeq \mu \Tr_A[\rho_A O_A^\dagger].
    \label{eq:widetilderhoA_condition}
\end{equation}
Now we can define $\mu$ by enforcing an exact equality for $\rho_A = O_A = \idop_A$, which gives
\begin{equation}
    \mu = \frac{1}{d_A}\Tr_A[\widetilde{\rho}_{A}^2],
    \label{eq:mudef_CtoA}
\end{equation}
which is the purity of the scrambled state $\lvert \Phi\rangle$, and therefore measures the entanglement generated by the scrambling unitary acting on the product state $\lvert \psi\rangle_A \otimes \lvert \phi\rangle_B$. Further, the average postselection probability over any orthonormal basis of states is
\begin{equation}
    p_{\text{avg}} \equiv \Tr_A[\widetilde{\rho}_{A} \rho_A \widetilde{\rho}_{A}]_{\rho_A = \idop_A/d_A} = \mu,
\end{equation}
which is consistent with our expectation in Sec.~\ref{subsec:orthogonalencoding} that the postselection probability is determined by the purity of the state generated by the scrambler.

Without loss of generality we may write \begin{equation}
    \widetilde{\rho}_A = \sum_k p_k |k\rangle\langle k|
\end{equation}
with $\sum_k p_k=1$.  Now consider the ``trial functions" $\rho_A = O_A = |k\rangle\langle k|$.  Eq.~\eqref{eq:widetilderhoA_condition} implies that $|p_k^2-\mu|<\epsilon$ or, for each $k$, \begin{equation}
    \sqrt{\mu-\epsilon}<p_k<\sqrt{\mu+\epsilon}.
\end{equation}
Crucially, this is both a necessary and sufficient condition for Eq.~\eqref{eq:widetilderhoA_condition} to hold with different ``trial functions''. We can reduce this to a necessary condition in terms of a more intuitive measure, such as purity, as follows.
Using $\sum_k p_k=1$ we find \begin{equation}
  \frac{1}{d_A^2}-\epsilon \le  \mu \le \frac{1}{d_A^2}+\epsilon.
\end{equation}
From our definition of $\mu$ in Eq.~\eqref{eq:mudef_CtoA}, we conclude that   \begin{equation}
    \Tr_A \left[\widetilde{\rho}_{A}^2\right] = d_A \mu  \le \frac{1}{d_A}+\epsilon d_A
    \label{eq:purityconstraint1_teleportation}
\end{equation}
is necessary for Eq.~\eqref{eq:widetilderhoA_condition} to hold.

Due to the natural scaling of $\epsilon$ being measured relative $\Tr_A[\widetilde{\rho}_{A}^2] \Tr_A[\rho_A^2] / d_A$, let us now write (implicitly taking $\rho_A$ to be a pure state):
\begin{equation}
    \epsilon = \frac{\widetilde{\varepsilon}}{d_A}\Tr_A[\widetilde{\rho}_{A}^2].
\end{equation}
A \textit{good} range of $\widetilde{\varepsilon}$ for which teleportation occurs successfully is $\widetilde{\varepsilon} \ll 1$. Keeping this in mind, Eq.~\eqref{eq:purityconstraint1_teleportation} gives a necessary condition for successful teleportation:
\begin{equation}
     \Tr_A[\widetilde{\rho}_{A}^2] \leq \frac{1}{d_A(1-\widetilde{\varepsilon})},
\end{equation}
which is very close to maximal entanglement, as $\Tr_A[\widetilde{\rho}_{A}^2] \geq 1/d_A$. This also sets our post-selection probability to be $p \simeq 1/d_A^2$.

\subsection{SWAP fidelity for Haar random scramblers}

Here we consider the fidelity of this SWAP operation if $U_{AB}$ is chosen at random from the Haar measure~\cite{Mehta, ChaosComplexityRMT}.\footnote{In fact we will only use the second moment of the Haar measure, so this analysis works for any 2-design~\cite{ChaosComplexityRMT}. Moreover, we expect the leading order behavior from this analysis to apply to a much broader class of systems than $2$-designs.}
Specifically, consider the operator $\tilde{S}_{AC}: \mathcal{H}_A \otimes \mathcal{H}_C \to \mathcal{H}_A \otimes \mathcal{H}_C$ given by
\begin{equation}
    \tilde{S}_{AC} = \frac{1}{\sqrt{p}} \langle \phi \rvert_B U^\dagger_{CB} U_{AB} \lvert \phi\rangle_{B}
\end{equation}
where $\lvert \phi\rangle_{B} \in \mathcal{H}_B$ is an arbitrary state, $U_{AB}$ is chosen at random from the Haar measure, and $U_{CB}$ is the same unitary but acting on $CB$ instead of $AB$.
The constant $\sqrt{p}$ is the square root of the postselection probability, defined by
\begin{equation}
    p \equiv \lvert \langle{\phi}_B \rvert U_{CB}^\dagger U_{AB} \lvert \chi \phi \psi \rangle_{ABC} \rvert^2
\end{equation}
We include $\sqrt{p}$ so that $\tilde{S}_{AC}$ maps normalized states to normalized states.


The main claim of this subsection is that for $d_B \gg d_A^2 \gg 1$, $\tilde{S}_{AC}$ acts like the SWAP operator with very small fluctuations,
\begin{equation}
    \overline{\lvert \langle \psi \chi \rvert \tilde{S}_{AC} \lvert \chi \psi \rangle_{AC} - 1 \rvert^2} = O(1/d_B)~.
\end{equation}
On the other hand, when $d_B \lesssim d_A^2$ the fluctuations are large and hence $\tilde{S}_{AC}$ is with high probability a bad implementation of SWAP.

To show this, we will use the following standard formulas for integrals over the Haar measure on $U(d)$,
\begin{equation}\label{eq:Haar_integrals}
\begin{split}
    &\int dU U_{i_1 j_1} U^\dagger_{j_1' i_1'} = \frac{1}{d} \delta_{i_1 i_1' } \delta_{j_1 j_1'}~, \\
    &\int dU U_{i_1 j_1}U_{i_2 j_2} U^\dagger_{j_1' i_1'}U^\dagger_{j_2' i_2'} = \frac{1}{d^2 -1 }\bigg[ \delta_{i_1 i_1' }\delta_{i_2 i_2' } \delta_{j_1 j_1'}\delta_{j_2 j_2'} + \delta_{i_1 i_2' }\delta_{i_2 i_1' } \delta_{j_1 j_2'}\delta_{j_2 j_1'} 
    \\&~~~~~~~~~~~~~~~~~~~~~~~~~~~~~~~~~~~~~~~~~~~~~~~~~~~~ - \frac{1}{d}\left( \delta_{i_1 i_1' }\delta_{i_2 i_2' } \delta_{j_1 j_2'}\delta_{j_2 j_1'} + \delta_{i_1 i_2' }\delta_{i_2 i_1' } \delta_{j_1 j_1'}\delta_{j_2 j_2'} \right)\bigg]~.
\end{split}
\end{equation}
An integral over an unequal number of $U$ and $U^\dagger$ will vanish. 
We will write e.g.
\begin{equation}
    U_{AB} = \sum_{i_a i_b j_a j_b} U_{i_a i_b, j_a j_b} \lvert i_a \rangle_A \lvert i_b \rangle_B \langle j_a \rvert_A \langle j_b \rvert_B~.
\end{equation}
For compactness, we write
\begin{equation}
    \lvert \langle \psi \chi \rvert \tilde{S}_{AC} \lvert \chi \psi \rangle_{AC} - 1 \rvert^2 = \lvert X - 1 \rvert^2 = \lvert X \rvert^2 - X - X^* + 1~.
\end{equation}
Evaluating the averages of each term is difficult with the unitaries in $p$ in the denominator.
Fortunately, as long as $d_B, d_A \gg 1$ then to a good approximation
\begin{equation}
     \overline{\frac{1}{\sqrt{p}} \langle \phi \rvert_B U^\dagger_{CB} U_{AB} \lvert \phi\rangle_{B}} \simeq  \frac{1}{\sqrt{\overline{p}}} \overline{\langle \phi \rvert_B U^\dagger_{CB} U_{AB} \lvert \phi\rangle_{B}}~.
\end{equation}
With this and \eqref{eq:Haar_integrals}, we evaluate the average of each term to find
\begin{equation}
\begin{split}
    &\overline{X} = \overline{X^*} = \frac{1}{\sqrt{\bar{p}} d_A}~~,~~~~~~ \overline{|X|^2} = \frac{1}{\bar{p} d_A^2} + \frac{1}{\bar{p} d_A^2 d_B} + O\left(\frac{1}{\bar{p} d_A^3 d_B}\right)
\end{split}
\end{equation}
We compute the average postselection probability again using \eqref{eq:Haar_integrals},
\begin{equation}
    \label{eq:probability_est}
    \bar{p} = \overline{\lvert \langle{\phi}_B \rvert U_{CB}^\dagger U_{AB} \lvert \chi \phi \psi \rangle_{ABC} \rvert^2} = \frac{1}{d_B} + \frac{1}{d_A^2} + O\left( \frac{1}{d_A d_B^2}\right) + O\left(\frac{1}{d_A^2 d_B}\right)
    \simeq 
    \begin{cases}
        1 / d_A^2 & d_B \gg d_A^2 \\
        1 / d_B & d_B \ll d_A^2
    \end{cases}
\end{equation}
Therefore as claimed, it holds that
\begin{equation}
    \overline{\lvert \langle \psi \chi \rvert \tilde{S}_{AC} \lvert \chi \psi \rangle_{AC} - 1 \rvert^2} \simeq 
    \left(1 + \frac{1}{d_B}\right)\left(\frac{1}{1 + \frac{d_A^2}{d_B}} \right) - \frac{2}{\sqrt{1 + \frac{d_A^2}{d_B}}} + 1
    =
    \begin{cases}
        O(1/d_B) & d_B \gg d_A^2 \\
        O(1) & d_B \ll d_A^2
    \end{cases}
\end{equation}
From here, for the fidelity $\mathcal{F}$ for the full $AC$ SWAP protocol as defined in the main text,
\begin{equation}
    \mathcal{F} = \frac{\Tr_{AC}[\rho^{\text{out}}_{AC} \rho_{CA}]}{\sqrt{\Tr_{AC}[(\rho_{AC}^{\text{out}})^2] \Tr_{AC}[\rho_{AC}^2]}},
    \label{eq:fidelities}
\end{equation}
where $\rho_{CA}$ is the ideal swapped version of $\rho_{AC}$ (i.e. given by an ideal SWAP gate acting on $AC$), we can estimate $\mathcal{F}$ for a typical Haar random unitary $U$ acting on pure states as (cf. Eq.~\eqref{eq:FidelityProbabilityBound}, which is saturated by this expression to leading order):
\begin{equation}
    \mathcal{F} \sim \overline{\lvert X\rvert^2} \simeq \frac{1}{d_A^2 \overline{p}} \approx \frac{1}{1+d_A^2/d_B}.
    \label{eq:Fidelity_Haar_Random}
\end{equation}
We therefore see that $d_B \gg d_A^2$ is typically necessary to have $\mathcal{F} \approx 1$. Further, on writing $d_B = d_D^2$ (which connects our protocol to Yoshida-Kitaev recovery, as discussed in Sec.~\ref{sec:connectionToYK}), we recover the fidelity estimate in Ref.~\cite{YoshidaKitaev} for unidirectional Yoshida-Kitaev recovery.


\subsection{General dimension bound}

Our protocol is a special case of a general class of operations involving two unitary gates.
In general, the unitaries on $AB$ and $CB$ need not be inverses.
One could try to implement the SWAP by acting a unitary $U_{AB}$ followed by a possibly unrelated unitary $W_{CB}$, i.e. acting the operator
\begin{equation}
    \langle \phi | W_{CB} U_{AB} | \phi \rangle_B : \mathcal{H}_A \otimes \mathcal{H}_C \to \mathcal{H}_A \otimes \mathcal{H}_C
\end{equation}
In this appendix we argue that considering this more general class of operators does not offer any improvement on at least one technical aspect. 

In our protocol, in which $U$ is a scrambling unitary and $W = U^\dagger$, we found high SWAP fidelity when $\dim \mathcal{H}_B \gg (\dim \mathcal{H}_A)^2$, and poor fidelity when $\dim \mathcal{H}_B \lesssim  (\dim \mathcal{H}_A)^2$.
We now prove that this behavior is natural within this general class of operations:
for any choice of $U$ and $W$ there would be poor SWAP fidelity in the regime $\dim \mathcal{H}_B \lesssim  (\dim \mathcal{H}_A)^2$. 
First a lemma.

{\bf Lemma:} Let $|\Phi \rangle$ be a normalized maximally entangled state with Schmidt rank $D$, i.e.
$$
    |\Phi\rangle = \frac{1}{\sqrt{D}}\sum_{k=1}^D |k\rangle \otimes |k\rangle
$$
Let $|\psi\rangle$ be any (not necessarily normalized) state with Schmidt rank $\le r$.
Then 
\begin{equation}
    |\langle \Phi|\psi \rangle| \le \sqrt{\frac{r}{D}}\sqrt{\langle \psi|\psi \rangle}
\end{equation}

{\bf Proof of lemma:}
Write the Schimdt decomposition of $|\psi \rangle$:
\begin{equation}
    |\psi \rangle = \sum_{k=1}^r s_k |a_k\rangle \otimes |b_k\rangle
\end{equation}
with $s_k \ge 0$.
Note $\langle \psi|\psi \rangle = \sum_k s_k^2$.
Then
$$
    \langle \Phi|\psi \rangle = \frac{1}{\sqrt{D}} \sum_{j=1}^D \langle j | \otimes \langle j | \left( \sum_{k=1}^r s_k |a_k\rangle \otimes |b_k \rangle \right) = \frac{1}{\sqrt{D}}\sum_{k=1}^r s_k \sum_{j=1}^D \langle j|a_k \rangle \langle j|b_k \rangle
$$
It  follows that
$$
    |\langle \Phi|\psi \rangle| = \frac{1}{\sqrt{D}} \left| \sum_{k=1}^r s_k \sum_{j=1}^D \langle j|a_k \rangle \langle j|b_k\rangle \right| \le \frac{1}{\sqrt{D}} \sum_{k=1}^r s_k \left| \sum_{j=1}^D \langle j|a_k \rangle \langle j|b_k \rangle\right|
$$
By Cauchy-Schwarz,
\begin{equation}
     \left|\sum_{j=1}^D \langle j|a_k \rangle \langle j|b_k \rangle\right| \le \left( \sum_j |\langle j|a_k \rangle|^2 \right)^{1/2} \left( \sum_j |\langle j|b_k \rangle|^2 \right)^{1/2} = 1
\end{equation}
and hence
\begin{equation}
    |\langle\Phi|\psi \rangle| \le \frac{1}{\sqrt{D}}\sum_{k=1}^r s_k
\end{equation}
Applying Cauchy-Schwarz again, we get
\begin{equation}
    \sum_{k=1}^r s_k \le \sqrt{r} \left( \sum_{k=1}^r s_k^2 \right)^{1/2} = \sqrt{r}\sqrt{\langle \psi|\psi \rangle}
\end{equation}
Therefore we finally get
\begin{equation}
    |\langle \Phi|\psi \rangle| \le \sqrt{\frac{r}{D}}\sqrt{\langle \psi|\psi \rangle}
\end{equation}

{\bf Main claim:}

Let $\mathcal{H}_A$, $\mathcal{H}_B$, and $\mathcal{H}_C$ be finite dimensional Hilbert spaces with 
\begin{equation}
    \dim \mathcal{H}_A = \dim \mathcal{H}_C = d~,~~~~ \dim \mathcal{H}_B = m
\end{equation}
Fix any normalized state $|\phi\rangle \in \mathcal{H}_B$ and any unitaries
\begin{align*}
    &U: \mathcal{H}_A \otimes \mathcal{H}_B \to \mathcal{H}_A \otimes \mathcal{H}_B \\
    &W: \mathcal{H}_C \otimes \mathcal{H}_B \to \mathcal{H}_C \otimes \mathcal{H}_B
\end{align*}
Define for any normalized state $|\Psi\rangle \in \mathcal{H}_A \otimes \mathcal{H}_C$ the postselection probability
\begin{equation}
    p(\Psi) := ||(\langle\phi|_B \otimes I_{AC}) W U (|\phi\rangle_B \otimes |\Psi\rangle_{AC})||^2
\end{equation}
and the postselected normalized amplitude
\begin{equation}
    F(\Psi) := \frac{|\langle\Psi|S^\dagger (\langle \phi |_B W U |\phi\rangle_B) |\Psi\rangle|}{\sqrt{p(\Psi)}}
\end{equation}
where $S$ is the SWAP operator exchanging $A$ and $C$.
Then there exists a normalized state $|\Psi\rangle \in \mathcal{H}_A \otimes \mathcal{H}_C$ with $p(\Psi) > 0$ such that
\begin{equation}
    F(\Psi) \le \min \left\{1, \sqrt{\frac{m + 1}{d^2 + 1}} \right\}
\end{equation}
In particular, if $m \ll d^2$ then the postselected state is very far from the state obtained by acting SWAP on $AC$.

{\bf Proof:}
First define the linear operator $ K: \mathcal{H}_A \otimes \mathcal{H}_C \to \mathcal{H}_A \otimes \mathcal{H}_C$ by
\begin{equation}
    K := (\langle\phi|_B \otimes I_{AC}) W U (|\phi\rangle_B \otimes I_{AC})
\end{equation}
We can write
$$
    p(\Psi) = \langle\Psi|K^\dagger K |\Psi\rangle
$$
and
$$
    F(\Psi)^2 = \frac{|\langle\Psi|  M |\Psi\rangle|^2}{\langle \Psi|K^\dagger K |\Psi\rangle}
$$
where $M := S^\dagger K$.

Choose an orthonormal bass $\{|b\rangle\}$ for $b \in \{1,...,m\}$ of $\mathcal{H}_B$ and define $F_b: \mathcal{H}_A \to \mathcal{H}_A$ and $E_b: \mathcal{H}_C \to \mathcal{H}_C$ as
\begin{equation}
    F_b := \langle b|_B U |\phi\rangle_B~,~~~~E_b := \langle\phi|_B W |b\rangle_B 
\end{equation}
Then we can write
\begin{equation}\label{eq:K_rewrite}
    K = \sum_{b=1}^m F_b \otimes E_b
\end{equation}
It then holds that 
\begin{equation}\label{eq:tr_bound}
    |\mathrm{tr}(S^\dagger K)|^2 \le m~ \mathrm{tr}(K^\dagger K)
\end{equation}
by the following argument.
We can regard $S$ and $K$ as states $|S\rangle$ and $|K\rangle$ in the Hilbert space of operators on $\mathcal{H}_A \otimes \mathcal{H}_C$ with inner-product $\langle S|K \rangle := \mathrm{tr}(S^\dagger K)$.
Let $|\tilde{S}\rangle$ denote the normalized version of $|S\rangle$. 
Then by the above lemma it holds
\begin{equation}
    |\langle\tilde{S}|K \rangle| \le \sqrt{\frac{m}{d^2}}\langle K|K \rangle
\end{equation}
because $|\tilde{S}\rangle$ is maximally entangled in this Hilbert space and $|K\rangle$ has Schmidt rank at most $m$ by \eqref{eq:K_rewrite}.
In other words,
\begin{equation}
    \frac{|\mathrm{Tr}(S^\dagger K)|}{\langle S|S \rangle} \le \sqrt{\frac{m}{d^2}} \sqrt{\langle K|K \rangle} \implies |\mathrm{Tr}(S^\dagger K)|^2 \le m~ \mathrm{Tr}(K^\dagger K)
\end{equation}
Now we bound the average of $F(\Psi)$ as follows.
Let $|\psi \rangle \in \mathcal{H}_A \otimes \mathcal{H}_C$ be Haar-random, and let $\rho := |\Psi \rangle \langle \Psi|$.
Standard Haar-integration identities give
\begin{equation}
    \mathbb{E}[\rho] := \frac{I}{d^2}~,~~~~\mathbb{E}[\rho^{\otimes 2}] = \frac{I + F_{AC}}{d^2(d^2 + 1)}
\end{equation}
where $F_{AC}$ swaps the two copies of $AC$.
Therefore
\begin{equation}
\begin{split}
    \mathbb{E}\left[\lvert\langle\Psi| M | \Psi\rangle\rvert^2\right] &= \mathbb{E} \mathrm{tr}[(M \otimes M^\dagger)(\rho \otimes \rho)] = \frac{\mathrm{tr}[M M^\dagger] + \lvert \mathrm{tr}M \rvert^2}{d^2(d^2+1)} \\
    \mathbb{E}\left[\langle\Psi| K^\dagger K | \Psi\rangle\right] &= \mathrm{tr}[K^\dagger K \mathbb{E}(\rho)] = \frac{\mathrm{tr}[K^\dagger K]}{d^2}
\end{split}
\end{equation}
Now using $\mathrm{Tr}[M M^\dagger] = \mathrm{Tr}[K^\dagger K]$ and \eqref{eq:tr_bound}, we have
\begin{equation}
    \mathbb{E}\left[\lvert\langle \Psi| M | \Psi \rangle \rvert^2\right] \le \frac{(m+1) \mathrm{tr}[K^\dagger K]}{d^2(d^2+1)}
\end{equation}
Therefore
\begin{equation}\label{eq:num_denom_bound}
    \frac{\mathbb{E}\left[\lvert\langle\Psi| M | \Psi \rangle\rvert^2\right]}{\mathbb{E}\left[\langle\Psi| K^\dagger K | \Psi\rangle\right]} \le \frac{m+1}{d^2 + 1}
\end{equation}
It follows that there exists a state $\ket{\Psi}$ such that $\langle\Psi| K^\dagger K | \Psi\rangle > 0$ and
\begin{equation}
    F(\Psi) \le \sqrt{\frac{m+1}{d^2 + 1}}~.
\end{equation}
To see this, assume for contradiction that no such state exists.
Then for every $|\Psi\rangle \in \mathcal{H}_A \otimes \mathcal{H}_C$ such that $\langle\Psi| K^\dagger K | \Psi\rangle > 0 $ it holds that
$$
   \lvert\langle\Psi| M | \Psi\rangle\rvert^2 > \frac{m+1}{d^2 + 1} \langle\Psi| K^\dagger K | \Psi\rangle
$$
Then taking the expectation of both sides over $|\Psi\rangle$ would give
$$
    \mathbb{E} \lvert\langle\Psi| M | \Psi\rangle\rvert^2 > \frac{m+1}{d^2 + 1} \mathbb{E} \langle\Psi| K^\dagger K | \Psi\rangle
$$
contradicting \eqref{eq:num_denom_bound}.
This finishes the proof.

\subsection{Comments on recovery protocols and scrambling}
\label{sec:connectionToYK}

Here, we situate our analysis in the context of general recovery protocols, and formally connect our protocol to the Yoshida-Kitaev recovery protocol. In particular, we show that the isometry condition captures a necessary and sufficient condition for recovery for \textit{individual states} in the Yoshida-Kitaev protocol (where previously only sufficient conditions were known~\cite{YoshidaKitaev, YoshidaYao}), and also analyze how the Yoshida-Kitaev protocol can be obtained through a series of circuit transformations from our schematic. Finally, we discuss the similarities and differences between the two scrambling mechanisms in our protocol (isometry and entanglement generation), relating them to operator entanglement in time~\cite{HosurQiRobertsYoshida} and space, and identify a necessary condition for them to co-occur (as in the Dicke model).

\subsubsection{Decoupling, error correction, and Petz maps}

In the general language of quantum information and error correction, the Hayden-Preskill protocol can be analyzed~\cite{HaydenPreskill} in terms of decoupling a subsystem from its initial state~\cite{DecouplingOneShot, HaydenPreskill, YoshidaKitaev}, in which case recovery can be performed~\cite{HPPetz_Nakayamaetal} using a complicated quantum channel called the Petz map~\cite{Petz, BarnumKnill, li2025optimality} (we note that decoupling is sufficient, but not necessary, for this recovery). The Petz map is successful in recovering information when the Knill-Laflamme conditions for quantum error correction are satisfied~\cite{KnillLaflamme, BarnumKnill}, but these conditions are often violated in many systems~\cite{li2025optimality}. Moreover, the Petz recovery channel is very difficult to implement in practice. However, the Yoshida-Kitaev protocol~\cite{YoshidaKitaev} provides a more efficient practical scheme for decoding, and has been argued~\cite{HPPetz_Nakayamaetal} to amount to the Petz map for sufficiently ``scrambling'' systems such as random unitaries and the Sachdev-Ye-Kitaev model.

Here, our perspective is somewhat different: instead of focusing on when scrambled information can in principle be recovered using a quantum channel, our focus is on the precise necessary and sufficient conditions for an efficient practical scheme involving only unitary evolution and projections (such as Yoshida-Kitaev recovery) to be able to recover scrambled information. This motivates our analysis in terms of isometric encoding in Sec.~\ref{subsec:pullbacksaredecoders}, which precisely provides such a necessary and sufficient condition in our case.

\subsubsection{Yoshida-Kitaev-type analysis for our protocol}

We can directly connect our analysis to that of Yoshida-Kitaev~\cite{YoshidaKitaev} for Hayden-Preskill recovery, obtaining a sufficient condition for recovery similar to the latter. We will show that when the encoding and reverse-decoding maps are identical (up to a constant of proportionality), our analysis recovers connections between the fidelity and (1) the postselection probability and (2) mutual information with a reference system obtained in previous analyses of the Yoshida-Kitaev protocol.

In particular, \cite{YoshidaKitaev} shows that if the postselection probability is near-minimal, i.e. near $1/d_A^2$, then the fidelity is guaranteed to approach $1$ (in our notation, their main result for a closely related fidelity is $\mathcal{F} \geq 1/(d_A^2 p)$, where the probability $p$ can be expressed in various ways in terms of the mutual information between subsystems or subsystem-averaged out-of-time-ordered correlators). For the fidelity in Eq.~\eqref{eq:AC_fidelity_def}, we can write using $\Tr_C[(\rho_C^{\text{out}})^2] \leq p^2$ and $\Tr_A[\rho_A^2] \leq 1$, \textit{independently of whether isometric encoding occurs}:
\begin{equation}
    \mathcal{F}_{A\to C}[\rho_A] \geq \frac{\Tr_B[\mathcal{M}^e_{BA}(\rho_A)\mathcal{M}^d_{BA}(\rho_A)^\dagger]}{p}.
    \label{eq:fidelitycrossmutualinformation1}
\end{equation}
In the special case $\gamma_C = \idop_C/d_C$ considered in Eqs.~\eqref{eq:AC_specialcase} and \eqref{eq:specialcase_purity}, which directly corresponds to the configuration of $C$ in the unidirectional Yoshida-Kitaev protocol, the numerator is $\mathcal{P}_B/d_A$, for which we have $\mathcal{P}_B \geq 1/d_A$. We therefore obtain the same bound~\footnote{Using notation that is closer to Ref.~\cite{YoshidaKitaev}, their bound on the fidelity is expressed as $\mathcal{F} \geq 1/(d_A d_R \Delta)$, where $\Delta = p$ is the postselection probability; $R$  is an auxiliary system used to control the initial state in $A$, for which way may set $d_R = d_A$ in our setting.} as Ref.~\cite{YoshidaKitaev},
\begin{equation}
    \mathcal{F}_{A\to C}[\rho_A] \geq \frac{1}{d_A^2 p}.
    \label{eq:FidelityProbabilityBound}
\end{equation}
The decay of $p$ to near $1/d_A^2$ is therefore a sufficient condition for Hayden-Preskill recovery with good fidelity, but not a necessary one. This inequality is also of interest in showing that the decay of \textit{nonlocal} OTOCs (which determine the decay of $p$) is sufficient for recovery~\cite{YoshidaKitaev}, in the context of connecting these quantities to scrambling (but note that local OTOCs, sometimes associated with scrambling~\cite{MSSotocBound}, need not determine recovery~\cite{TezukaSYKScrambling}). It is not clear if a similar inequality follows for $\gamma_C \neq \idop_C/d_C$ as in our SWAP protocol.



\begin{figure}
    \centering
    \includegraphics[scale=1.1]{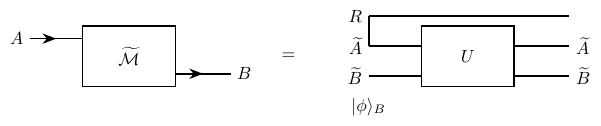}
    \caption{An example of an encoder containing a scrambler with a reference system.}
    \label{fig:referencesystem}
\end{figure}

More generally, the fidelity in the Yoshida-Kitaev protocol can be constrained in terms of the mutual information between a reference system $R$ (say, of dimension $d_A$) prepared in an EPR state with $A$ and an output subsystem ($B$ in our case). To construct this setting in our framework (again, specializing to the reverse decoder being identical to the encoder), let us take $\mathcal{H}_A = \mathcal{H}_{R} \otimes \mathcal{H}_{\widetilde{A}}$ at the input with the pure initial state
\begin{equation}
    \lvert \psi\rangle_{R\widetilde{A}} = \frac{1}{\sqrt{d_A}}\sum_{k=1}^{d_A} \lvert k\rangle_{R} \lvert k\rangle_{\widetilde{A}},
    \label{eq:initstate_for_YK_analysis}
\end{equation}
and $\mathcal{H}_B = \mathcal{H}_R \otimes \mathcal{H}_{\widetilde{B}}$ at the output. 
Our encoding map and reverse-decoding map for this setting are both essentially given by a direct adaptation of Eq.~\eqref{eq:encoder} with the $R$ subsystem just ``going along for the ride'' without any dynamics of its own (see Fig.~\ref{fig:referencesystem}):
\begin{equation}
    U_{AB} = \idop_R \otimes U_{\widetilde{A}\widetilde{B}}.
\end{equation}
The output density operator for the state \eqref{eq:initstate_for_YK_analysis} is given by:
\begin{equation}
    \rho_{R\widetilde{B}} = \mathcal{M}^{e}_{BA}(\lvert \psi\rangle_{R\widetilde{A}}\langle \psi\rvert) = d_A \mathcal{M}^{d}_{BA}(\lvert \psi\rangle_{R\widetilde{A}}\langle \psi\rvert).
    \label{eq:rhomidstateRB}
\end{equation}
Moreover, in terms of the ``reduced'' encoding map $\mathcal{M}_{\widetilde{B}\widetilde{A}}$ given by
\begin{equation}
    \widetilde{\mathcal{M}}_{\widetilde{B}\widetilde{A}}(O_{\widetilde{A}}) = \Tr_A(U_{\widetilde{A}\widetilde{B}} O_{\widetilde{A}} \otimes \lvert\phi\rangle_{\widetilde{B}}\langle\phi\rvert U_{\widetilde{A}\widetilde{B}}^\dagger)
    \label{eq:encoderreduced}
\end{equation}
we can write the postselection probability using Eq.~\eqref{eq:postselectionprobability} as
\begin{equation}
    p = \Tr_{\widetilde{B}}\left[\mathcal{M}_{\widetilde{B}\widetilde{A}}\left(\frac{\idop_{\widetilde{A}}}{d_A}\right)\mathcal{M}_{\widetilde{B}\widetilde{A}}\left(\frac{\idop_{\widetilde{A}}}{d_A}\right)\right] = \Tr_{\widetilde{B}}\left[\rho_{\widetilde{B}}^2\right],
\end{equation}
where $\rho_{\widetilde{B}} = \Tr_R[\rho_{R\widetilde{B}}]$, with the latter defined as in Eq.~\eqref{eq:rhomidstateRB}.

Eqs.~\eqref{eq:fidelitycrossmutualinformation1} and \eqref{eq:rhomidstateRB} imply that:
\begin{equation}
    \mathcal{F}_{A\to C}\left[\vphantom{\sum}\rho_A = \lvert \psi\rangle_{R\widetilde{A}}\langle \psi\rvert\right] \geq \frac{\Tr_{R\widetilde{B}}\left[\rho_{R\widetilde{B}}^2\right]}{d_A \Tr_{\widetilde{B}}\left[\rho_{\widetilde{B}}^2\right]} =\frac{1}{d_A^2} \exp\left(I^{(2)}(R, \widetilde{B})\right),
\end{equation}
where  we recognize the 2nd R\'{e}nyi mutual information between $R$ and $\widetilde{B}$ after the encoding step~\cite{HorodeckiEntanglementReview, HosurQiRobertsYoshida}:
\begin{equation}
    I^{(2)}(R, \widetilde{B}) \equiv S^{(2)}_{R} + S^{(2)}_{\widetilde{B}} - S^{(2)}_{R\widetilde{B}},
\end{equation}
in which $S^{(2)}_X = -\ln \Tr_X[\rho_X^2]$ is the 2nd R\'{e}nyi entropy~\cite{HorodeckiEntanglementReview} of the post-encoding state in the subsystem $X$ (note that $S^{(2)}_R = \ln d_A$ as $R$ is in a maximally mixed state throughout the encoding process). This recovers the connection to mutual information with a reference system for this fidelity (implicitly ``averaged'' in some sense over reference states) obtained in \cite{HosurQiRobertsYoshida, YoshidaYao} when the initial state is maximally entangled with this system. For our more general SWAP protocol with an arbitrary initial state in $C$, we may instead have a form of ``cross-mutual-information'', analogous to the cross-purity in Eq.~\eqref{eq:crosspurity_def}, that may constrain fidelity with a reference system. However, it appears that one formally needs the full isometry discussion in Sec.~\ref{subsec:pullbacksaredecoders} to analyze recovery in individual initial states.

\subsubsection{Isometry analysis for the Yoshida-Kitaev protocol, and entanglement generation}
\label{sec:isometry_to_entanglement}

Conversely, we can identify a precise necessary and sufficient condition for Yoshida-Kitaev decoding to be successful for specific initial states in terms of our isometry condition. This is to be contrasted with Ref.~\cite{YoshidaYao}, in which only a sufficient condition on par with (but slightly different from) Eq.~\eqref{eq:FidelityProbabilityBound} is obtained for individual states.


First, we will consider a general schematic of encoding and decoding that captures the structure of the Yoshida-Kitaev protocol, and then treat the latter as a special case of this schematic. As with the prior subsection, we will assume that the encoder and reverse-decoder maps are identical. A distinguishing feature of the Yoshida-Kitaev protocol is that one has two different partitions into subsystems before and after the scrambling unitary $U$. Specifically, let $\mathcal{H}$ be the Hilbert space that the scrambler acts on. Before scrambling, one has the decomposition
\begin{equation}
    \mathcal{H} = \mathcal{H}_A \otimes \mathcal{H}_B,
\end{equation}
where $A$ is the \textit{input} subsystem, and $B$ is the \textit{output} subsystem for the scrambling operation, both on the \textit{same} side (as to why will be discussed shortly).
After scrambling, we have an (in principle) entirely different decomposition:
\begin{equation}
    \mathcal{H} = \mathcal{H}_C \otimes \mathcal{H}_D,
\end{equation}
where $D$ is a subsystem on which we can control the initial state or project onto the final state, while $C$ is left completely uncertain (maximally mixed). As $B$ is the output subsystem and we have control over the state in $C$, it is convenient to consider the dynamics of the inverse scrambler:
\begin{equation}
    U^\dagger: \mathcal{H}_C \otimes \mathcal{H}_D \to \mathcal{H}_A\otimes \mathcal{H}_B,
\end{equation}
where $CD$ is initialized to the state
\begin{equation}
    \rho_{CD}(0) = \frac{\idop_C}{d_C} \otimes \lvert \phi\rangle_D\langle\phi\rvert.
\end{equation}
To motivate why both the input and output subsystems are on the same side, we note that this is a standard feature of teleportation protocols~\cite{BennettTeleportation, NielsenChuang}: for example, when one prepares an EPR state~\cite{NielsenChuang} between $A$ and $B$, the operator transfer property that enables teleportation,
\begin{equation}
    O_A\lvert \text{EPR}\rangle_{AB} = O_B^T \lvert \text{EPR}\rangle_{AB},
    \label{eq:EPRidentity}
\end{equation}
can be regarded as taking $A$ to be an input subsystem and $B$ to be an output subsystem, both on the same side (i.e., future) of the EPR state. In this setting, $C$ and $D$ are involved in the initial state for whichever algorithm is used to prepare the EPR state.

\begin{figure}
    \centering
    \includegraphics[scale=1]{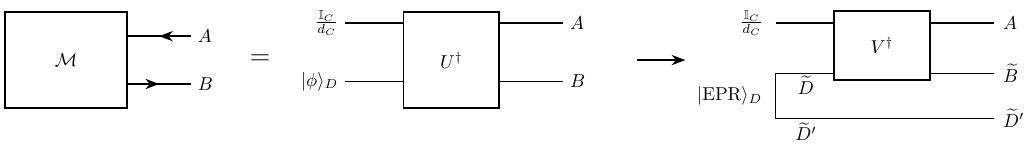}
    \caption{Encoding between subsystems on the same side, and the Yoshida-Kitaev encoder as an instance of a scrambling encoding map with an auxiliary EPR pair.}
    \label{fig:YKencoder}
\end{figure}

The corresponding encoding map is (with Hermitian conjugates $\dagger$ introduced for formal convenience; see Fig.~\ref{fig:YKencoder}):
\begin{equation}
    \mathcal{M}_{BA}(O_A^\dagger)^\dagger = \Tr_A\left(U^\dagger \frac{\idop_C}{d_C}\otimes \lvert \phi\rangle_D\langle\phi\rvert U O_A^\dagger\right).
\end{equation}
For simplicity, let us take the reverse-decoder to be identical to the encoder. 
The complex conjugate of the isometry condition \eqref{eq:isometric_encoding_pair} then reads:
\begin{equation}
    \Tr_B[\mathcal{M}_{BA}(O_{1A}^\dagger)^\dagger \mathcal{M}_{BA}(O_{2A}^\dagger)] \simeq p_{\lambda} \Tr_A[O_{1A}^\dagger O_{2A}].
    \label{eq:YKisometry}
\end{equation}

To get some intuition for this condition, let us consider the full density operator $\rho_{AB}(U^\dagger)$ after the action of $U^\dagger$ on the initial state $(\idop_C/d_C) \otimes \lvert \phi\rangle_D\langle \phi\rvert$. A basis expansion of $\rho_{AB}(U^\dagger)$ with respect to the bipartition $AB$ in terms of an orthonormal basis $\lvert k\rangle_A\langle j\rvert$ of operators can be written as:
\begin{equation}
    \rho_{AB}(U^\dagger) = U^\dagger \frac{\idop_C}{d_C}\otimes \lvert \phi\rangle_D\langle\phi\rvert U = \sum_{k,j = 1}^{d_A}\lvert k\rangle_A\langle j\rvert \otimes M_{B}^{(kj)},
    \label{eq:YK_SchmidtDecomposition}
\end{equation}
where it follows that $M_B^{(kj)} = \mathcal{M}_{BA}(\lvert j\rangle \langle k\rvert)^\dagger$. The isometry condition now becomes:
\begin{equation}
    \Tr_B\left[M_B^{(k_1j_1)} M_B^{(k_2 j_2)\dagger}\right] \simeq p_{\lambda} \delta_{k_1 k_2} \delta_{j_1 j_2},
    \label{eq:YK_schematic_isometry}
\end{equation}
which is a statement that the Schmidt decomposition~\cite{NielsenChuang} of \eqref{eq:YK_SchmidtDecomposition} is essentially of maximal rank (i.e. maximal operator entanglement for the output density operator across the $AB$ bipartition), having $d_A^2$ nearly identical eigenvalues that are approximately $\sqrt{p_{\lambda}}$.

If $A$ is taken to be a reference system $R$ (so as to capture isometric encoding), the mutual information between $R$ and $B$ is given by the logarithm of
\begin{equation}
    \frac{d_A \sum_{k,j} \Tr_B[M^{(kj)}_B M_B^{(jk)}]}{\sum_{k,j} \Tr_B[M^{(kk)}_B M_B^{(jj)}]},
\end{equation}
which illustrates how mutual information captures some aspects of operator entanglement in this instance (i.e. the denominator is small given that $M_B^{(jj)}$ and $M_B^{(kk)}$ are nearly orthogonal for $j=k$, leading to large mutual information).


As a sanity check, let us consider the case where $C$ is a trivial subsystem (dimension $1$), and the entire Hilbert space is represented by $\mathcal{H}_D$. In this case, the final state $\rho_{AB}(U^\dagger)$ is a pure state, and the isometry condition essentially states that this is a maximally entangled EPR state:
\begin{equation}
    \left.\rho_{AB}(U^\dagger)\right\rvert_{\mathcal{H}_D = \mathcal{H}} \propto \sum_{k,j=1}^{d_A} \lvert k\rangle_A\lvert \widetilde{k}\rangle_B \otimes {{_A}\langle j\rvert} {{_B}\langle \widetilde{j}\rvert},
\end{equation}
where $\lbrace\lvert \widetilde{k}\rangle\rbrace_{k=1}^{d_A}$ is a suitable orthonormal set in $\mathcal{H}_B$. This is just the standard quantum teleportation protocol, where the encoding operation is an EPR measurement and the (formal) decoding operation is preparing an EPR state, by acting with $U^\dagger$ on the pure state $\lvert \phi\rangle_D$. Of course, this state may itself be a highly entangled (or EPR) state, but if we demand that the initial state is unentangled, then we see that we \textit{require} $U^\dagger$ to generate entanglement for recovery to be successful. The same applies for more general choices of $C$ and $D$ if we demand that the state in $D$ is a product state with respect to its constituent subsystems, as $C$ is already in a product state (which is maximally mixed, but not entangled in the sense of having correlations between subsystems). From this viewpoint, our discussion of the $C \to A$ mechanism in Sec.~\ref{sec:CtoAmechanism} is a more specialized analysis of this form of recovery.

\begin{figure}
    \centering
    \includegraphics[width=0.96\linewidth]{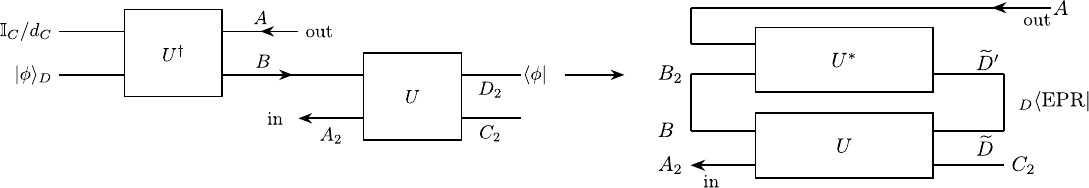}
    \caption{The Yoshida-Kitaev protocol as a special case of a scrambling encoder-decoder system with an additional EPR pairs. The EPR operator transfer identity Eq.~\eqref{eq:EPRidentity} has been used to transform $U^\dagger$ to $U^*$ separately in $A$ and $B$, using the partial transpose properties depicted in Fig.~\ref{fig:partialtranspose}. 
    }
    \label{fig:YK_recovery_manips}
\end{figure}

At the other extreme, if we take $C$ to be the full system $\mathcal{H}_C = \mathcal{H}$, then we are initially in a maximally mixed state and no encoding is possible as $U^\dagger \idop_C/d_C U = \idop_C/d_C$. More generally, a nontriviality indicated by the Yoshida-Kitaev protocol is that we can even be very close to this case: $D$ can in principle be much smaller than $C$, and recovery may in general succeed as long as $D$ remains much larger than $A$.


We can specifically recover the Yoshida-Kitaev protocol as an instance of our general schematic, by splitting $D$ into two subsystems $\widetilde{D}$ and $\widetilde{D}'$, where $\widetilde{D}'$ is an ``inert'' subsystem that is not subject to any dynamics. We then require that the unitary $U^\dagger$ factorizes as (depicted in Fig.~\ref{fig:YKencoder}):
\begin{equation}
    U^\dagger = (V^\dagger)_{C\widetilde{D}} \otimes \idop_{\widetilde{D}'},
\end{equation}
and the initial state $\lvert \phi\rangle_D$ is a maximally entangled state between $\widetilde{D}$ and $\widetilde{D}'$:
\begin{equation}
    \lvert \phi\rangle_D = \frac{1}{\sqrt{d_{\widetilde{D}}}}\sum_{k=1}^{d_{\widetilde{D}}} \lvert k\rangle_{\widetilde{D}} \lvert k\rangle_{\widetilde{D}'}.
\end{equation}
Moreover, $B$ is also split into $\widetilde{B}$ and $\widetilde{D}'$ such that
\begin{equation}
    U = (V)_{A\widetilde{B}} \otimes \idop_{\widetilde{D}'}.
\end{equation}
This is depicted in Fig.~\ref{fig:YK_recovery_manips}, where $V^\dagger$ has been transformed to $V^*$ via the transpose property of maximally entangled states ($O_{\widetilde{D}}\lvert \phi\rangle_{\widetilde{D}\widetilde{D'}} = O^T_{\widetilde{D}'} \lvert \phi\rangle_{\widetilde{D}\widetilde{D'}}$). 
Here, as $\lvert \phi\rangle_D$ is already an entangled state, it may be sufficient for $V$ to transfer this entanglement to $A$ from $\widetilde{D}$ (without generating any entanglement by itself, e.g. via a SWAP gate) to generate operator entanglement in the output state (see also Ref.~\cite{YKrecoveryClifford} for a related transformation that avoids merging $\widetilde{D}$ and $\widetilde{D}'$ into $D$). 

We therefore find that isometric encoding reduces to operator entanglement when $A$ and $B$ are both on the ``same side'' of the unitary gate. Moreover, it is the generation of density operator entanglement, rather than decoupling, mutual information or the decay of out-of-time-ordered correlators as considered in previous discussions of this protocol~\cite{YoshidaKitaev, YoshidaYao}, that is a precise necessary and sufficient condition for successful recovery in this setting. As an aside, in the context of fast scrambling bounds (which constrain the earliest time after which ``scrambling'' may occur, either instantaneously or over an extended time interval), this verifies entanglement based bounds~\cite{LashkariFastScrambling, BentsenGuLucasScrambling, dynamicalqspeedlimit, dynamicalqfastscrambling} as being more directly relevant to such protocols than those on \textit{local} out-of-time-ordered correlators~\cite{MSSotocBound} (which do not seem operationally connected to recovery from large subsystems~\cite{TezukaSYKScrambling}).


\subsubsection{Relating isometric encoding and entanglement generation via the partial transpose}

\begin{figure}
    \centering
    \includegraphics[width=0.95\linewidth]{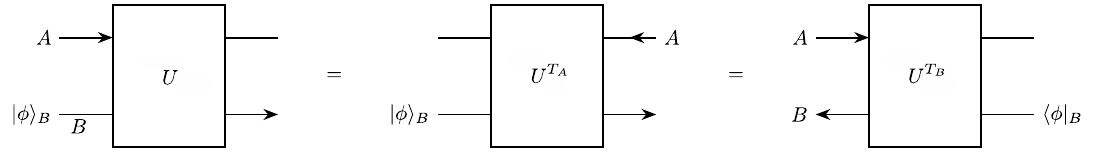}
    \caption{Depiction of the effect of partial transpose on different subsystems in circuit diagrams. Applying the transformation from the diagram on the middle to the one on the left converts the entanglement mechanism to the isometric encoding mechanism. Applying this transformation to the general teleportation circuit on the left side of Fig.~\ref{fig:YK_recovery_manips} (with some relabeling of subsystems, including identifying $A$ and $C$ as similar subsystems, together with swapping the ``in'' and ``out'' labels) gives the isometric encoding circuit in Fig.~\ref{fig:encoderdecoder}. Conversely, applying this procedure to the isometric encoder of Fig.~\ref{fig:encoderdecoder} (again identifying $A$ and $C$ as similar subsystems) gives the Yoshida-Kitaev circuit after applying the transformations in Fig.~\ref{fig:YK_recovery_manips}.}
    \label{fig:partialtranspose}
\end{figure}

Having connected isometric encoding to entanglement in the above section, it is also worth connecting our $A \to C$ mechanism to the Yoshida-Kitaev protocol. In short, our $A \to C$ mechanism is a variant of the Yoshida-Kitaev schematic discussed above with $A$ being an input subsystem and $B$ being an output subsystem. In fact, we can construct the Yoshida-Kitaev protocol within our isometric encoding mechanism for $A \to C$ provided that the scrambling unitary $U$ is replaced by the (in general, nonunitary) completely positive map:
\begin{equation}
    U_{AB} \rho_{AB} U_{AB}^{\dagger} \to U_{AB}^{T_B} \rho_{AB} U_{AB}^{T_B \dagger},
\end{equation}
where $T_B$ denotes a partial transpose~\cite{NielsenChuang} with respect to the $B$ subsystem. This is depicted in Fig.~\ref{fig:partialtranspose}.

The partial transpose $T_B$ plays a crucial role here, in converting between the two encoding mechanisms: (1) isometric encoding by 
$U$ from an input subsystem to an output subsystem B and (2) entanglement generation by $U^\dagger$  between output subsystems. For operators, the latter can be identified with conventional operator entanglement in space, as in Sec.~\ref{sec:isometry_to_entanglement}; correspondingly, the former can then be identified with operator entanglement in time~\cite{HosurQiRobertsYoshida}.

To illustrate this in a simple example, let $U = \text{SWAP}_{AB}$ between two subsystems $A$ and $B$ (not to be confused with our full SWAP gate between $A$ and $C$), which performs perfect isometric encoding from the input $A$ to the output $B$ (and, incidentally, from input $B$ to output $A$). Then $U^{T_B} = \lvert \text{EPR}\rangle_{AB}\langle \text{EPR}\rvert$ is a projector onto an EPR state, whose output is $\lvert EPR\rangle_{AB}$ that has maximal entanglement between $A$ and $B$. In this sense, SWAP can be understood as generating an entangled state in time, between input $A$ and output $B$. However, neither operation exhibits the other scrambling mechanism (i.e., SWAP is non-entangling, and the EPR projection generates entanglement but does not preserve the original state of $A$). From this point of view, our full SWAP protocol in the main text manages to combine both mechanisms in a single circuit by swapping the role of $A$ and $C$ as input/output subsystems for each gate.

Finally, we consider a special circumstance that is particularly relevant to the Dicke model. We note that the Dicke model Hamiltonian is invariant under such a partial transpose (say, in the Fock basis of bosons):
\begin{align}
    H_{D}^{T_B} &= \frac{2g}{\sqrt{N}}\,(a_B^{T_B}+a_B^{\dagger T_B})S^{z}_A+\delta\, a_B^{T_B} a_B^{\dagger T_B} +\Omega\,S^{x}_A \nonumber \\
    &= \frac{2g}{\sqrt{N}}\,(a_B^{\dagger *}+a_B^{*})S^{z}_A+\delta\, a_B^{\dagger *} a_B^{*} +\Omega\,S^{x}_A \nonumber \\
    &= H_D
\end{align}
as $a_B^{\ast} = a_B$ in the Fock basis. This means that isometric encoding and entanglement generation must \textit{co-occur} in the Dicke model, as well as any other scrambling interaction that is invariant under a partial transpose. In other words, we expect that there is no regime in the Dicke model where either component $A \to C$ or $C \to A$ of our SWAP operation succeeds and the other fails, up to errors of order $\epsilon$ contained in the above analyses.

Finally, to motivate Fig. 1b in the main text, we consider the question of the gravitational analogue of our protocol. Our $C \to A$ protocol where $U_{AB}$ is the partial transpose of a unitary is shown above, via its connection to quantum teleportation with a partial projection, to be closely related to the Yoshida-Kitaev protocol. The latter is connected to the Hayden-Preskill thought experiment, in which one considers information recovery from a black hole $B$ that has radiated more than half its initial information content in the form of Hawking radiation (corresponding to the EPR pair involving $B$ and $B_2$), i.e. after the Page time~\cite{HaydenPreskill, SekinoSusskind, PageSubsystem, PageTime}. It is also worth asking if there is any such direct gravitational analogue of our $A \to C$ protocol without the partial transpose relation to the Yoshida-Kitaev decoder, i.e. if $U_{AB}$ is a unitary. As there are no EPR pairs in this version, we believe that our protocol best represents the original black hole information recovery thought experiment of Susskind and Thorlacius~\cite{SusskindThorlaciusGedanken} (see also \cite{SekinoSusskind}), where one attempts recovery from Hawking radiation without any previously emitted radiation that it may be entangled with\footnote{In the context of Fig.~\ref{fig:encoderdecoder}, we can qualitatively think of the subsystems as follows: (1) Before scrambling: $A$ - infalling matter, $B$ - black hole; (2) after scrambling: $A$ - black hole remnant, $B$ - Hawking radiation, $C$ - raw material for recovery; (3) after unscrambling: $B$ - projected Hawking radiation, $C$ - reconstructed information originally contained in the infalling matter. See also Fig. 1b in the main text.}.



\section{Numerical details}

\subsection{Primary Hamiltonian implementation}

For simulating the Dicke model dynamics, we work in the fully symmetric spin sector $S=N_A/2$ with dimension $d_A=N_A+1$ and represent initially the bosonic mode in a fixed, shifted Fock window centered around the target mean occupation $\mu=|\alpha_0|^2$. The truncated bosonic ladder operator is constructed directly in the window $\{n_{\min},\dots,n_{\max}\}$, where $n_{\min}=\lfloor\mu\rfloor-\Delta$ and $n_{\max}=\lfloor\mu\rfloor+\Delta$ with a half–width $\Delta$ chosen proportional to $|\alpha_0|$, physically this is motivated by the fact that the dynamics observed lead to a thermalization of the quadratures of the bosonic mode while we observe an increment in the variance of the bosonic population. The coherent reference state $\lvert\alpha\rangle$ is projected into this window using its exact number-state amplitudes and then renormalized; the neglected tail probability $\sum_{n\notin[n_{\min},n_{\max}]} e^{-|\alpha|^2}|\alpha|^{2n}/n!$ is kept below numerical noise and monitored for checking convergence \cite{LewisSwan2021}.

With this basis, the Hamiltonian used in time propagation, assuming that the first and second ensemble correspond to spin and boson, is:

\begin{equation}
\label{eq:Lab_frame}
H_{\mathrm D}
=\frac{2g}{\sqrt{N_A}}\; S_z\otimes( b+ b^\dagger)
+\delta\,\idop_A \otimes  n_{\mathrm{rel}}
+\Omega_A\, S_x\otimes\idop_B,
\end{equation}
where $ b$ and $ n_{\mathrm{rel}}$ act in the truncated window and the global phase $\delta n_{\min}$ is dropped. Internal diagnostics include norm conservation below a given threshold, and stability of $\{\overline{\mathcal{F}}_{AC},\overline{p}\}$ under increases of window width and time-step refinement to ensure convergence. 

We see that the Eq.~(\ref{eq:Lab_frame}) and the Dicke model in the main text are connected by a $\pi/2$-rotation around $\hat{S}_y$, so provided that we perform an average over the pure states manifold this average is invariant under local rotations.

To characterize typical performance we average the metrics over an ensemble of $n_s$ randomly chosen over the set of pure states that for $N$ atoms in the Dicke manifold has the general form $\left|\psi\left(0\right)\right>=\sum_{j=0}^N c_j \left|J=\frac{N}{2},J_z=-\frac{N}{2}+j\right>$ using a Haar-random state sampling in the symmetric \((N+1)\)-dimensional Dicke subspace.  This is accomplished in the following way:
We view the Dicke coefficients $c_j$ as a vector in \(\mathbb{C}^{d}\) with \(d=N+1\) (equivalently, a real vector in \(\mathbb{R}^{2d}\)). Then, we draw a real Gaussian vector \(\vec{x} \sim \mathcal{N}(0,I_{2d})\), where $I_{2d}$ is the identity matrix with size $2d$ which corresponds to the multivariate covariance matrix of the $2d$ random variables such that each variable has mean equal to zero and variance 1, after we normalize the vector to \(\tilde{\vec{x}}=\vec{x}/\|\vec{x}\|\) (which is uniform on the \((2d-1)\)-sphere \(S^{2d-1}\)), and then pair entries into complex coefficients by setting \(c_k=\tilde{x}_k + i\,\tilde{x}_{k+d}\) for \(k=0,\ldots,d-1\). The resulting coefficient vector \(c=(c_0,\ldots,c_{d-1})\) satisfies \(\sum_{k=0}^{d-1} |c_k|^2 = 1\) and is distributed according to the unitary-invariant (Fubini–Study) measure, i.e., Haar-random in the symmetric subspace spanned by \(\{|J=\tfrac{N}{2},\,J_z=-\tfrac{N}{2}+j\rangle\}_{j=0}^{N}\) \cite{Maziero2015}. This sampling scheme provides a way to explore generic pure states in the symmetric Dicke manifold—going beyond spin-coherent (polarized) states—and including paradigmatic nonclassical states such as nontrivial Dicke states \(\big|J=\tfrac{N}{2},\,J_z=m\big\rangle\), the \(N\)-qubit GHZ (cat) state
$\left|\psi_{\mathrm{GHZ}}\right\rangle
= \frac{\left|0\right\rangle^{\otimes N} + \left|1\right\rangle^{\otimes N}}{\sqrt{2}}
= \frac{\left|J=N/2,\,J_z=-N/2\right\rangle + \left|J=N/2,\,J_z=N/2\right\rangle}{\sqrt{2}}$, among others.

After setting the sampling, we determine parameter scans sweep $(\delta,\Omega_A)$ on a rectangular grid at fixed $g=1$ (without loss of generality) that provides a way to characterize the dependence of the fidelity with the parameters of the Dicke model. At given times of interest, we record the ensemble averages of the overall fidelity $\overline{\mathcal{F}}_{AC}$, where we use the Uhlmann fidelity defined as $\mathcal{F}(\rho,\sigma)\equiv\left(\mathrm{Tr}\sqrt{\sqrt{\sigma}\rho\sqrt{\sigma}}\right)^2$, and the probability of detection $\overline{p}$ as well as full time traces as needed. As our numerical simulations in the main text deal with pure states, this reduces precisely to the fidelity measure used in the main text for these states.
For a (possibly mixed) recovered state $\rho_{AC}(t)$ and initial state $\rho_{AC}(0)$ we calculate the fidelity of the SWAP gate such that:

\begin{equation}
\mathcal{F}_{AC}(t)\equiv F\big(\mathrm{SWAP}[\rho_{AC}(t)],\rho_{AC}(0)\big)
\label{eq:fid_def}
\end{equation}
which reduces to $\mathcal{F}_{AC}(t)=\langle\psi_{\rm tar}|\mathrm{SWAP}[\rho_{AC}(t)]|\psi_{\rm tar}\rangle$ when $\rho_{AC}(0)=|\psi_{\rm tar}\rangle\!\langle\psi_{\rm tar}|$ \cite{NielsenChuang}. 
The probability of detection (success) associated with a given outcome modeled by the projector $ P_{\mathrm{succ}}=|\alpha\rangle\!\langle\alpha|\!\ge0$ is given by
\begin{equation}
p(t)=\mathrm{Tr}\big[ P_{\mathrm{succ}}\,\rho_{AC}(t)\big]=\left<\alpha|\rho_{AC}(t)|\alpha\right>,
\label{eq:prob_def}
\end{equation}

The numerical estimators (time- and sample-averaged over Haar draws and discrete times $t_\ell\in[2t_{\min},2t_{\max}]$, here 2 accounts for the total time in the 2-gates implementation) of the Uhlmann fidelity and the probability of detection are
\begin{equation}
\overline{\mathcal F}_{AC}=\frac{1}{N_{\mathrm{samp}}N_t}\sum_{i=1}^{N_{\mathrm{samp}}}\sum_{\ell=1}^{N_t}
\mathcal F_{AC}^{(i)}(t_\ell),
\qquad
\overline{p}=\frac{1}{N_{\mathrm{samp}}N_t}\sum_{i=1}^{N_{\mathrm{samp}}}\sum_{\ell=1}^{N_t}
p^{(i)}(t_\ell).
\label{eq:fid_prob_estimators}
\end{equation}

where $N_{\text{samp}}$ is the number of sampled states in the Haar-random sampling and the $N_t$ is the number of equidistant sampled points in $[2t_{\min},2t_{\max}]$.

To characterize dynamical spin--boson entanglement we monitor the mixedness generated in the spin subsystem \(A\) during the time evolution. For each initial condition \(|\Psi_0\rangle=|\psi_A\rangle\otimes|\alpha_0\rangle\) (with \(|\psi_A\rangle\) drawn according to the Haar-random sampling in the symmetric Dicke subspace described above), we evolve to time \(t\) and form the reduced state
\begin{equation}
    \rho_A(t)\;=\;\mathrm{Tr}_B\Big[\,U(t)\,|\Psi_0\rangle\!\langle\Psi_0|\,U^\dagger(t)\Big],
\end{equation}
from which we compute the purity
\begin{equation}
    \mathcal P_A(t)\;=\;\mathrm{Tr}\big[\rho_A(t)^2\big]\;\in\;\big[1/d_A,\,1\big].
\end{equation}
We aggregate over an ensemble of \(N_{\mathrm{samp}}\) Haar-random samples \(|\psi_A\rangle\) and a discrete time grid \(\{t_\ell\}_{\ell=1}^{N_t}\subset[t_{\min},t_{\max}]\) via
\begin{equation}
    \overline{\mathcal P}_A\;=\;
    \frac{1}{N_{\mathrm{samp}}N_t}\sum_{i=1}^{N_{\mathrm{samp}}}\sum_{\ell=1}^{N_t}
    \mathrm{Tr}\Big(\rho_{A,i}(t_\ell)^2\Big),
    \qquad
    \overline{S}_{2,A}\;=\;-\log_{d_A}\!\big(\overline{\mathcal P}_A\big).
\end{equation}
Here \(\overline{S}_{2,A}=0\) indicates that the evolution keeps \(A\) pure (no spin--boson entanglement generated), while \(\overline{S}_{2,A}\to 1\) signals strong and persistent entanglement production on the accessible spin system. The window time is in the regime of large-time in order to ensure faithful recovery.

\medskip
\noindent\textit{Numerical checks.}
We verify convergence of $\overline{S}_{2,A}$ by increasing the boson window (so that the neglected coherent–state tail is below numerical precision) and tightening diagonalization tolerances.

After analyzing the role of entanglement, we focus on early–time recovery. We set a target fidelity \(\mathcal F_\star=0.9\) and identify the "minimal" recovery time \(t_r\) such that \(\overline{\mathcal F}_{AC}(t_r)\ge \mathcal F_\star\), being $t_r$ the minimum recovery time, holds for all \(t\in[t_r,\,t_r+\Delta t]\) with \(g\,\Delta t=1\). The control parameters \((\Omega_r,\delta_r)\) achieving this define our nominal (imperfect) recovery. To assess robustness, we then probe parameter detunings
\[
\delta \rightarrow \delta_r+\epsilon_\delta,\qquad
\Omega \rightarrow \Omega_r+\epsilon_\Omega,
\]
by scanning the shift plane \((\epsilon_\delta,\epsilon_\Omega)\), and quantify the sensitivity of \(\overline{\mathcal F}\) to errors in detuning and drive. This reveals regions resilient to control imperfections, and, moreover, provides a route to metrology by leveraging the measured sensitivities of the subsystems.

\subsection{\edit{Floquet phase modulation of the Dicke evolution}}
\label{sec:supp_floquet_phase_modulation}

\edit{The numerical results in the main text are obtained from evolution under the time-independent Dicke Hamiltonian in Eq.~\eqref{eq:Lab_frame}. To improve the fidelity using state of the art implementations we consider a minimal Floquet modification in which only the phase of the spin-dependent force is varied. The detuning, coupling strength, and transverse field are kept fixed, so the modulation selects different motional quadratures of the same native interaction. The Hamiltonian applied in each interval is
\begin{equation}
H_{\mathrm D}(\varphi)
=
\frac{2g}{\sqrt{N_A}}\; S_z
\left(
e^{-i\varphi}b
+
e^{i\varphi}b^\dagger
\right)
+
\delta\, n_{\mathrm{rel}}
+
\Omega_A\, S_x .
\label{eq:floquet_phase_hamiltonian}
\end{equation}
The static evolution of Eq.~\eqref{eq:Lab_frame} corresponds to $\varphi=0$. The value $\varphi=\pi/2$ couples the spin to the conjugate bosonic quadrature.}

\edit{The alternating sequence step used in the simulation is
\begin{equation}
U_F(2\tau_{\mathrm{sw}})
=
\exp\left[-iH_{\mathrm D}(\pi/2)\tau_{\mathrm{sw}}\right]
\exp\left[-iH_{\mathrm D}(0)\tau_{\mathrm{sw}}\right],
\label{eq:floquet_step}
\end{equation}
so that the applied Hamiltonians are
\begin{equation}
H_{\mathrm D}(0),\;
H_{\mathrm D}(\pi/2),\;
H_{\mathrm D}(0),\;
H_{\mathrm D}(\pi/2),\ldots
\label{eq:floquet_sequence}
\end{equation}
in consecutive intervals of duration $\tau_{\mathrm{sw}}$, denoting the switching time. After $m$ periods, corresponding to a scrambling time $t=2m\tau_{\mathrm{sw}}$, the unitary is
\begin{equation}
U_{\mathrm{scr}}(t)
=
\left[
U_F(2\tau_{\mathrm{sw}})
\right]^m .
\label{eq:floquet_scrambling_unitary}
\end{equation}
The reverse evolution in the decoding step is obtained by applying the corresponding inverse sequence.}

\begin{figure}[!ht]
    \centering

    \begin{minipage}[t]{0.48\linewidth}
        \centering
        \includegraphics[width=\linewidth]{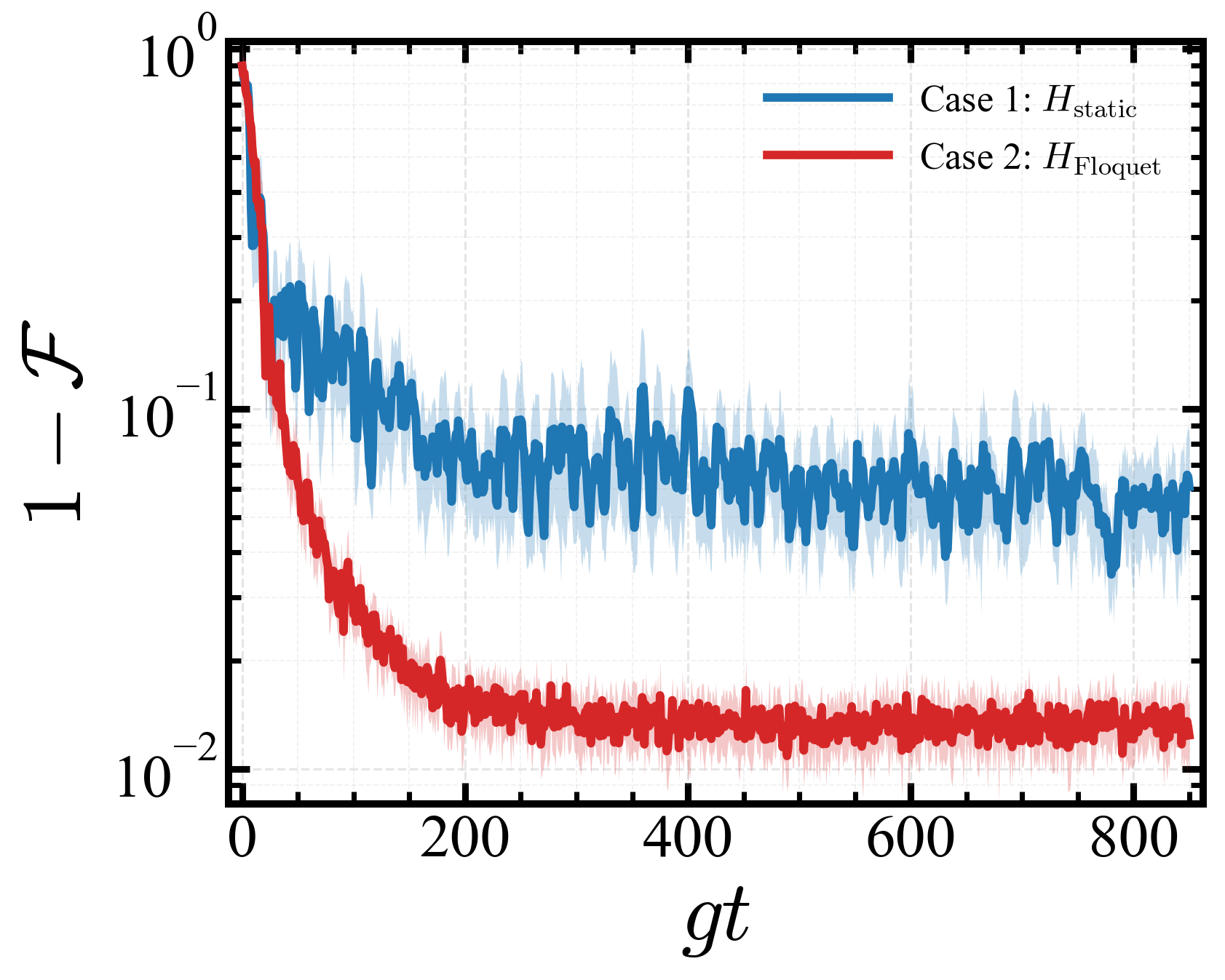}
        \centerline{\small (a)}
    \end{minipage}
    \hfill
    \begin{minipage}[t]{0.48\linewidth}
        \centering
        \includegraphics[width=\linewidth]{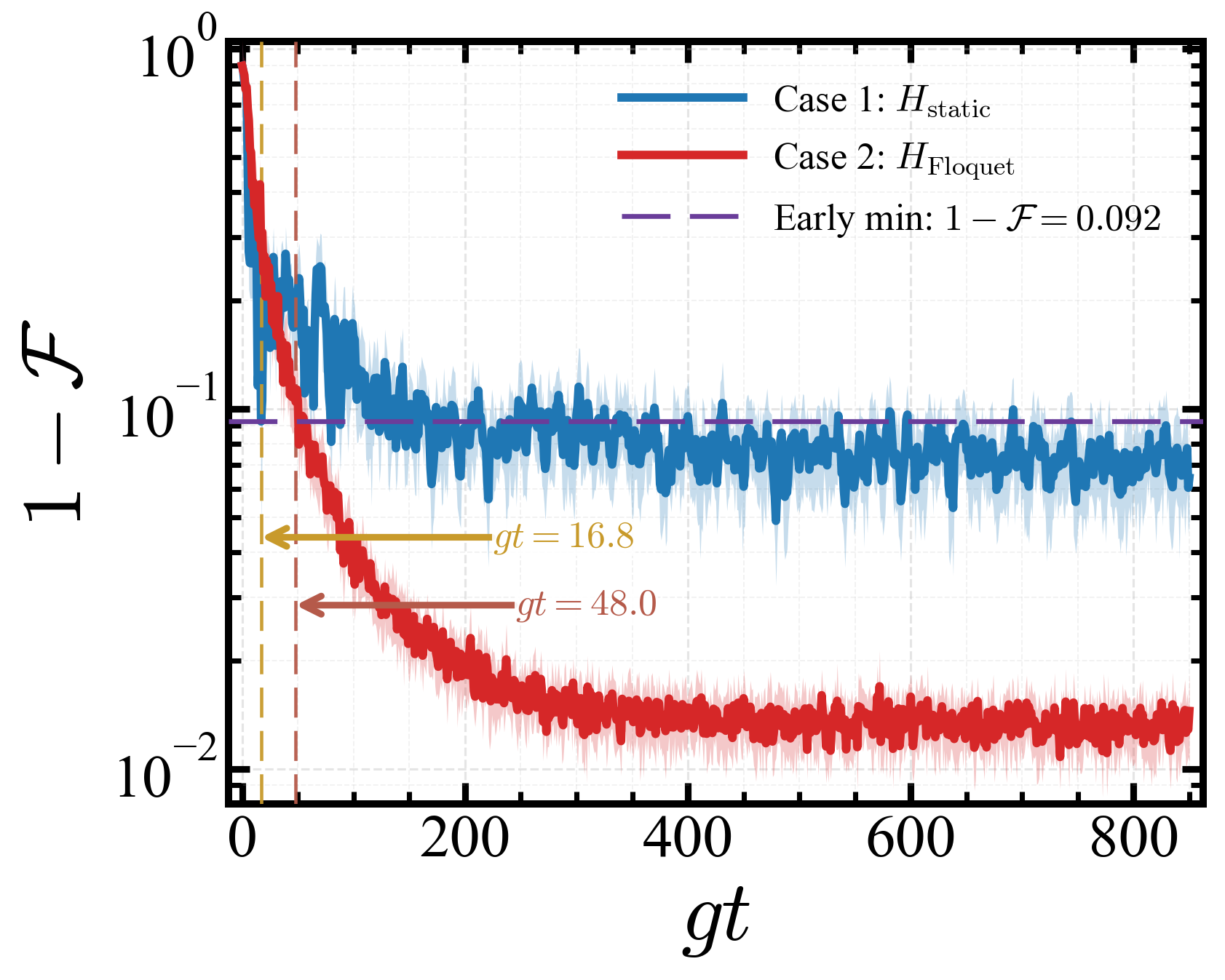}
        \centerline{\small (b)}
    \end{minipage}
    
    \caption{\edit{
    Infidelity $1-\mathcal{F}$ for the static Dicke evolution of Eq.~\eqref{eq:Lab_frame}, $H_{\mathrm{static}}$, and the alternating Floquet sequence of Eq.~\eqref{eq:floquet_step}, $H_{\mathrm{Floquet}}$. The Haar-state sampling and fidelity estimator are the same as in Eqs.~\eqref{eq:fid_def} and \eqref{eq:fid_prob_estimators}. 
    (a) Long-time data for $N_A=4$, $\phi=30$, $\delta/g=0.267$, and $\Omega_A/g=3.2$. 
    (b) Short-time data for $N_A=4$, 
    $\phi=30$,
    $\delta/g=0.40$, and $\Omega_A/g=3.78$. The violet dashed horizontal line in (b) marks the minimum value of $1-\mathcal{F}$ in the Hamiltonian case found for $gt<50$. The gold and brick-red dashed vertical lines indicate, respectively, the time of this minimum, $gt=16.8$, and the first time at which the Floquet sequence reaches $1-\mathcal{F}\leq 0.1$, $gt=48.0$.}
    }
    \label{fig:supp_floquet_xp_NA4}
\end{figure}

\edit{The phase modulation changes the scrambling dynamics because the two quadrature forces do not commute. For small enough switching time with period $2\tau_{\mathrm{sw}}$, Eq.~\eqref{eq:floquet_step} gives the effective hamiltonian
\begin{equation}
H_{\mathrm{eff}}
=
\frac{
H_{\mathrm D}(0)
+
H_{\mathrm D}(\pi/2)
}{2}
-
\frac{i\tau_{\mathrm{sw}}}{4}
\left[
H_{\mathrm D}(\pi/2),
H_{\mathrm D}(0)
\right]
+
O(\tau_{\mathrm{sw}}^2).
\label{eq:floquet_effective_hamiltonian}
\end{equation}
Since $[b+b^{\dagger},i(b^{\dagger}-b)]=2i$, the commutator in Eq.~\eqref{eq:floquet_effective_hamiltonian} contains a collective nonlinear contribution proportional to $S_z^2$, along with additional terms from the transverse field and the bosonic rotation. The alternating sequence therefore uses noncommuting spin--boson generators rather than a single static force.}

\edit{Figure~\ref{fig:supp_floquet_xp_NA4} compares the static Dicke evolution with the alternating Floquet sequence for $N_A=4$. Panel (a) shows the long-time point, $\delta/g=0.267$ and $\Omega_A/g=3.2$, where the alternating sequence reaches a lower infidelity than the static evolution over the late-time window. Panel (b) shows the short-time point, $\delta/g=0.40$ and $\Omega_A/g=3.78$. In this case the orange dashed line marks the minimum infidelity found for $gt<50$. The improvement is therefore not tied only to a late-time recurrence of the static Hamiltonian, but can also be obtained in an earlier operating window when the drive parameters are changed.}

\section{Experimental implementation details}
\label{sec:supp_impl_details}

We show how the trapped-ion Hamiltonian
\begin{equation}
H=\delta\, b_B^\dagger b_B
+\sum_{l=A,C,M}\Big(\Omega_l S_l^x+\Delta_l S_l^z
+\frac{2\widetilde g_l}{\sqrt{N_T}}(b_B+b_B^\dagger)S_l^z\Big)
\label{eq:supp_H_total}
\end{equation}
implements the protocol operations: a forward Dicke evolution $U_{AB}(t)$ on $(A,B)$ and an inverse Dicke evolution
$U^\dagger_{CB}(t)$ on $(C,B)$. We consider three collective spin ensembles $l\in\{A,C,M\}$ (spin-$1/2$ ions) coupled to a
shared COM mode $B$. The protocol uses $A$ and $C$ (with $N_A=N_C\equiv N$, fully symmetric manifolds), while $M$ is an
auxiliary ensemble used only for boson readout (frozen during the unitaries). Here
\begin{equation}
N_T\equiv N_A+N_C+N_M=2N+N_M
\end{equation}
is the total number of ions. It is convenient to define
\begin{equation}
g_l \equiv \widetilde g_l\sqrt{\frac{N_l}{N_T}}
\qquad\Rightarrow\qquad
\frac{2\widetilde g_l}{\sqrt{N_T}}=\frac{2g_l}{\sqrt{N_l}},
\label{eq:supp_g_def}
\end{equation}
so that for $l\in\{A,C\}$ one has $(2\widetilde g_l/\sqrt{N_T})=(2g_l/\sqrt{N})$. We assume $\Omega_l,\Delta_l$ are independently
tunable while $\widetilde g_l$ are fixed (equivalently $g_l$ are fixed once $N_T$ is fixed), furthermore $g_A=g_C$.

\subsection{Displacing the bosonic mode (frozen $M$)}
\label{sec:supp_displace_boson}

Set $\Omega_M=0$ and prepare $M$ in $\ket{\downarrow}^{\otimes N_M}$, so $S_M^z=-N_M/2$ and
\begin{equation}
\frac{2\widetilde g_M}{\sqrt{N_T}}(b_B+b_B^\dagger)S_M^z
\to -\frac{\widetilde g_M N_M}{\sqrt{N_T}}(b_B+b_B^\dagger)
\equiv -g_M\sqrt{N_M}(b_B+b_B^\dagger),
\end{equation}
where in the last step we used $g_M=\widetilde g_M\sqrt{N_M/N_T}$. Thus (up to a constant),
\begin{equation}
H_{B,M}=\delta\, b_B^\dagger b_B-g_M\sqrt{N_M}(b_B+b_B^\dagger).
\label{eq:supp_boson_with_linear}
\end{equation}
Define a displacement
\begin{equation}
b_B=\tilde b_B+\beta,\qquad \beta\in\mathbb{R},
\label{eq:supp_b_shift_def}
\end{equation}
which removes the linear term for
\begin{equation}
\beta=\frac{g_M\sqrt{N_M}}{\delta}
=\frac{\widetilde g_M N_M}{\delta\sqrt{N_T}}
\qquad (\delta\neq 0).
\label{eq:supp_beta}
\end{equation}
Then, for $l\in\{A,C\}$,
\begin{align}
\frac{2\widetilde g_l}{\sqrt{N_T}}(b_B+b_B^\dagger)S_l^z
&=\frac{2\widetilde g_l}{\sqrt{N_T}}(\tilde b_B+\tilde b_B^\dagger)S_l^z
+\frac{4\widetilde g_l\beta}{\sqrt{N_T}}S_l^z,
\end{align}
so the shift is absorbed into
\begin{equation}
\Delta_l^{\rm eff}
=\Delta_l+\frac{4\widetilde g_l\beta}{\sqrt{N_T}}
=\Delta_l+\frac{4\widetilde g_l\widetilde g_M}{\delta}\frac{N_M}{N_T},
\qquad l\in\{A,C\}.
\label{eq:supp_Delta_eff}
\end{equation}
(Equivalently, using \eqref{eq:supp_g_def},
$\Delta_l^{\rm eff}=\Delta_l+\frac{4g_l\beta}{\sqrt{N_l}}$.)

The Hamiltonian used for the unitary operations is
\begin{equation}
H_{\rm op}
=\delta\,\tilde b_B^\dagger\tilde b_B
+\sum_{l=A,C}\Big(\Omega_l S_l^x+\Delta_l^{\rm eff}S_l^z
+\frac{2\widetilde g_l}{\sqrt{N_T}}(\tilde b_B+\tilde b_B^\dagger)S_l^z\Big).
\label{eq:supp_H_op_displaced}
\end{equation}
If the inverse step uses $\delta\to-\delta$, then $\beta\to-\beta$ unless the force also flips; applying a fast
$R_x(\pi)$ pulse on $M$ between steps sends $S_M^z\to -S_M^z$ and keeps $\beta$ unchanged.

\subsection{Forward Dicke evolution on $(A,\tilde b_B)$ }
\label{sec:supp_AB_dicke_refocus_C}

The spectator coupling is
\begin{equation}
V_{C\tilde B}
=\frac{2\widetilde g_C}{\sqrt{N_T}}(\tilde b_B+\tilde b_B^\dagger)S_C^z
=\frac{2g_C}{\sqrt{N}}(\tilde b_B+\tilde b_B^\dagger)S_C^z.
\end{equation}
We lock $C$ by setting $\Omega_C=\Omega_{\rm lock}$ and $\Delta_C^{\rm eff}=0$, so that $H_{\rm lock}=\Omega_{\rm lock}S_C^x$.
In the interaction picture,
\begin{equation}
S_C^z(t)=S_C^z\cos(\Omega_{\rm lock}t)+S_C^y\sin(\Omega_{\rm lock}t),
\end{equation}
so the first Magnus term is bounded ($\sim g_C/\Omega_{\rm lock}$). The leading secular correction comes from the second Magnus term
and yields an effective spectator Hamiltonian
\begin{equation}
H_{\rm corr}^{(1)}
\simeq
-\frac{2g_C^2}{N\,\Omega_{\rm lock}}(\tilde b_B+\tilde b_B^\dagger)^2\,S_C^x,
\label{eq:supp_lock_correction}
\end{equation}
i.e. suppressed as $g_C^2/\Omega_{\rm lock}$.

With $C$ locked, the active Hamiltonian is
\begin{equation}
H_{A\tilde B}
=\delta\,\tilde b_B^\dagger\tilde b_B+\Omega_A S_A^x+\Delta_A^{\rm eff}S_A^z
+\frac{2\widetilde g_A}{\sqrt{N_T}}(\tilde b_B+\tilde b_B^\dagger)S_A^z
=\delta\,\tilde b_B^\dagger\tilde b_B+\Omega_A S_A^x+\Delta_A^{\rm eff}S_A^z
+\frac{2g_A}{\sqrt{N}}(\tilde b_B+\tilde b_B^\dagger)S_A^z.
\end{equation}
Apply $R_A=e^{-i\frac{\pi}{2}S_A^y}$ so that $S_A^x\mapsto S_A^z$ and $S_A^z\mapsto -S_A^x$, giving
\begin{equation}
H_{A\tilde B}^{\rm rot}
=\delta\,\tilde b_B^\dagger\tilde b_B+\Omega_A S_A^z-\Delta_A^{\rm eff}S_A^x
-\frac{2g_A}{\sqrt{N}}(\tilde b_B+\tilde b_B^\dagger)S_A^x.
\end{equation}
Setting $\Delta_A^{\rm eff}=0$ yields the Dicke form
\begin{equation}
H_{A\tilde B}^{\rm Dicke}
=\omega_z S_A^z+\delta\,\tilde b_B^\dagger\tilde b_B
-\frac{2g_A}{\sqrt{N}}S_A^x(\tilde b_B+\tilde b_B^\dagger),
\qquad \omega_z\equiv\Omega_A,
\label{eq:supp_AB_Dicke}
\end{equation}
and $U_{AB}(t)=e^{-itH_{A\tilde B}^{\rm Dicke}}$.

\subsection{Inverse Dicke evolution on $(C,\tilde b_B)$}
\label{sec:supp_inverse_dicke_refocus_A}

We need
\begin{equation}
U_{CB}^\dagger(t)=e^{+itH_{C\tilde B}^{\rm Dicke}}=e^{-it(-H_{C\tilde B}^{\rm Dicke})}.
\end{equation}
Lock $A$ by choosing $\Omega_A=\Omega_{\rm lock}^{(A)}$, $\Delta_A^{\rm eff}=0$; the residual spectator effect is
$\sim g_A^2/\Omega_{\rm lock}^{(A)}$ (as above).

The active Hamiltonian on $(C,\tilde b_B)$ is
\begin{equation}
H_{C\tilde B}
=\delta\,\tilde b_B^\dagger\tilde b_B+\Omega_C S_C^x+\Delta_C^{\rm eff}S_C^z
+\frac{2\widetilde g_C}{\sqrt{N_T}}(\tilde b_B+\tilde b_B^\dagger)S_C^z
=\delta\,\tilde b_B^\dagger\tilde b_B+\Omega_C S_C^x+\Delta_C^{\rm eff}S_C^z
+\frac{2g_C}{\sqrt{N}}(\tilde b_B+\tilde b_B^\dagger)S_C^z.
\end{equation}
Rotate $C$ by $R_C=e^{-i\frac{\pi}{2}S_C^y}$ to obtain (for $\Delta_C^{\rm eff}=0$)
\begin{equation}
H_{C\tilde B}^{\rm Dicke}
=\omega_z S_C^z+\delta\,\tilde b_B^\dagger\tilde b_B
-\frac{2g_C}{\sqrt{N}}S_C^x(\tilde b_B+\tilde b_B^\dagger),
\qquad \omega_z\equiv\Omega_C.
\end{equation}
Thus we engineer
\begin{equation}
-\;H_{C\tilde B}^{\rm Dicke}
= -\omega_z S_C^z-\delta\,\tilde b_B^\dagger\tilde b_B
+\frac{2g_C}{\sqrt{N}}S_C^x(\tilde b_B+\tilde b_B^\dagger),
\label{eq:supp_neg_dicke_target}
\end{equation}
using: (i) $\delta\to-\delta$ (opposite force detuning from the COM mode); (ii) $\Omega_C\to-\Omega_C$ (carrier phase shift by $\pi$);
(iii) conjugation by $Z_C=e^{-i\pi S_C^z}$ so $S_C^x\to -S_C^x$.
Local rotations ($R_A,R_C$, and any auxiliary $\pi$ pulses such as $Z_C$ and the $M$ flip used to keep $\beta$ fixed) can be undone
after each step.

\subsection{Effective dispersive (QND) measurement Hamiltonian}
\label{sec:supp_disp_meas_derivation}

To project onto the reference coherent state, note that in the lab frame the target boson state is $|\phi+\beta\rangle_B$.
Apply $D[-(\phi+\beta)]$ so $|\phi+\beta\rangle_B\mapsto|0\rangle_B$, and postselect on vacuum. We work with the lab operator $b_B$.

During the measurement, keep $A$ and $C$ idle via spin locking (as above), and engineer a dispersive coupling between $(M,B)$.
With $\Delta_M=0$,
\begin{equation}
H_{MB}
=
\delta\, b_B^\dagger b_B
+\Omega_M S_M^x
+\frac{2\widetilde g_M}{\sqrt{N_T}}(b_B+b_B^\dagger)S_M^z
=
\delta\, b_B^\dagger b_B
+\Omega_M S_M^x
+\frac{2g_M}{\sqrt{N_M}}(b_B+b_B^\dagger)S_M^z .
\label{eq:supp_H_MB_lab}
\end{equation}
Rotate $M$ by $R_M=e^{-i\frac{\pi}{2}S_M^y}$ so that $S_M^x\mapsto S_M^z$ and $S_M^z\mapsto -S_M^x$:
\begin{equation}
H_{MB}^{\rm rot}
=
\delta\, b_B^\dagger b_B
+\Omega_M S_M^z
-\frac{2g_M}{\sqrt{N_M}}(b_B+b_B^\dagger)S_M^x,
\label{eq:supp_H_MB_rot}
\end{equation}
i.e.
\begin{equation}
V
=
-\frac{g_M}{\sqrt{N_M}}(b_B+b_B^\dagger)(S_M^+ + S_M^-).
\end{equation}
In the interaction picture of $H_0=\delta\,b_B^\dagger b_B+\Omega_M S_M^z$,
\begin{align}
V(t)
&=
-\frac{g_M}{\sqrt{N_M}}
\Big[
b_B S_M^+\,e^{+i(\Omega_M-\delta)t}
+b_B^\dagger S_M^-\,e^{-i(\Omega_M-\delta)t}
\nonumber\\[-1mm]
&\hspace{2.6cm}
+b_B S_M^-\,e^{-i(\Omega_M+\delta)t}
+b_B^\dagger S_M^+\,e^{+i(\Omega_M+\delta)t}
\Big],
\end{align}
with detunings $\Delta_-=\Omega_M-\delta$ and $\Delta_+=\Omega_M+\delta$. In the dispersive regime
\begin{equation}
|\Delta_\pm|\gg \frac{g_M}{\sqrt{N_M}}\sqrt{n_{\max}+1},
\label{eq:supp_disp_condition_meas}
\end{equation}
a second-order Schrieffer--Wolff transformation yields
\begin{equation}
H_{\rm eff}^{\rm rot}
\simeq
\delta\,b_B^\dagger b_B
+\Omega_M S_M^z
+\chi_M\Big(b_B^\dagger b_B+\frac{1}{2}\Big)S_M^z
+\chi_M^{(2)}(b_B^2+b_B^{\dagger 2})S_M^z,
\label{eq:supp_Heff_rot_general}
\end{equation}
where
\begin{equation}
\chi_M=\frac{2g_M^2}{N_M}\Big(\frac{1}{\Delta_-}+\frac{1}{\Delta_+}\Big)
=\frac{2\widetilde g_M^2}{N_T}\Big(\frac{1}{\Delta_-}+\frac{1}{\Delta_+}\Big),
\qquad
\chi_M^{(2)}=\frac{g_M^2}{N_M}\Big(\frac{1}{\Delta_-}-\frac{1}{\Delta_+}\Big)
=\frac{\widetilde g_M^2}{N_T}\Big(\frac{1}{\Delta_-}-\frac{1}{\Delta_+}\Big).
\label{eq:supp_chi_coeffs}
\end{equation}
For $|\Omega_M|\gg|\delta|$ one has $|\chi_M^{(2)}|/|\chi_M|=|\delta|/(2|\Omega_M|)\ll 1$, and in the interaction picture of
$\delta\,b_B^\dagger b_B$ the residual $(b_B^2+b_B^{\dagger 2})S_M^z$ term rotates at $\pm 2\delta$.
Neglecting it and moving to the rotating frame $\delta\,b_B^\dagger b_B
+(\Omega_M +\chi_M/2)S_M^z
$ gives a number-QND interaction:
\begin{equation}
H_{\rm meas}^{\rm rot}\simeq \chi_M\, b_B^\dagger b_B\,S_M^z.
\label{eq:supp_Hmeas_rot_QND}
\end{equation}
Undoing the basis rotation, the lab-frame form is
\begin{equation}
H_{\rm meas}^{\rm lab}\simeq \chi_M\, b_B^\dagger b_B\,S_M^x.
\label{eq:supp_Hmeas_lab_QND}
\end{equation}
In the main text we denote $g_M^{\rm disp}\equiv \chi_M$.

\section{Implementing the postselection step: projecting onto a boson coherent state}

In the Dicke model implementation of our protocol, a key step is the final postselection step, involving a projective measurement onto the initial coherent state $\lvert \phi\rangle_B$ of the boson mode. Such a measurement is difficult to carry out for a bosonic mode in many experiments, due to lacking direct boson projection capability. Here, we will describe a way to indirectly perform such a projective measurement by transferring information to a spin systems, in which projective measurements are more straightforward.

Our strategy may be summarized as follows:
\begin{enumerate}
    \item Displace the coherent state to a vacuum state, to set up a postselection onto the vacuum state.
    \item Perform a nondemolition measurement of the boson number via a coupling to a large external spin system, such that the spin state is left unaltered by the boson vacuum but not by the other Fock states.
    \item Perform a postselection onto the initial state of spins, which amounts to a postselection on to the vacuum state of bosons, and therefore onto the coherent state before the displacement operation.
\end{enumerate}

The first step amounts to the displacement operation $D(-\phi)$, so that
\begin{equation}
    {_B}\langle \phi\rvert = {_B}\langle 0\rvert D(-\phi).
\end{equation}

The second step involves a nondemolition coupling~\cite{QNDspin, QNDspinQDOT, SFFmeas} between the boson mode and a large ``measurement'' spin system $M$ with $N_M$ spin-$1/2$ particles, which we take to be:
\begin{equation}
    H_{\text{meas}} = g_M S_M^z b_B^\dagger b_B.
\end{equation}
It is convenient to write $S_M^z = \sum_j \sigma^z_j/2$, where $\sigma_j^z$ is the Pauli-$z$ operators for the $j$-th measurement spin.

Let us say that the measurement spin system is initialized to be completely polarized in the $x$-direction (a maximal $S_M^x$ eigenstate), in particular:
\begin{equation}
    \lvert \eta\rangle_M = \bigotimes_{j=1}^{N_M} \frac{\lvert 0\rangle_j + \lvert 1\rangle_j}{\sqrt{2}},
\end{equation}
where $\lvert s \rangle_j$ with $s \in \lbrace 0,1\rbrace$ denotes a $\sigma^z$ eigenstate of the $j$-th measurement spin. This state is an eigenstate of $S_M^x$ with eigenvalue $N_M/2$.
The boson mode may be in an arbitrary state $\lvert \varphi\rangle_B$, that we can expand in the eigenbasis $\lvert n\rangle_B$ of $b^\dagger_B b_B$ as:
\begin{equation}
    \lvert \varphi\rangle_B = \sum_{n=0}^{\infty} \varphi_n\lvert n\rangle_B.
\end{equation}

Our measurement procedure amounts to evolving the $BM$ system under $H_{\text{meas}}$ for a time $t_M$:
\begin{equation}
    e^{-i H_{\text{meas}} t_M}\lvert \varphi\rangle_B\otimes \lvert \eta\rangle_M = 2^{-N_M/2}\sum_{n=0}^{\infty} \sum_{s_j \in \lbrace 0, 1 \rbrace} \varphi_n e^{-i g_M t_M n \sum_j (2s_j-1)/2} \lvert n\rangle_B \otimes \lvert s_1 \ldots s_{N_M}\rangle_M.
    \label{eq:measurementstate_postselection}
\end{equation}

Finally, we project the state in Eq.~\eqref{eq:measurementstate_postselection} onto the initial state $\lvert \eta\rangle_M$ of the measurement spins (which can be implemented as an $S_M^x$ measurement), which leaves us with the boson state:
\begin{equation}
    \lvert \varphi^{\text{out}}\rangle_B = {_M}\langle \eta\rvert e^{-i H_{\text{meas}} t_M}\lvert \varphi\rangle_B\otimes \lvert \eta\rangle_M = \sum_{n=0}^{\infty}\varphi_n \cos^{N_M}\left(\frac{g_M t_M n}{2}\right) \lvert n\rangle_B.
    \label{eq:measuredoutputstate}
\end{equation}
with overall success probability:
\begin{equation}
    P_{\eta} = \langle \varphi^{\text{out}}\vert \varphi^{\text{out}}\rangle_B = \sum_{n=0}^{\infty} \lvert \varphi_n\rvert^2 \cos^{2 N_M}\left(\frac{g_M t_M n}{2}\right) \geq \lvert \varphi_0\rvert^2.
    \label{eq:projection_technique_success}
\end{equation}

For our projective measurement to succeed, we want that $\lvert \varphi^{\text{out}}\rangle_B \approx \lvert 0\rangle_B$, i.e., we want to maximize the overlap:
\begin{equation}
    P_0 \equiv \frac{\lvert \langle 0\vert \varphi^{\text{out}}\rangle_B\rvert^2}{\langle \varphi^{\text{out}}\vert \varphi^{\text{out}}\rangle_B} = \frac{\lvert \varphi_0\rvert^2}{P_{\eta}} \leq 1,
\end{equation}
which corresponds to minimizing $P_{\eta}$ by Eq.~\eqref{eq:projection_technique_success}.

Let us assume that $\lvert \varphi_n\rvert^2 = 0$ for $n > n_{\max}$, corresponding to some maximum allowed occupancy $n_{\max}$ of the boson mode (in practice, we want $\sum_{n > n_{max}} \lvert \varphi_n\rvert^2 \leq \delta$ for some small $\delta \geq 0$). Then, we can make $P_{\eta} \approx \lvert \varphi_0\rvert^2$ by choosing parameters $g, N_M, t_M$ such that the weight function is negligible (or ideally, $0$) for all relevant $n \neq 0$ (noting that at $n=0$, it is identically equal to $1$):
\begin{equation}
    \cos^{2 N_M}\left(\frac{g_M t_M n}{2}\right) \approx 0,\;\;\ \text{ for } n \in [1,n_{\max}] \cap \mathbb{Z}.
\end{equation}
To be more precise, we want to impose that $P_\eta \leq \lvert \varphi_0\rvert^2 + \epsilon$. Using $\sum_n \lvert \varphi_n\rvert^2 = 1$ in Eq.~\eqref{eq:projection_technique_success}, this amounts to imposing:
\begin{equation}
    (1-\lvert\varphi_0\rvert^2) \max_{n \in [1,n_{\max}] } \cos^{2 N_M}\left(\frac{g_M t_M n}{2}\right) \leq \epsilon.
    \label{eq:measurementconstraint}
\end{equation}

We can ensure that this is the case by choosing $n_{\max}$ to occur within one period of the (squared) cosine, in particular with the cosine having a magnitude no larger than for $n=1$ [in other words, the period of the squared cosine must be at least $n_{\max} + 1$], and by imposing the above constraint for $n=1$ (which is where the cosine attains a maximum in the desired range of $n$ in this case) and $\lvert \varphi_0\rvert^2 = 0$ (which will give the most stringent constraint among the range of values of $\varphi_0$). This imposes a choice of $g_M t_M$ such that
\begin{equation}
    0 < g_M t_M \leq \frac{\pi}{n_{\max} + 1}, 
\end{equation}
and a choice of $N_M$ such that $n=1$ satisfies Eq.~\eqref{eq:measurementconstraint}, i.e.
\begin{equation}
    N_M \geq \frac{\log \epsilon}{\log \cos^2(g_M t_M/2)}.
\end{equation}
Qualitatively, this means that we need short times (or a weak interaction strength) and a large number of measurement spins. In particular, on carrying out a small-$g_M t_M$ expansion, we get
\begin{equation}
    N_M \gtrsim \frac{1}{g_M t_M} \log \frac{1}{\epsilon} \geq \frac{n_{\max}+1}{\pi} \log \frac{1}{\epsilon}.
    \label{eq:NMconstraint1}
\end{equation}

\subsection{An alternative protocol}

An alternative protocol with a different initial state $\lvert \xi\rangle_M$ of the measurement spins (that, however, is more difficult to prepare in practice) allows $N_M$ to be smaller than the above constraint \eqref{eq:NMconstraint1} for small values of $\epsilon$. In particular, we have
\begin{equation}
    \lvert \xi\rangle_M = \frac{1}{\sqrt{N_M+1}}\sum_{s=0}^{N_M} \lvert s\rangle_{M},
\end{equation}
where $\lvert s\rangle_M$ denotes an eigenstate of $S_M^z$ with $s$ excitations above the lowest value:
\begin{equation}
    S^z_M \lvert s\rangle_M = \left(s-\frac{N_M}{2}\right)\lvert s\rangle_M.
\end{equation}
Here, in analogy with Eqs.~\eqref{eq:measuredoutputstate} and \eqref{eq:projection_technique_success}, we have the final boson state
\begin{equation}
    \lvert \varphi^{\text{out}}\rangle_B = {_M}\langle \xi\rvert e^{-i H_{\text{meas}} t_M}\lvert \varphi\rangle_B\otimes \lvert \xi\rangle_M = \sum_{n=0}^{\infty}\varphi_n e^{-i n g_M t_M N_M} \frac{\sin\left(\frac{g_M n (N_M+1) t_M}{2}\right)}{(N_M+1)\sin\left(\frac{g_M n t_M}{2}\right)} \lvert n\rangle_B.
\end{equation}
with success probability
\begin{equation}
    P_{\xi} = \langle \varphi^{\text{out}}\vert \varphi^{\text{out}}\rangle_B = \sum_{n=0} \lvert \varphi_n\rvert^2 \frac{\sin^2\left(\frac{g_M n (N_M+1) t_M}{2}\right)}{(N_M+1)^2\sin^2\left(\frac{g_M n t_M}{2}\right)} \geq \lvert \varphi_0\rvert^2.
\end{equation}

Let us again assume that $\lvert \varphi_n\rvert^2 = 0$ for $n > n_{\max}$, corresponding to some maximum allowed occupancy $n_{\max}$ of the boson mode. Then, we can make $P_{\xi} \approx \lvert \varphi_0\rvert^2$ by choosing parameters $g, N_M, t_M$ such that the weight function is negligible (or ideally, $0$) for all relevant $n \neq 0$:
\begin{equation}
    \frac{\sin^2\left(\frac{g_M n (N_M+1) t_M}{2}\right)}{(N_M+1)^2\sin^2\left(\frac{g_M n t_M}{2}\right)} = 0,\;\;\ \text{ for } n \in [1,n_{\max}] \cap \mathbb{Z}.
    \label{eq:MeasurementProbabilitiesAlternate}
\end{equation}
We can ensure that this is the case by choosing $N_M$ to satisfy:
\begin{equation}
    g_M t_M (N_M+1) = 2\pi,
    \label{eq:meas_alternateconstraint1}
\end{equation}
so that the numerator is always $0$ at $n \in \mathbb{Z}$, and choosing $g_M t_M$ to satisfy
\begin{equation}
    g_M t_M n_{\max} < 2 \pi,
    \label{eq:meas_alternateconstraint2}
\end{equation}
so that the denominator is nonzero for $0 < n \leq n_{\max}$, ensuring that $P_\xi = \lvert \varphi_0\rvert^2$. Therefore, if the above conditions can be met ``exactly'' (to within errors) in an experiment, our steps lead to an exact projection $\lvert \phi\rangle_B\langle \phi\rvert$ onto the boson coherent state. From Eqs.~\eqref{eq:meas_alternateconstraint1} and \eqref{eq:meas_alternateconstraint2}, the number of measurement spins required to satisfy Eq.~\eqref{eq:MeasurementProbabilitiesAlternate} is:
\begin{equation}
    N_M \geq n_{\max},
\end{equation}
which is smaller than Eq.~\eqref{eq:NMconstraint1} for $\epsilon \lesssim e^{-\pi}  \approx 0.0432$.

\subsection{\edit{Amplitude amplification of the probabilistic decoder}}

\edit{The $A\to C$ direction of the protocol is postselected by projecting the auxiliary
system $B$ back onto its initial state. Under the isometric-encoding condition
discussed above, the corresponding success probability is approximately
input-state independent, $p\simeq p_{\lambda}$. For maximally scrambling
dynamics one typically has $p_{\lambda}\sim d_A^{-2}$, so direct repetition of
the postselected decoder incurs an average sampling overhead
$\mathcal{O}(p^{-1})\sim\mathcal{O}(d_A^2)$.}

\edit{It is nevertheless useful to ask whether this sampling overhead can be traded
for coherent circuit depth. As in the deterministic recovery construction of
Yoshida and Kitaev \cite{YoshidaKitaev}, this can be done by replacing repeated
postselection with coherent amplitude amplification. Consider a fixed pure input
state $\ket{\psi}_A$, a pure auxiliary state $\ket{\phi}_B$, and a pure initial
state $\ket{\gamma}_C$ of the receiving system without loss of generality. Define
\begin{equation}
W_{ABC}=(I_A\otimes U_{CB}^{\dagger})(U_{AB}\otimes I_C),
\label{eq:W_amp}
\end{equation}
together with
\begin{equation}
\ket{\Psi_0}=\ket{\psi}_A\otimes\ket{\phi}_B\otimes\ket{\gamma}_C,
\qquad
\ket{\Psi_1}=W_{ABC}\ket{\Psi_0}.
\label{eq:Psi01_amp}
\end{equation}
The successful branch is postselected by projecting $B$ back onto
$\ket{\phi}_B$. We therefore introduce
\begin{equation}
P_{\phi}=I_A\otimes\ket{\phi}\bra{\phi}_B\otimes I_C,
\qquad
p=\bra{\Psi_1}P_{\phi}\ket{\Psi_1}.
\label{eq:p_amp}
\end{equation}
Here $p$ is the bare success probability of the postselected decoder. The
corresponding normalized success and failure states are
\begin{equation}
\ket{\Psi_{\mathrm{succ}}}
=\frac{P_{\phi}\ket{\Psi_1}}{\sqrt{p}},
\qquad
\ket{\Psi_{\mathrm{fail}}}
=\frac{(I_{ABC}-P_{\phi})\ket{\Psi_1}}{\sqrt{1-p}},
\label{eq:succ_fail_amp}
\end{equation}
so that
\begin{equation}
\ket{\Psi_1}
=
\sqrt{p}\ket{\Psi_{\mathrm{succ}}}
+
\sqrt{1-p}\ket{\Psi_{\mathrm{fail}}}.
\label{eq:decomp_amp}
\end{equation}
Writing
\begin{equation}
\sin\theta=\sqrt{p},
\qquad
\theta\in(0,\pi/2),
\label{eq:theta_amp}
\end{equation}
the state $\ket{\Psi_1}$ lies in the two-dimensional subspace spanned by
$\ket{\Psi_{\mathrm{succ}}}$ and $\ket{\Psi_{\mathrm{fail}}}$.}

\edit{The circuit implementation uses a reflection about the known postselected
subspace,
\begin{equation}
R_{\phi}=I_{ABC}-2P_{\phi},
\label{eq:Rphi_amp}
\end{equation}
and its conjugate under the coherent decoder. The Grover iterate is
\begin{equation}
G=W_{ABC}R_{\phi}W_{ABC}^{\dagger}R_{\phi}.
\label{eq:G_amp}
\end{equation}
The rightmost $R_{\phi}$ marks the postselected branch after the bare decoder has
been applied. The factor $W_{ABC}R_{\phi}W_{ABC}^{\dagger}$ is the same
reflection conjugated by the decoder. Thus each amplification step requires one
application of $W_{ABC}$, one application of $W_{ABC}^{\dagger}$, and two
reflections about the known auxiliary state $\ket{\phi}_B$. This is the form
used in the numerical circuit, and it avoids a reflection about the unknown input
state $\ket{\psi}_A$.}

\edit{After $k$ amplification steps, the state before the final postselection is
\begin{equation}
\ket{\Psi_k}=G^kW_{ABC}\ket{\Psi_0},
\label{eq:Psik_amp}
\end{equation}
with success probability
\begin{equation}
p_k=\bra{\Psi_k}P_{\phi}\ket{\Psi_k}.
\label{eq:pk_def_amp}
\end{equation}
For the bidirectional SWAP benchmark, the ideal postselected state is
\begin{equation}
\ket{\Psi_{\mathrm{tar}}}
=\ket{\gamma}_A\otimes\ket{\phi}_B\otimes\ket{\psi}_C.
\label{eq:target_amp}
\end{equation}
The conditional fidelity after $k$ amplification steps is evaluated by comparing
the normalized postselected state with $\ket{\Psi_{\mathrm{tar}}}$. Equivalently,
when $p_k$ is nonzero,
\begin{equation}
F_k=
\frac{\left|\bra{\Psi_{\mathrm{tar}}}P_{\phi}\ket{\Psi_k}\right|^2}{p_k}.
\label{eq:Fk_amp}
\end{equation}
}

\edit{In the ideal amplitude-amplification limit, the two reflections generate a
rotation in the success--failure subspace. Up to the conventional sign choice of
the reflections,
\begin{equation}
G^k\ket{\Psi_1}
=
\sin\bigl((2k+1)\theta\bigr)\ket{\Psi_{\mathrm{succ}}}
+
\cos\bigl((2k+1)\theta\bigr)\ket{\Psi_{\mathrm{fail}}}.
\label{eq:rotation_amp}
\end{equation}
Accordingly, after $k$ amplification steps the success probability becomes
\begin{equation}
p_k
=
\|P_{\phi}G^k\ket{\Psi_1}\|^2
=
\sin^2\bigl((2k+1)\theta\bigr).
\label{eq:pk_amp}
\end{equation}
Away from the ideal isometric limit, different input components can acquire
slightly different amplification angles. In that case, Eq.~\eqref{eq:pk_amp}
should be read as a benchmark formula rather than an exact identity.}

\edit{A near-optimal choice is
\begin{equation}
k_{\mathrm{opt}}
\approx
\left\lfloor \frac{\pi}{4\theta}-\frac{1}{2} \right\rfloor,
\label{eq:kopt_amp}
\end{equation}
which gives $p_k\approx 1$ in the ideal coherent limit. In the small-$p$ regime,
$\theta\simeq\sqrt{p}$, and therefore
\begin{equation}
k_{\mathrm{opt}}=\mathcal{O}\left(\frac{1}{\sqrt{p}}\right).
\label{eq:k_scaling_amp}
\end{equation}
Combining $k_{\mathrm{opt}}=\mathcal{O}(p^{-1/2})$ with
$p\simeq p_{\lambda}\sim d_A^{-2}$, amplitude amplification can convert the
direct sampling overhead $\mathcal{O}(p^{-1})\sim\mathcal{O}(d_A^2)$ into a
coherent-depth overhead
\begin{equation}
k_{\mathrm{opt}}\sim\mathcal{O}(p^{-1/2})\sim\mathcal{O}(d_A).
\label{eq:dA_scaling_amp}
\end{equation}
The dimensional cost is therefore not removed, but it is reduced quadratically
in the standard Grover sense.}

\edit{In the ideal coherent limit, the amplification procedure increases the weight of
the postselected branch without changing the conditional state inside that
branch. Its usefulness is controlled by the balance between the conditional
fidelity of the decoder, the bare success probability $p$, and the coherent cost
of implementing $W_{ABC}$, $W_{ABC}^{\dagger}$, and $R_{\phi}$ repeatedly. For
moderate $d_A$, a few amplification steps can already provide a significant
increase in the success probability. For larger $d_A$, the required coherent
depth grows linearly with $d_A$, so the benefit must be weighed against finite
coherence time and control errors.}

\begin{figure}[!ht]
    \centering

    \begin{minipage}[t]{0.47\linewidth}
        \centering
        \includegraphics[width=\linewidth]{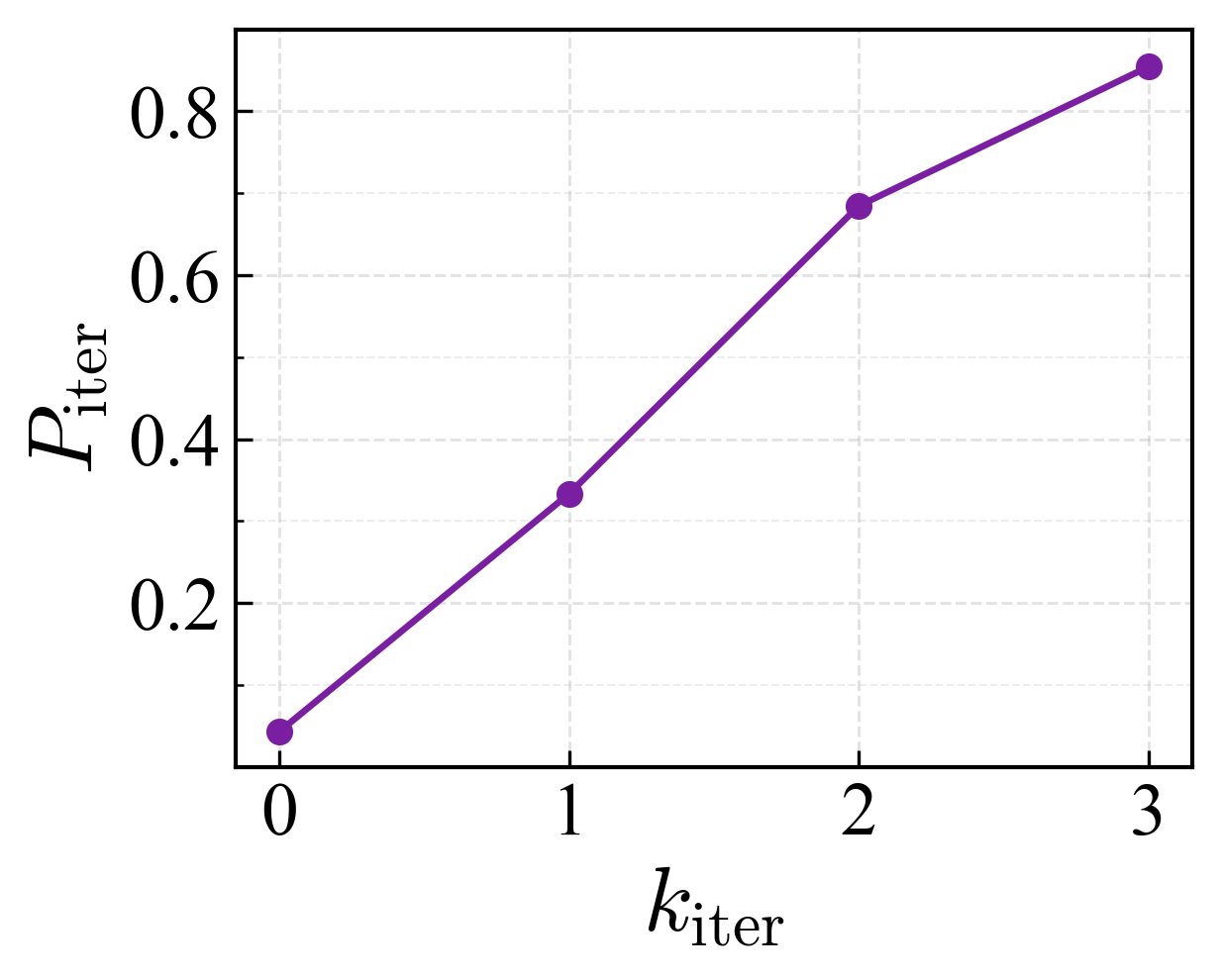}
        \vspace{-2mm}
        \centerline{\small (a)}
    \end{minipage}
    \hfill
    \begin{minipage}[t]{0.47\linewidth}
        \centering
        \includegraphics[width=\linewidth]{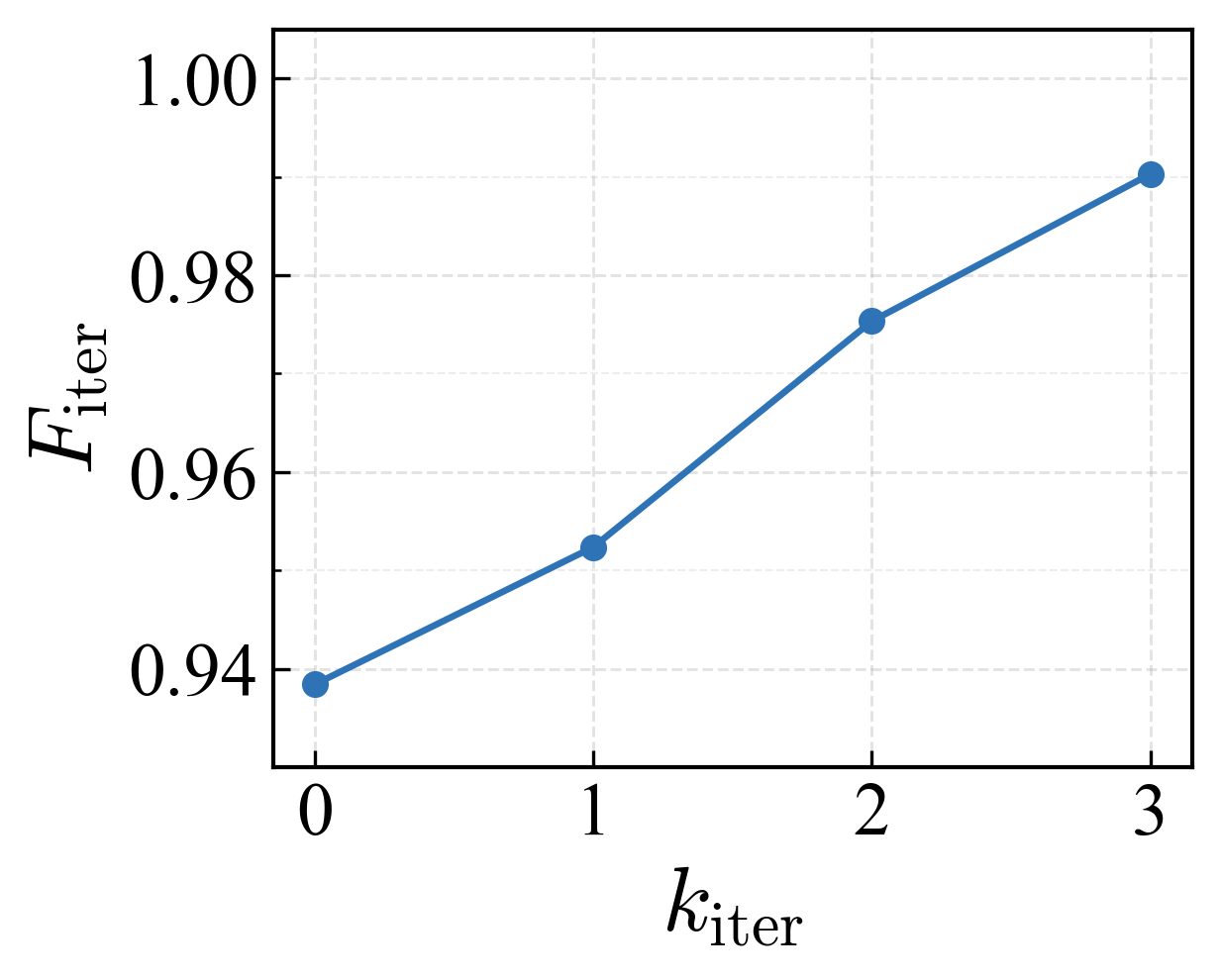}
        \vspace{-2mm}
        \centerline{\small (b)}
    \end{minipage}

    \caption{\edit{Grover-amplified probabilistic decoder at the late-time Dicke
    operating point. The data are obtained at
    $\delta/g=0.267$, $\omega_z/g=3.20$, and coherent-state displacement
    $\phi=30$, using the same late-time window $700 \leq gt \leq 850$ as in
    Fig.~2 in the main text. For each value of the number of
    amplification steps $k_{\rm iter}$, the plotted value is averaged over the
    sampled times in this window and over the input-state ensemble.
    (a) Heralding probability $P_{\rm iter}$ of the postselected decoder.
    The point $k_{\rm iter}=0$ corresponds to the bare probabilistic decoder,
    while increasing $k_{\rm iter}$ coherently amplifies the successful
    postselection branch. (b) Conditional decoder fidelity $F_{\rm iter}$
    evaluated on the same amplified branch. The amplification substantially
    increases the detection probability while boosting the conditional fidelity
    close to unity over the shown operating points. This demonstrates that,
    in the coherent limit, amplitude amplification can reduce the overhead
    from repeated postselection while improving the quality of
    the decoded branch.}}
    \label{fig:amp_amp_decoder}
\end{figure}

\edit{Figure~\ref{fig:amp_amp_decoder} shows a finite-dimensional numerical benchmark
of this construction. The bare decoder has a relatively small postselection
probability, but a few coherent amplification steps substantially increase
$p_{\rm iter}$ as a function of the number of iterations $k_{\rm iter}$. At the
same time, the conditional fidelity of the postselected branch increases monotonically
over the amplified operating points. This is consistent with the ideal coherent
picture in which amplitude amplification primarily changes the weight of the
successful branch in the final state.}

\bibliography{Scrambling_and_recovery}